\documentclass[11pt,a4paper,notitlepage]{report}
\usepackage{amsfonts}
\usepackage{graphics}  

\newcommand{\Rpi}{\ensuremath{{\cal R}_{\pi}}}

\begin{document}
\bibliographystyle{prsty}
\title{Frustrated Quantum Magnets}
\author{
Claire Lhuillier \\ Laboratoire de Physique Th{\'e}orique des
  Liquides,\\ Universit{\'e} P. et M. Curie and UMR 7600 of CNRS,\\ case 121, 4 Place
Jussieu, \\75252 Paris Cedex \\
email: claire.lhuillier@lptl.jussieu.fr
}
\date{\today}
\maketitle

\begin{abstract}
A description of different phases of two dimensional  magnetic
 insulators is given.

 The first chapters are devoted to the
understanding of the symmetry breaking mechanism in the
semi-classical N\'eel phases. Order by disorder selection is
illustrated.
All these phases break $SU(2)$ symmetry and are gapless 
phases with $\Delta S_z =1$ magnon excitations.

 Different gapful quantum phases  exist in two dimensions:
the Valence Bond Crystal phases (VBC) which have long range order in
local S=0 objects (either dimers in the usual Valence Bond
acception or quadrumers..), but also Resonating Valence Bond Spin
Liquids (RVBSL), which have no long range order in any local order
parameter and an absence of susceptibility to any local probe.
 VBC have gapful $\Delta S = 0, or 1$ excitations,
RVBSL on the contrary have deconfined spin-1/2 excitations.
Examples of these two kinds of quantum phases are given in
chapters 4 and 5.
A special class of magnets (on the kagome or pyrochlore lattices)
has an infinite local degeneracy in the classical limit: they
give birth in the quantum limit to different behaviors which are
illustrated and questionned in the last lecture.
\end{abstract}

\tableofcontents

\begin{chapter} {Introduction}

In this first chapter, we rapidly describe the basic 
knowledge on Heisenberg magnets to set the frame
and the notations of the next developments. Different excellent
text books can be used for a wider and slower introduction~\cite{m81,
c89a,auer94}.

\begin{section}{History}
The first microscopic model for magnetism goes back to Heisenberg
 when he realized in 1928, that the exchange energy between
electrons (introduced by Dirac and himself to explain
  the singlet triplet separation of the gaseous helium spectrum)
was also responsible for ferromagnetism.
Heisenberg and Dirac have first  suggested that exchange of electrons
 could be written in an
effective way in spin space through the use of  a spin Hamiltonian,
 which reads:
\begin{equation}
h(i,j)  = {\bf {S}}_i.{\bf {S}}_j
\label{Heis}
\end{equation}
where  ${\bf {S}}_i$, $ {\bf {S}}_j$ are the spin-1/2 operators 
of electrons i and j \footnote{It may be remembered that $h(i,j)$
is directly related to the spin-1/2 permutation operator $P(i,j)$ by:
\begin{equation}
P(i,j)= \frac{1}{2} + 2 {\bf {S}}_i.{\bf {S}}_j .
\end{equation}
Interesting from the conceptual point of view, this relationship
is also extremely useful for computational purposes.}.

This Hamiltonian allows the basic description of the 
magnetism of insulators on a lattice (Heisenberg and Van Vleck). In its simplest form, it is written:
\begin{equation}
{\cal H} = J \sum _{<i,j>} {\bf {S}}_i.{\bf {S}}_j
\label{Heis.Latt}
\end{equation}
 where the sum $<i,j>$ runs on pairs of next neighbor sites and
$J$ measures the 
strength of the effective coupling ( related to the tunnel frequency
 of a pair of electrons on two neighboring sites).

The Heisenberg Hamiltonian, as the more complex schemes of
interactions that will be studied in these lectures, are all
$SU(2)$ invariant (i.e. invariant in a global spin rotation).
On a given sample of $N$ spins, ${\cal H}$ commutes with the
total spin of the sample $S_{tot}$:
\begin{equation}
\left[{\cal H}, S^2_{tot}\right]=\left[{\cal H}, S^z_{tot}\right]=0.
\end{equation}
The eigen-states of ${\cal H}$  can thus be characterized by their
total energy and their total spin $S_{tot}$. Eigen-states of ${\cal H}$
with different $S_{tot}$ are a priori non degenerate. 
\end{section}

\begin{section}{$J<0$: the ferromagnet}
If $J$ is $<0$, the ground-state of (\ref{Heis.Latt}) can be 
readily written as:
\begin{equation}
 |F_z> = \prod_{i} |i, \, +>
\label{ferro}
\end{equation}
where the ket $|i, \, +>$
 indicates that spin at site i is in the eigen-state of $S^z_i$, 
with eigen-value $+\hbar /2 $ (in the following we will use $\hbar=1 $)
and the tensorial product involves each lattice spin.

It is easy to show that this state (\ref{ferro}) is an eigen-state of
 (\ref{Heis.Latt})
with the total energy:
\begin{equation}
E_{ferro} = N \frac{z}{2}  \; \frac{J}{4},
\end{equation}
where $N$ is the number of spins of the sample, and $z$ the
coordination number of the lattice.
It is also an eigen-state of  the total spin
$S^2_{tot}$  and its $z$-component $S^z_{tot}$with eigen-values
 $N/2 (N/2+1)$ and $M^z_{tot} = N/2$.
It minimizes the energy of each bond and is thus the ground-state of (\ref{Heis.Latt}).
This state can be written without ambiguity  as:
\begin{equation}
 |F_z> = |S_{tot} =N/2,\, M^z_{tot}=N/2>
\label{ferro2}
\end{equation}
It is degenerate with the $2N$ other eigen-values of $S^z_{tot}$,
 running from $N/2-1$ to $-N/2$.

You should notice that in the thermodynamic limit such a degeneracy 
is  negligible: 
the associate entropy of the extensive ground-state is 
${\cal O}(Ln(N))$,
 which does not contradict Nernst Theorem.

On a macroscopic sample, $|F_z>$ describes a system with a
 macroscopic magnetization pointing in the $z$ direction.
Using the total spin operator this state could be rotated
in any direction $\bf u$
defined by the Euler angles $(\theta, \phi)$ with respect to the
reference frame
\footnote{Remember that the third Euler angle $\chi$ 
 measures an overall degree
of  rotational freedom (``gauge freedom''), that can be  put  to 0 in
 this context.}.
We thus obtain the quantum description of a coherent state with
a magnetization in the  $\bf u$ direction:
\begin{equation} 
|F_{{\bf u}}> = e^{i S^z_ {tot} \phi}e^{i S^y_ {tot} \theta} |F_z>.
\label{coherentstate}
\end{equation}
Let me underline that this coherent state remains in the
ferromagnetic ground-state multiplicity and that its total
spin is well defined and equal to $N/2$.
 
The {\it semi-classical character of such states }
is embodied in the following property:
 the quantum overlap of two states pointing in
 different directions decreases exponentially with $S_{tot}$,
 that is with the system size $N$ \cite{auer94}. For macroscopic
 samples, the state of a ferromagnet can thus be described with
 classical words and concepts.
This can be said in another way:
the macroscopic spin, understood as a quantum observable,
 obeys quantum
 commutation relationships:

\begin{equation}
[S^x_{tot}, S^y_{tot}]= i S^z_{tot} 
\end{equation}
as $S_{tot}$ is proportional to $N$, the relative value of the ``quantum fluctuations''
(measured by the commutator/$N^2$) becomes negligible in the thermodynamic limit.

The selection of a special state as (\ref{ferro}) to describe the
 ferromagnetic ground-state is a
``minor (or trivial?) symmetry breaking'' of the problem.
By $SU(2)$ rotations this eigen-state generates all
the ground-state multiplicity, and this multiplicity only 
\footnote{Some authors deny
the use of the word ``symmetry breaking'' in that case, where the
 ground-state, does not involve any mixture of eigen-states with
different symmetries. They are certainly right from 
the theoretical point of view. In view of the experimental possibility
 in a macroscopic ferromagnet to point a given direction,
we nevertheless use this expression with an appropriate qualifier.}.
\end{section}

\begin{section}{$J>0$:  N{\'e}el antiferromagnet and spin gapped
Phases}
\begin{subsection}{A few historical markers}

If J is $>0$ the ground-state of (\ref{Heis.Latt}) is absolutely not
 obvious. In 1932, N{\'e}el
suggested that the description of experiments was consistent
 with a picture of the 
ground-state as a special arrangement of ferromagnetic sublattices 
with a zero total 
magnetization.

Let us examine the simplest case of the Heisenberg problem on the square lattice. 
This lattice
 may be partitioned in two sublattices A and B with a double unit cell.
Each spin of the A (resp. B) lattice  is exclusively coupled to the B
 (resp. A) lattice .
Such a problem is said bipartite \footnote{This is a class of problems,
for which exact results 
are available: Marshall (Peierls) theorem, and Lieb and coll theorems.
 See for example ref~\cite{auer94}.}.
In this case we usually write N{\'e}el's wave-function as:
\begin{equation} 
|Cl.\; N\acute{e}el\; w.f.> = \prod _{i \in A, \,j \in B}  |i\;,\;+>\, |j\;,\;->
\label{Nwf}
\end{equation}
This state has indeed a zero component of the total spin in the $z$
direction, but it is non zero in the $xy$ plane. This Ising state
has  maximal sublattice magnetizations: 
 $S_{A,B}=\frac{N}{4}, \; M_{A,B}= \pm \frac{N}{4}$.
The Ising state with a zero component of the total spin in the ${\bf u}$
direction, defined by the Euler angles $\theta$ and $\phi$, is indeed:
\begin{equation} 
|Cl.\; N\acute{e}el\; w.f.; {\bf u}> = e^{i S^z_ {tot} \phi}e^{i S^y_ {tot} \theta}|Cl.\; N\acute{e}el\; w.f.>
\label{Nwfu}
\end{equation}
 In this antiferromagnetic case the idea that it is possible to restore
the overall symmetry of the problem by  rotation of (\ref{Nwf})
and averaging
has  more far fetching consequences that is usually thought!

In his biography N{\'e}el told that he had to face strong skepticism and objections 
specially from  C.J. G\"{o}rter (colloquium in Leyden 
at the Kammerlingh Onnes Lab; 1932).
 It seems that L. Landau equally rapidly
discarded this special variational wave-function with the same
 objections as C. J. G\"{o}rter.

I have not had access to authenticated sources, but
  the objections were probably of two kinds:
\begin{itemize}
\item The N{\'e}el state strongly breaks the $SU(2)$ symmetry of
the Hamiltonian and cannot be a good candidate to describe an
eigen-state,
\item The existence of ferromagnetic sublattices is not proved
and elementary quantum mechanics seems in strong disagreement
with this assumption.
\end{itemize}

\end{subsection}

\begin{subsection} {$SU(2)$ symmetry breaking of the N{\'e}el states.}
\label{SBneel}

As I will  explicit below and explain in details in the next chapter,
 the N{\'e}el states breaks the 
$SU(2)$ symmetry of the Heisenberg Hamiltonian (and the lattice geometrical symmetries) in 
a radical way (quite different from the ferromagnetic
case).
 

{\it The classical N{\'e}el state is not an eigen-state of the total
spin, and as such it can only be described as a linear
combination of many eigenstates of (\ref{Heis.Latt})}.

In order to have an elementary  view of this question let us rephrase
 N{\'e}el wave-function in simple quantum terms:
two ferromagnetic sublattices A and B, defined by their total
spins $S_{A,B} = N/4$ are to be coupled in such a way that:
\begin{equation}
(S^z_A + S^z_B)|Cl.\; N\acute{e}el\; w.f.> =0
\end{equation}
We know from elementary spin algebra that there are $(N/2+1)$
$SU(2)$ invariant ways to do this: the different states resulting
of this coupling can be labeled in an unique way
by their total spin $S_{tot}$, which can range (for even
N) from $S_{tot}=0$ to $S_{tot}=N/2$. They can be written  in an
unambiguous way under the form $|\frac{N}{4},\frac{N}{4},S_{tot},M_S>$.
 In any of these subspaces,
one can indeed select the $M_S=0$ component of the total spin,
thus fulfilling N{\'e}el prescription.

Elementary spin algebra leads to the following expression for the
classical N{\'e}el state wave-function:
\begin{equation} 
|Cl.\; N\acute{e}el\; w.f.> = \sum_{S,M_S} {\frac{(-1)^{M_S}}
{\sqrt{ 2S + 1}}}
\left(   \begin{array}{ccc}
         N/4 & N/4 & S\\
	 N/4 & - N/4 &  M_S\\
          \end {array}  \right)
 |S_A,S_B,S,M_S>
\label{Nwf3}
\end{equation}
where $S$ runs through the $N/2 +1$ possible values of $S_{tot}$,
and in general for each value of $S$, $M_S$ runs from $-S$ to
$+S$. Here the selection rule on the $z$ components of the spins
 implies that $M_S=0$.
In this expression the coefficients
$ \left( \begin{array}{ccc}
         S_A & S_B & S\\
	 M_A &  M_B &  M_S\\
          \end {array} \right)$
 are known as  Wigner
``3j'' symbols.  These
coefficients are the coefficients of the unitary transformation
which transforms the uncoupled sublattice spins $S_{A,B}$ 
 to the $SU(2)$ invariant coupled combinations.
 The Wigner ``3j'' symbols can be calculated by
elementary algebra, they are tabulated in books and in 
computer libraries.

Comparison of this antiferromagnetic coherent state (\ref{Nwf3}) to the
ferromagnetic one (\ref{ferro}), (\ref{ferro2})  shows explicit
qualitative differences: the ferromagnetic state is a state with a
definite total spin $N/2$, whereas (\ref{Nwf3}) involves components
with total spin ranging from $0$ to $N/2$. This shows that N{\'e}el 
wave-function can at best be described as a linear combination
of a large number of eigen-states of ${\cal H}$.

For bipartite lattices a theorem originally
due to Hulthen (1938)\cite{h38}, Marshall (1955)\cite{m55}
 and strengthened by Lieb and
Mattis (1962)\cite{lm62} states that the absolute
ground-state of the antiferromagnetic Heisenberg
(\ref{Heis.Latt}) (and of more general antiferromagnetic models
respecting the bipartition of the lattice) is unique and has
total spin zero. Moreover the ground-state energies in each $S$
sector are ordered accordingly to $S_{tot}$:
\begin{equation}
\forall S^{'}_{tot} > S_{tot} \hspace*{1cm} E_{0}(S^{'}_{tot}) > E_{0}(S_{tot}).
\label{lm_ordering}
\end{equation}

From that point, we might infer that  the $|S_A=N/4, S_B =N/4,
0, 0>$ state would be  a good starting point to
describe the absolute ground-state $|\Psi_0>$ of (\ref{Heis.Latt}),
 and forget all the other components of the classical N{\'e}el state
(\ref{Nwf3}).  But in such a point of view, we
lose the foundations for the semi-classical
approaches: a state with total spin 0 does not allow to point any
direction in spin space. According to Wigner Eckart theorem,
  the
three components of the sublattice magnetizations (as the
components of any vector) are simultaneously and exactly
zero in such a state:
\footnote{P. W. Anderson and many authors have written that this
exact property seems paradoxical and contradictory with
observations and with the semi-classical approaches (either the
simplest spin wave approach, as well as, the more sophisticated
field theoretical approaches laying upon a 
description of the ground-state by a coherent state).
 The second assumption is
theoretically correct, PW. Anderson knew indeed the answer to
the paradox and I
will describe in the next chapter a simple way to reconcile both
approaches. The second assumption about experimental observations
seems more questionable! This too will be briefly discussed in chapter 2.}
\begin{eqnarray}
<{\Psi}_{0}|{\bf S}_{A,B}|\Psi_0>\;\;\; \propto \;\; \;
<\Psi_0|{\bf S}_{tot}|\Psi_0> \;\; \;
= \;\;0.
\end{eqnarray}

To answer  G\"{o}rter and Landau objection, and support N{\'e}el
picture for quantum antiferromagnets, it is thus necessary to show that
eigen-states with different $S_{tot}$ appear in the exact spectrum
as the different $SU(2)$ invariant components of the supposed-to-be
quantum N{\'e}el
state and  are degenerate in the thermodynamic limit. In such a
limit, a quantum superposition of these eigen-states embodies the
``strong'' symmetry breaking associated to N{\'e}el's
scenario\footnote{This corresponds to the strict definition of a
symmetry breaking situation where the macroscopic order parameter
does not commute with the Hamiltonian. Technically this can happen
only by a mixing of different Irreducible Representations (IR) of the
broken symmetry group. (Elementary example of the broken
left-right symmetry in a one dimensional problem with an
Hamiltonian invariant under reflection).}. Such a mechanism has
been described in full length by P. W. Anderson in two books~\cite{a52,a63,a84}.

In this chapter and for the sake of simplicity only bipartite
lattices and collinear N{\'e}el states are studied.\footnote{ Qualitatively
the 3-sublattice N{\'e}el state on the triangular lattice has the
same properties as the collinear state~\cite{blp92,bllp94,le95}
with the minor difference  that
the SU(2) invariant components of the 3-sublattice N{\'e}el states
originate from the coupling of three macroscopic spins of
length $N/6$. The ground-state multiplicity is thus somewhat
larger, and of dimension ${\cal O}(N^3)$. In this case the
demonstration of Lieb-Mattis theorem on the quantum ordering
 of the ground-states
energy in each $S$ sector fails: the positive sign
property of the ground-state wave-function (Marshall property)
 is no more true.
Nevertheless, empirically we have  observed that the
ordering property
(\ref{lm_ordering}) was realized in exact spectra of
 most systems for large enough
sizes: the only restrictions come from systems with competing
interactions, very near a quantum critical transition to a ferromagnetic
state, where we have sometimes observed  some violations
of relation (\ref{lm_ordering}) for large $S$.} 

\end{subsection}

\begin{subsection}{Ferromagnetic sublattices, ``quantum
fluctuations'' and dimer pairing}

The second difficulty with N{\'e}el's scenario, the 
existence of ferromagnetic sublattices cannot be supported by 
quantum mechanics without specific calculations.
 In fact the sublattice magnetizations
are not good quantum numbers, they are decreased and eventually wiped out
by "quantum fluctuations". This point is common knowledge today.
 An  essential stone mark to this understanding is  the first 
spin-wave calculations done in 1952 by P.W.A. Anderson~\cite{a52}
 and R. Kubo~\cite{k52}
\footnote{Even if you are quite
 familiar with the modern formalism of spin-waves, this paper develops
 a global
 physical understanding of the subject, and remains an impressive piece
 of work.
The conclusion of the 1952 paper of Anderson also describes
 (in an elusive way) the hint toward the solution of the
symmetry breaking problem. 
}.

In this approach one clearly sees that the transverse term of the
Heisenberg Hamiltonian:
\begin{eqnarray}
{\bf {S}}_i.{\bf {S}}_j |i, +> |j, -> 
& = &  {S}^{z}_i\,{S}^{z}_j \;\;|i, +> |j, -> \; +\nonumber\\
&&  +\; 1/2 \; \left[{S}^+_i\,{S}^-_j  + {S}^-_i\,{S}^+_j  \right]\;|i, +> |j, ->
\nonumber\\
&=& -\frac{1}{4}\;\;|i, +> |j, -> + \frac{1}{2}\;\;|i, -> |j, +> 
\end{eqnarray}
induces spin-flips, decreasing the sublattice magnetizations.
These low energy excitations (spin-waves) can be described as
quantum oscillators: they have  zero point quantum fluctuations, which
renormalize and stabilize the Ising energy and decrease the
sublattice magnetization. This
spin-wave calculation lays on an $\frac{1}{S}$ expansion and its
validity for spins$-1/2$ has often be questioned. It appears to
be qualitatively valid when compared to exact results (when they
exist) or to more sophisticated numerical work (see Table 1), I
will explain in the next chapter the physical reason of
this "good" behavior.

To my knowledge {\bf exact results} exist for 1-dimensional
systems (Bethe problem, Majumdar -Gosh $J_1-J_2$ problem) where
they predict the absence of N{\'e}el long range order (and
algebraic spin-spin decaying correlations in the first case,
exponentially decaying ones in the second case). For larger
lattice dimensionality, only the case of the cubic lattice has
been shown to be N{\'e}el ordered~\cite{kls88}. On the other hand,
the Mermin-Wagner theorem precludes existence of N{\'e}el long
range order (NLRO) at $T \neq 0$ for lattices with dimension $d
<3$. This theorem does not give any indications for the $T=0$
behavior of 2-dimensional magnets which are the central point of
these lectures. (Rigorous proofs of order exist for  spin 1 and
larger~\cite{dls78,jnfp86,aklt88}.)

{\bf N{\'e}el order versus dimer pairing}: naive approach and
numerical results.

The classical N{\'e}el wave-function (\ref{Nwf}), is a
variational solution with an  energy per bond 
$(- \frac{J}{4})$.   Whereas the quantum
ground-state of (\ref{Heis}) is:
\begin{equation}
|(i,j)> =\frac{1}{\sqrt2} \left[ |i, +>|j, ->  - |i, -> |j, +>\right]
\label{dimere} 
\end{equation}
with the energy $-\frac{3}{4} J$. This state that we will call
either a singlet state or a dimer realizes a very important
stabilization of a pair of spins (if compared to the classical
state) but it  does not allow to point any
direction in spin space (it is a state with a total spin zero).
At this microscopic scale quantum mechanics in its radicalism
does not favor  the idea of an $SU(2)$ symmetry breaking.
 The controversy on the existence of N{\'e}el long range order,
 specifically in frustrated ($\equiv$ triangular) geometry or 
with competing interactions has been a long lasting debate
opened by P.W.A. Anderson and P. Fazekas \cite{a73,fa74} and
 fueled again in 1987 with the discovery of
High Temperature Superconductors in cuprates.

When looking in a simple-minded way at a lattice of coordination
number $z$, the energy balance between the classical Ising-like
N{\'e}el state and the quantum dimer covering is not so clear.
  The classical N{\'e}el state has an energy
\begin{equation}
 E_{cl} =  \frac{N}{2} z \frac{J}{4} cos(\theta)
\label{Isingenergy}
\end{equation}
(where $\theta$ is the angle between sublattice magnetizations)
 to be compared to the quantum energy of a dimer
covering 
\begin{equation}
E_{dim} =- \frac{N}{2} 3 \frac{J}{4}.
\end{equation}
 This simple
 approach predicts N{\'e}el order on the square lattice, it is
inconclusive for the hexagonal lattice, or the triangular lattice
 (which have N{\'e}el long
range order) and it predicts that the Heisenberg model on
the kagome lattice is disordered  (which
is correct, see Table~\ref{ener-_param}).
\begin{table}
\begin{center}
\begin{tabular}{|c||c|c|c|c|c|c|}
\hline
&Coordination &$2<{\bf S}_i.{\bf S}_j>$	&&\\
Lattices&number&per bond&$M/M_{cl}$&\\
\hline
\hline
dimer &1&-1.5&& \\
1 square &2&-1&&\\
\hline
 1 D Chain&2&-0.886&0&\\

honeycomb~\cite{fsl01}&3 &-0.726& 0.44& bipartite\\
sq-hex-dod.~\cite{tr99}&3&-0.721 &0.63&lattices\\
square~\cite{tc90}&4&-0.669& 0.60&\\
classical value&&-0.5&1&\\
\hline
\hline
 one triangle&2&-0.5 &&\\
\hline
 kagome~\cite{web98}&4&-0.437&0&frustrating\\
triangular~\cite{bllp94} &6&-0.363&.50&lattices\\
classical value&&-0.25&1&\\
\hline
\hline
1 tetrahedron&3&-0.5&&\\
\hline
checker-board~\cite{fmsl03}&6&-0.343&0&frustr. latt.\\
\hline
\hline

\end{tabular}
\end{center} 
\caption[99]{ Quantum energy per bond and sublattice magnetization 
in the
ground-state of the spin-1/2 Heisenberg Hamiltonian on various
simple cells and lattices. The sq-hex-dod. is a bipartite lattice formed with
squares, hexagons and dodecagons.}
\label{ener-_param}
\end{table}

Indeed this approach is naive in both limits.

In the
classical limit we have neglected the "fluctuation effects"
  generated by the transverse coupling:
these fluctuations effectively contribute noticeably to the
stabilization of the ground-state of ordered systems
(see Table~\ref{ener-_param}).

 In the quantum disordered limit,
 the dimer covering solutions do not take into account
the resonances between different non orthogonal
coverings which are very numerous and are an  essential concept for
 understanding the Resonating Valence Bond Spin Liquids 
(concept introduced in the present context
  by P. W. Anderson in 1973  and  named in
 honor to Linus Pauling). The existence of this second kind of 
phases remained speculative until the end of the nineties. We
 now think that these different scenarios  
can be realized in two dimensional spin-1/2 quantum antiferromagnets
(see Table~\ref{4phases}).

\begin{table}
\begin{center}
\begin{tabular}{|c||c|c||}
\hline
Phases& G.-S. Symmetry Breaking&Order Parameter\\
\hline
\hline
&SU(2)& \\
 Semi-class. N{\'e}el order & Space Group&Staggered Magnet.\\
& Time Reversal& \\
\hline
& &dimer-dimer LRO  {\it or}\\
Valence Bond Crystal&Space Group& S=0 plaquettes LRO \\
\hline
 R.V.B. Spin Liquid && No local \\
 (Type I)&topological degeneracy&order parameter\\
\hline
R.V.B. Spin Liquid&&No local	\\
(Type II) &topological degeneracy	&order parameter\\
\hline

\end{tabular}
\end{center} 
\caption[99]{The four 2-dimensional phases described in these
lectures.
}
\label{4phases}
\end{table}

In the first part of these lectures, I will try to extract the
generic features of the Quantum N{\'e}el phase in a fully 
quantum $SU(2)$ invariant framework. In so doing I hope to be able
to convince you that the symmetry breaking mechanism implemented
in the N{\'e}el state could be understood from a completely quantum
and rather simple approach.

 In the second part of the lectures we
will discuss the new quantum phases where the ground-state does
not break $SU(2)$ symmetry and has no long range order
in spin-spin correlations. We will see that at least two or three
different phases with these general properties have been
exhibited in realistic spin models. The differences between these
quantum phases depend on the pattern of dimer-dimer correlations:
either they display long range order and the system is a Valence
Bond Crystal or any correlation functions are short ranged and it
is a liquid (Resonating Valence Bond Liquid). We will try to describe
 the
generic properties of their  excitations, discuss some experimental
prescriptions and recent results.
 \end{subsection} 
\end{section}
\begin{section}  {Miscellaneous remarks on the use of the
words ``quantum
fluctuations'', ``quantum disorder''}

In  antiferromagnets the word "quantum fluctuations" is
often used with different acceptions, depending on the
context.

When we say in the spin-wave approach of the antiferromagnet that
the sublattice magnetization can be wiped out by ``quantum
fluctuations'' let be conscious that it is a model dependent
concept! In that case these  ``quantum
fluctuations'' do not describe a real microscopic, 
time-dependent mechanism: it is just a way to describe a
renormalization (we might say a dressing) of the Ising-like
states. 

On the other hand, when we say in the RVB spin liquid state, that
the system can fluctuate between different dimer covering
configurations this correspond to true excitations of the system
which may be gapped  or not.

Third, when neutronists say that they  measured longitudinal or
transverse spin fluctuations, they use the word in its strictest
acceptance!  
The root mean square fluctuations of the sublattice
magnetization is defined as $ \sqrt{<\Psi_0|S^2_A|\Psi_0>}$ .
It has the value $ \sqrt{\frac{N}{4}(\frac{N}{4}+1)}$, in the
classical Ising-like N{\'e}el state (\ref{Nwf3}), and is still of
order ${\cal O}(N)$ in any quantum ground-state with long range
N{\'e}el order (as for example, on the square, hexagonal or
triangular lattice). The same is true of the total spin
fluctuations as we will understand in the next chapter!
 And this is observable!
The staggered susceptibility is the experimental
quantity that can be measured experimentally: it is related to
 the Fourier
transform of the above correlation function. It is non zero in
NLRO systems and zero in spin gapped ones. 

In fact the fluctuations of the total spin are zero in a "quantum
disordered" system with a spin gap\footnote{"quantum disordered
is another awkward expression. The Valence Bond Crystals and
standard (type I) resonating Valence Bond Spin Liquids are not
"disordered systems". The degeneracy of their ground-state is
lower than the degeneracy of the N{\'e}el state and they have well
defined excitations. Their order is not of the N{\'e}el type but
they have specific order, as we will see in the following lectures.}. But
local fluctuations of a configuration of spins and dimers in a
``quantum disordered'' spin liquid can also be observed with local
probes: as for example muons \cite{ukkll94}....

Only a few examples of situations that can be uncovered by the
loose expression ``quantum fluctuations''!  
\end{section}
\end{chapter}

\begin{chapter} {The semi-classical N{\'e}el phase: quantum
mechanics and symmetry breakings}

In this chapter we want to uncover in a very simple quantum
mechanical point of view, the nature of the semi-classical phase
and the ingredients of the $SU(2)$ symmetry breaking. This is
grounded in the existence of a ``tower'' of $SU(2)$ invariant
states which collapse in the thermodynamic limit in a
ground-state multiplicity that can be described either in the
$SU(2)$ invariant basis, or in an (overcomplete) basis of
semi-classical coherent N{\'e}el states. We will do it first in a
pedestrian calculus approach and then phrase is again in a more
basic and conceptual point of view parallelizing the
translational symmetry breaking of solids. The space symmetry
breaking will be quickly discussed in this chapter. A  more detailed
study will be done in the next chapter where we analyze
the mechanism of ``order by disorder'' in these semi-classical
antiferromagnets.

\begin{section} {Calculus approach}

Let us consider the Heisenberg problem on a lattice of $N$ sites
with periodic boundary conditions.
It is interesting to look first to an exactly solvable model,
that emerges easily from the expression of the Heisenberg
Hamiltonian ( Eq. \ref{Heis.Latt}) in reciprocal space:
\begin{equation}
{\cal H} = 2 J \sum _{k \,\in BZ}\gamma_{k } {\bf S_k.S_{-k}}.
\label{Heis.rec}
\end{equation}
In this expression:
\begin{equation}
{\bf {S}_k} = \frac{1}{\sqrt N}\sum _{i} {\bf {S}_i}\; exp(-i {\bf k}.{\bf R}_i)
\end{equation}
where ${\bf R}_i$ is the coordinate of spin i,  $N$ the (even) number
of lattice sites and ${\bf k}$ runs on the reciprocal points of the
lattice in the first Brillouin zone (BZ). 
$\gamma_{k}$ is the structure factor of the lattice: 
\begin{equation}
\gamma_k = \frac{1}{2} \sum _{i=1,2} cos({\bf k}.{\bf e}_i)
\label{lattice_stru}
\end{equation}
 with ${\bf e}_j,\, (j=1,2)$,  the
unit vectors generating the lattice.
On this lattice the N{\'e}el state is invariant by 2-step  translations
associated to wave-vectors
 ${\bf k} = (0,0)$ and $\bf k = {\bf k}_0 = (\pi, \pi)$.
Let us select these special components in the Heisenberg
Hamiltonian and rewrite it as:
\begin{equation}
{\cal H}  =  {\cal H}_0 + {\cal V}
\end{equation}
{\rm with}
\begin{equation}
 {\cal H}_0  = 2J ({\bf S}_0^2 - {\bf S}_{{\bf k}_0} . {\bf S}_{-{\bf k}_0})
\end{equation}
\begin{equation}
{\cal V} = 2J \sum _{{\bf k} \in BZ^*}\gamma_{k} {\bf S}_k.{\bf S}_{-k} 
\label{pertqf}
\end{equation}
where $BZ^*$ is to be understood as the  first Brillouin zone
minus the $\bf k =0$ and  ${\bf k}_0$ points.

Simple algebra leads to:

\begin{eqnarray}
{\cal H}_0 & = &2J ({\bf S}_0^2 - {\bf S}_{{\bf k}_0} . {\bf S}_{-{\bf k}_0})\nonumber\\
&=& \frac{4J}{N} ({\bf S}_{tot}^2  - {\bf S}_A^2 - {\bf S}_B^2),
\label{lieb-mattis_ham}
\end{eqnarray}
where ${\bf S}_{tot}$ is the total spin of the sample and ${\bf S}_{A,B}$
the total spins of the $A, \; B$ sublattices.

You might recognize in ${\cal H}_0$  the toy model used by Lieb and 
Mattis in the demonstration of the ordering theorem~\cite{lm62}:
it describes a
problem with constant long range interactions between spins on
different sublattices and no interactions between spins on the
same sublattice. This model can be solved exactly. \footnote{The same kind of
toy model can be introduced in the problem of the 3-sublattice
N{\'e}el state on a triangular lattice: in that last case it
 involves the Fourier
components of the spins at the three soft points (which are the
center and the two non equivalent corners of the Brillouin zone)
and reads:
\begin{equation}
{\cal H}^{tri}_0 = \frac{9J}{2N} ({\bf S}_{tot}^2  - {\bf S}_A^2 
- {\bf S}_B^2 - {\bf S}_C^2) 
\label{3sub}
\end{equation}
 where ${\bf S}_{A,B,C}$ are 
the total spins of the $A, \; B, \; C $ sublattices. Such a model
allows the same developments as those done below except indeed
the comments on the Lieb-Mattis ordering theorem~\cite{bllp94}.}

\begin{subsection}{The Ising-like N{\'e}el state in an
$SU(2)$ invariant model}

Hamiltonian ${\cal H}_0$  (Eq. 
\ref{lieb-mattis_ham}) is an $SU(2)$ invariant
 Hamiltonian which commutes
with ${\bf S}^2_{tot}$ and ${\bf S}^z_{tot}$. It also commutes with  
${\bf S}^2_{A,B}$, which in this model are conservative quantities
(good quantum numbers). All these observables commute two by two
 and with ${\cal H}_0$ . Eigen-states of ${\bf S}^2_{tot}$,
 ${\bf S}^z_{tot}$, ${\bf S}^2_A$,  ${\bf S}^2_B$ are also
eigen-states of (\ref{lieb-mattis_ham}), with eigen-values:
\begin{equation}
E( S, S_A, S_B) = \frac{4J}{N} \left[ S (S + 1) - S_A (S_A + 1)
 -S_B (S_B + 1)\right]
\label{lmspectrum}
\end{equation}
The quantum numbers for a sample with an (even) number $N$ of sites
 are:
\begin{itemize}
\item $ S_A, S_B \in [0, 1, .., N/4]$,
\item For a given set of values of the sublattice magnetizations
 $ (S_A, S_B ) \;$
the value of the total spin $ S \in [|S_A - S_B|,...,S_A+ S_B]$,
\item For a given value $S$ of the total spin of the sample, its $z$
component $M_S  \; \in [-S, -S + 1, ...., S - 1, S]$.
\end{itemize}

{\bf The ground-state in each $S$ sector $ E_{0}(S)$} is obtained for the
 maximum
sublattice magnetization $N/4$. The energies of these
low energy states obey the following relation:
\begin{equation}
E_{0}(S) = - \frac{J}{2} (N+4) + \frac{4J}{N} \left[ S (S + 1) \right]
\end{equation}
\footnote{This is a special illustration of Eq.~(\ref{lm_ordering}). 
Let us remark that this ordering property is shared by the toy
model (\ref{3sub}) associated to the 3-sublattice N{\'e}el order on the
triangular lattice.}.

The eigen-states associated to these eigen-values are the $SU(2)$
invariant components of the Ising-like N{\'e}el  state 
$|S_A =N/4,S_B=N/4, S\,, M_S>$ introduced in section \ref{SBneel}
(Eq.~\ref{Nwf3}). These eigen-states have four essential
properties:
\begin{itemize}
\item their number and their spatial symmetries are uniquely
defined by the coupling of the sublattice magnetizations~(these
are exact necessary requirements),
\item their sublattice magnetization is N/4,
\item they collapse to the absolute ground-state as
${\cal O} (\frac {1}{N})$.
\end{itemize}
 These levels form a set that has been
described by Anderson as a ``tower'' of states~\cite{a52,a63,a84}:
we have called them in our original paper QDJS (for quasi
degenerate joint states). In this lectures we will refer to this
set as the Anderson tower.

On a finite size lattice the classical N{\'e}el  state
 (\ref{Nwf3}) is a non stationary state of ${\cal H}_0$ 
(eq.~\ref{lieb-mattis_ham}). But, its
precession rate decreases as ${\cal O} (\frac {1}{N})$ with the
system size and becomes infinitely slow  in the
thermodynamic limit.

The coherent N{\'e}el states described by Eq.(\ref{Nwfu}), 
form  an (overcomplete)
basis of this ground-state multiplicity. 
The present
study of their $SU(2)$ invariant representation shows that the
multiplicity of this subspace is ${\cal O} (N^{\alpha})$ where
 $\alpha$ is the number of sublattices of the classical N{\'e}el
state\cite{bllp94,le95}.
 This gives
a non extensive entropy of the ground-state at $T=0$ 
 in agreement with Nernst theorem.

{\bf Excitations}

In this model an excited state is obtained by flipping a single spin
of a sublattice. From equation~(\ref{lmspectrum}) one sees that
these excitations  are localized  and
 have an energy:
\begin{equation}
 E^{exc}_{Ising} = 2J \left[ 1 + \frac{4 (S + 1)}{N} \right].
\label{excising}
\end{equation}
For any size these excitations are gapful and ${\cal O}(J)$.

{\bf Conclusion}

 ${\cal H}_0$ describes an Ising magnet in an $SU(2)$
invariant framework: its spectrum  has  the  very simple
structure schematized in Fig.~\ref{Isingspectra}.
\begin{figure}
\hspace{2.5cm}\resizebox{7cm}{!}{\includegraphics*{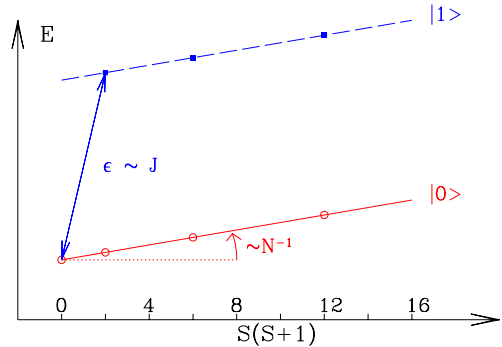}}
    \caption[99]{ Typical spectrum of a finite size 
collinear Ising magnet.
The tower of eigen-levels joined by the continuous line and noted
 $\left|0\right>$ is the Anderson tower of states needed to form
a symmetry breaking Ising ordered ground-state (Eq.~\ref{Nwf4}):
 such a state is non stationary on a finite size sample. 
The second set  $\left|1\right>$ (dashed line)
 is associated with the lowest excitations, which are highly
degenerate and non dispersive.
 }
    \label{Isingspectra}
\end{figure}
 In the thermodynamic
limit this magnet can be described either in an $SU(2)$ invariant
language with the help of the $|S_A,S_B,S,M_S>$ states or with the
coherent semi-classical N{\'e}el states described in
Eq.(\ref{Nwfu}). The two basis are connected by exact
transformation laws, Eq.(\ref{Nwf3}) and its inverse:
\begin{equation}
|N/4,N/4, S, M_S> = (2S+1) \int d\tau D^{\dagger}_s(\phi,\theta)|Cl.\; N\acute{e}el\; w.f.; {\bf u}>,
\end{equation}
where the differential integration volume reads
$d\tau = \frac{1}{4\pi}d\phi d(cos\theta)$, where $\phi \in [0,2\pi]$,
$\theta \in [0,\pi]$ and $D_S$ is the  rotation matrix in the
$S$ subspace. 
In the thermodynamic limit, the symmetry breaking point of view
is as valid as the $SU(2)$ invariant approach.

\end{subsection}
\begin{subsection}{``Quantum fluctuations'' in the Heisenberg model}

Modification of this picture in an Heisenberg magnet with next
neighbor exchange comes from the effect of the perturbation term
${\cal V}$ described in Eq.~(\ref{pertqf}). ${\cal V}$ does not
commute with ${\bf S}_A^2$ and ${\bf S}_B^2$: at first order in
perturbation each component of ${\cal V}$  couples
 the ground-state  of ${\cal H}_0 $ in each $S$ sector
 (which are characterized by maximum uniform
sublattice magnetizations)  
to states where the sublattice
magnetization is decreased by one unit in some modulated way. 
The analytical treatment of this perturbation is uneasy in
the $SU(2)$ invariant formalism,
 but indeed we recognize all the concepts at the basis of  
the usual algebraic spin wave approach: that is {\it renormalization of the
ground-state energy and of the sublattice magnetization by 
 the zero point quantum fluctuations of the spin waves
excitations}.

If the structure of the tower of states i.e.:
\begin{itemize}
\item  number and spatial symmetries of the states in each $S$ sector,
\item  existence in each of these states of a macroscopic sublattice
 magnetization (i.e.  $\sqrt{<|S^2_A|>} \propto {\cal O} (N)$),
\item  scaling as ${\cal O}(\frac{J S (S+1)}{N})$ with respect to
the ground-state,
\label{criteria}
\end{itemize}
resists to this renormalization,
  the nature of the ground-state multiplicity in the
thermodynamic limit gives a new foundation to the spin-wave symmetry
 breaking point of view~\footnote{As an example, all these criteria have
been thoroughly checked in the N{\'e}el ordered phase of the
Heisenberg model on the triangular lattice\cite{blp92,bllp94}.}.
The quantum N{\'e}el wave function thus emerges from the classical
picture (Eq.~\ref{Nwf3}) by the renormalization of the
eigenstates $|S_A,S_B,S,M_S>$ of ${\cal H}_0 $ under the action
of ${\cal V}$. We can then
write the quantum N{\'e}el wave-function as:
\begin{equation} 
|Qu.\; N\acute{e}el\; w.f.> = \sum_{S,M_S} {\frac{(-1)^{M_S}}
{\sqrt{ 2S + 1}}}
\left(   \begin{array}{ccc}
         S_A & S_B & S\\
	 S_A & - S_B &  M_S\\
          \end {array}  \right)
\widetilde{ |S,M_S>}_{0}
\label{Nwf4}
\end{equation}
where the kets $\widetilde{|S,M_S>}_{0}$ are now the exact low lying states of
the Anderson tower of ${\cal H}$ (Eq.~\ref{Heis.Latt}).

\end{subsection}

\begin{subsection} {The spin-wave algebraic approach} 

In order to gather all the material needed for a full
understanding of the symmetry breaking mechanism in N{\'e}el
antiferromagnets, let us recall the main results of a spin-wave
calculation. (For the derivation of the spin-wave approach in
antiferromagnets, see the above mentioned text-books~\cite{m81,
c89a,auer94}.)  

Departing from the Ising configuration~(Eq.\ref{Nwf}),
 the transverse terms
 of the Heisenberg Hamiltonian create $\Delta S^z
=1$ spin flips, which are mobile excitations.
\begin{itemize}
\item In an harmonic approximation these excitations are simply
described as spin-waves, with frequencies:
\begin{equation}
\omega_{\bf q} = 2 J\sqrt{1-\gamma^2_{\bf q}}
\end{equation}
where $\gamma_{\bf q}$ is the structure factor of the lattice
defined in Eq.(\ref{lattice_stru}). The spin flips excitations are
then dispersive, their frequency goes to zero when going to the
two soft points ${\bf k = 0, \; k_0}$. Around these points the
dispersion law is linear in ${\bf k}$ (resp. $({\bf k - k_0})$).
\item
The zero point energy of these excitations (which are oscillator- like)
renormalizes the Ising classical energy of the
 ground-state~(\ref{Isingenergy}).
To first order, this spin wave calculation gives the ground-state
energy of the Heisenberg Hamiltonian on the square lattice as: 
\begin{equation}
E^{s-w} =  - \frac{N}{2} z \frac{J}{4} - N J +
\sum_{{\bf q} \in BZ^{*}} \frac{\omega_{\bf q}}{2}
\label{esw}
\end{equation}
\item  These `` quantum fluctuations'' also renormalize the
sublattice magnetization. Let us define the order parameter $m$ in
the ground-state $|0>$ of this symmetry breaking representation by:
\begin{equation}
m= \frac{2}{N S}<0|S^z_A|0>.
\end{equation}
The first order spin-wave calculation leads to:
\begin{equation}
m^{s-w}= 1 -  {\frac{1}{N}}  \sum_{{\bf q} \in BZ^{*}}\left[ {\frac{1}{\omega_{\bf q}}} -1 \right]
\label{msw}
\end{equation}
The renormalization of the order parameter is dominated by the
fluctuations in the low energy modes.
The linear asymptotic behavior of $\omega_{\bf q}$ around the soft
points, implies that the spin-waves correction to the order parameter 
diverges in 1D. It gives finite corrections at $T=0$ on most of
the 2-dimensional lattices (square, triangular, hexagonal..
\footnote{The  exceptions: the checker-board and the kagome lattice will be
studied in a forthcoming chapter.}).
\item {\bf Finite Size Effects:}
The spin-wave approach  allows a
direct understanding of the finite size effects 
in a problem with N{\'e}el long range order. Let us first remind 
that on a finite size lattice of linear length $L$, the allowed wave
 vectors are quantized and of the form $\frac{2\pi}{L}$.
This introduces a cut-off of the long wave-length fluctuations
which is progressively relaxed as the size of the sample goes to
$\infty$. As $\omega_{\bf q}$ is linear in ${\bf q}$ around the
soft points, we thus expect that the ground-state energy $E^{s-w}$ 
(Eq.~\ref{esw}) and
the order parameter $m^{s-w}$ (Eq.~\ref{msw})
 on a lattice of finite size
$L$ will differ from the $L \to \infty$ limits by factors
of  order ${\cal O}(\frac{1}{L})$. This is exactly the result
obtained in more sophisticated approaches~\cite{gsvs89a,gsvs89b,nz89,f89,adm93,hn93}.

As we have already underlined the excitations of this model now
differ from those of ${\cal H}_0 $: they are itinerant and have
acquired dispersion. On a finite lattice the energy needed to
create the softest excitation is no more of order  $J$, but of
order $\frac{J}{L} \propto \frac{J}{N^{1/d}}$.
\end{itemize}
\end{subsection}

\begin{subsection} {Self-consistency of the N{\'e}el picture for an
Heisenberg magnet in an $SU(2)$ invariant picture:
 spectrum and finite size effects.}

If the structure of the tower of states is essentially preserved
by the quantum fluctuations due to ${\cal V}$, the semi-classical
picture of coherent states is preserved (see
subsection~\ref{criteria}), 
the spin-wave approach is a reasonable one and the essential
results of this approach should appear in the full spectra of
Eq.(\ref{Heis.Latt}).
Beyond the criteria already described to support the $SU(2)$
symmetry breaking, the following size effects
should be present:
\begin{itemize}
\item  The energy per site of the states of the low lying Anderson 
tower should converge to the thermodynamic limit with a leading
correction term going as $\frac{1}{N \,L} \propto
\frac{1}{L^{d+1}}$,
\item  The sublattice magnetization  $\sqrt{<|S^2_A|>}$
in each of these states
 should remain ${\cal O} (N)$, with a 
leading term to the finite size corrections
 of  order  ${\cal O}\frac{1}{L}$,
\item The low lying softest excitations with wave-vector
$\frac{2\pi}{L}$ should be described by a second tower of states
issued from the tower of excited states of the Ising model with
one spin-flip (Eq.~\ref{excising}). But contrary to the Ising
model, these states are  now  dispersive and the lowest
excitation is now distant from the ground-state tower
of states by an energy of the order of $\frac{J}{L}$: it is 
the Goldstone mode of the broken $SU(2)$ symmetry.
\end{itemize}

Some of these properties are summarized in the supposed-to-be
spectrum of a N{\'e}el antiferromagnet described in
Fig.~\ref{Neelspectra} .
\begin{figure}
\begin{center}
	\resizebox{8cm}{!}{\includegraphics{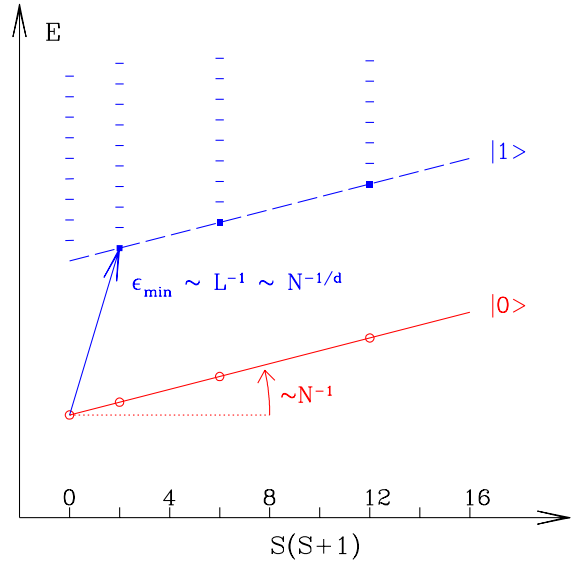}}  \end{center}
    \caption[99]{ Typical spectrum of a finite size 
collinear antiferromagnet with  N{\'e}el order.
The tower of eigen-levels joined by the continuous line and noted
 $\left|0\right>$ is the Anderson tower of states needed to form
a symmetry breaking N{\'e}el ordered ground-state (Eq.~\ref{Nwf4}):
 such a state is non stationary on a finite size sample. 
The second set  $\left|1\right>$ (dashed line)
 is associated with the lowest magnon.
 }
    \label{Neelspectra}
\end{figure}
This is to be compared to an exact spectrum of the Heisenberg
Hamiltonian on a square lattice (Fig.~\ref{sqN})\cite{d02} or on an
hexagonal lattice (Figs.~\ref{specthex},~\ref{hexN},~\ref{fig-pure-gspin}
)\cite{fsl01}.
\begin{figure}
\begin{center}
	\resizebox{8cm}{!}{\includegraphics{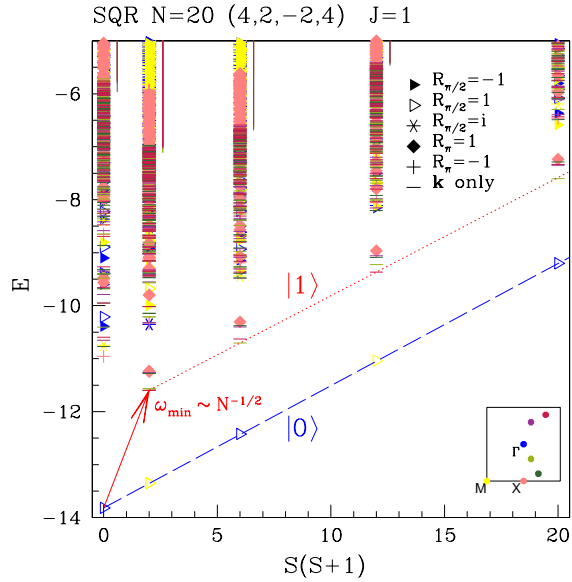}}  \end{center}
    \caption[99]{ Antiferromagnetic Heisenberg model
on the square lattice: eigen-energies vs eigen-values of ${\bf S}^2$.
The dashed-line is a guide to the eyes for the QDJS of the
symmetry breaking quantum N{\'e}el state (Eq.~\ref{Nwf4}). The dotted
line joins the states associated to the first magnon.
There is one QDJS for each $S$
(as expected for a collinear antiferromagnet):
they are ${\bf k=0}$ states, and ${\bf k=(\pi,\pi)}$ states,
invariant in $C_4$ rotations.
 }
    \label{sqN}
\end{figure}
\begin{figure}
\begin{center}
	\resizebox{8cm}{!}{\includegraphics{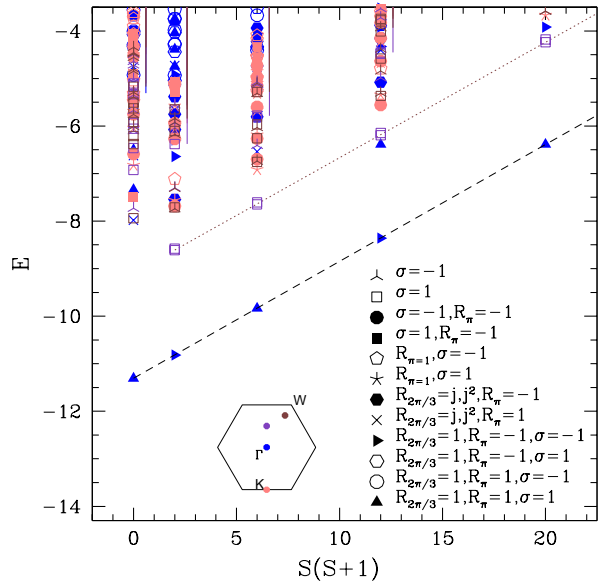}}  \end{center}
    \caption[99]{ Antiferromagnetic Heisenberg model
on the honeycomb lattice: eigen-energies vs eigen-values of ${\bf S}^2$.
The dashed-line is a guide to the eyes for the QDJS. The dotted
line joins the states associated to the first magnon.
There is one QDJS for each $S$
(as expected for a collinear antiferromagnet):
they are ${\bf k=0}$ states,
invariant under a $2\pi/3$ rotation
around an hexagon center, even (odd) under inversion,
odd (even) under a reflection with respect to an axis going through
nearest neighbor hexagon centers for $S$ even (odd) (taken from
ref.~\cite{fsl01}).}
\label{specthex}
\end{figure}
\begin{figure}
 \begin{center}
	\resizebox{8cm}{!}{\includegraphics{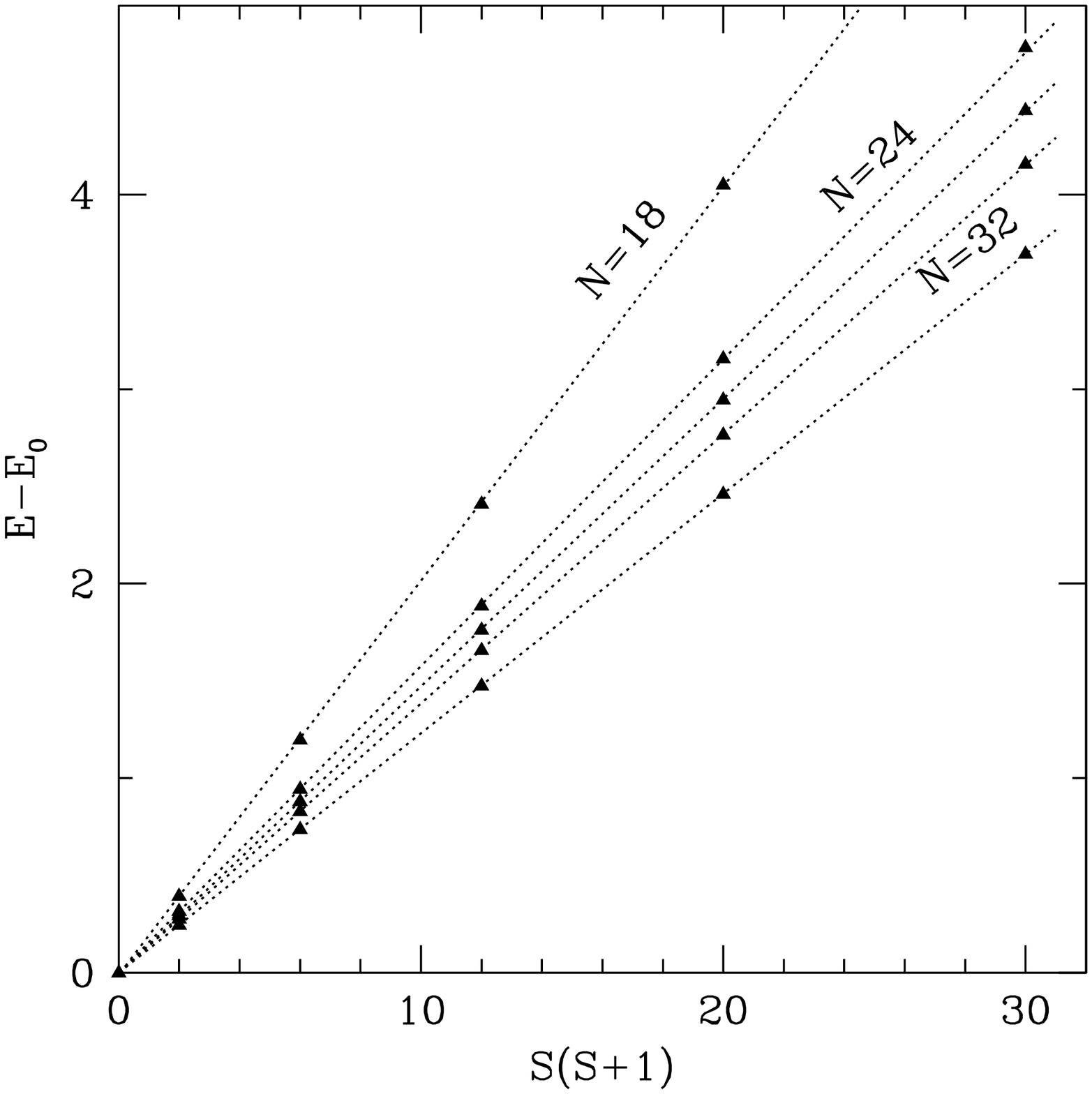}}  \end{center}
\caption[99]{AF Heisenberg model on the honeycomb lattice,
 scaling of the QDJS with $S$
 and  $N$ for $N=18,24,26,28,32$ (taken from
ref.~\cite{fsl01}).}
    \label{hexN}
	\begin{center}
	\resizebox{8cm}{!}{
	\includegraphics{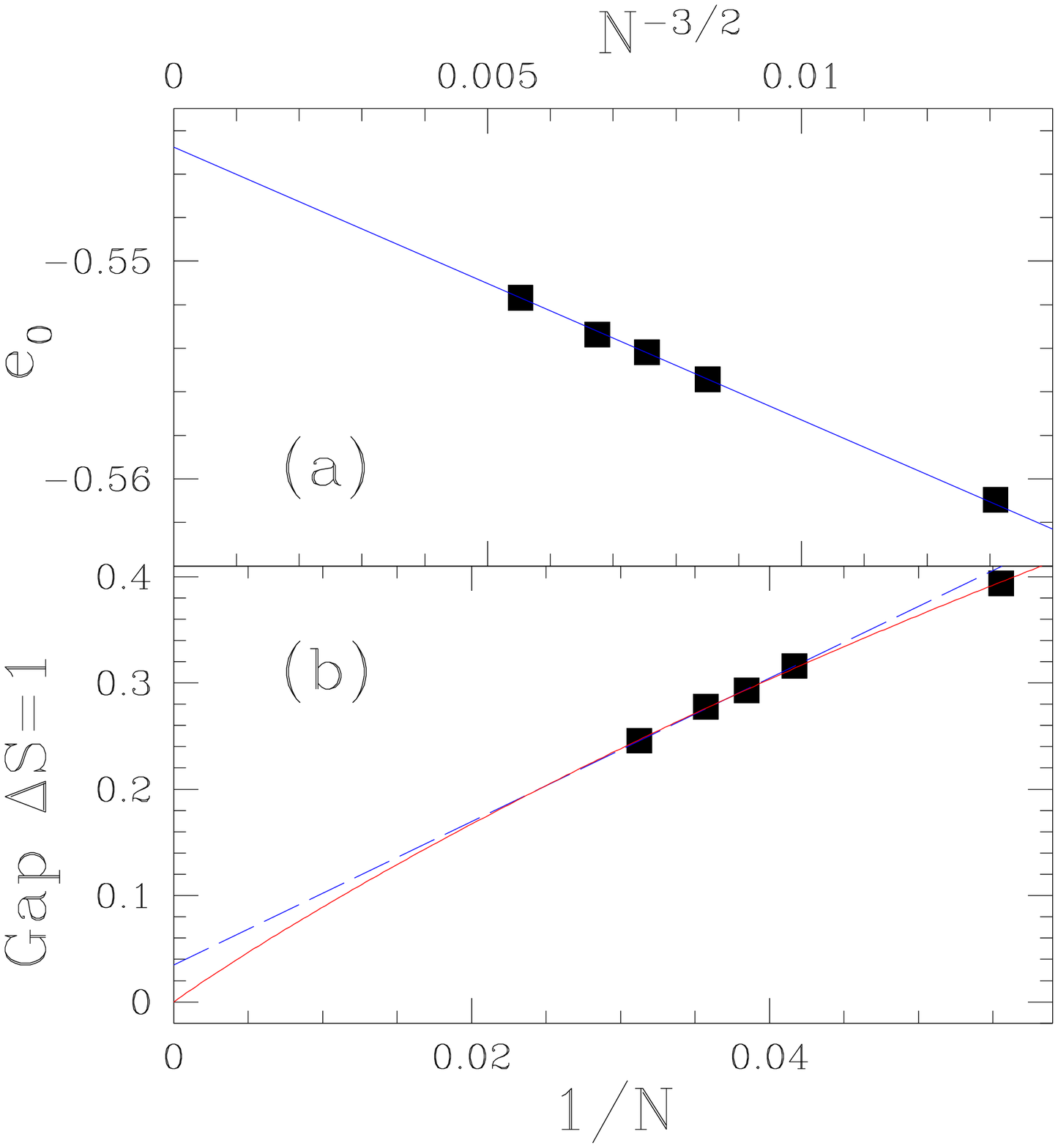}}
	\end{center}
    \caption[99]{ AF Heisenberg model on the honeycomb lattice, (a)~energy per site $e_0$
versus $N^{-\frac {3}{2}}$
(b)~spin-gap:
The  dashed line is a linear fit in $1/N$: for the sizes of
interest the restriction to the 
leading term of the finite size expansion is insufficient.
The full line is a fit to eq.~\cite{f89,hn93}:
$\Delta (N)= \frac{1}{4 \chi N} (1 - \beta\frac{c}{\rho \sqrt{N}}) 
+ {\cal O} (\frac{1}{N^2})$
where  $\chi$ is the spin susceptibility, $c$ is the spin-wave velocity, $\rho$ the spin stiffness
and $\beta$ is a number of order one (taken from
ref.~\cite{fsl01}).}
    \label{fig-pure-gspin}
\end{figure}

This global understanding of the spectra of finite size samples
of antiferromagnets is a
very useful tool to analyze exact  spectra of spin
models that can be obtained with present computer facilities
\footnote{Historically
the first authors to have looked for the Anderson tower of
states were probably A. S\"{u}t\"{o} and P. Fazekas in 1977~\cite{sf77}, and with
the modern computational facilities M. Gross, E. Sanchez-Velasco
and E. Siggia~\cite{gsvs89a,gsvs89b}.}.
 It
seems that it may equally help to understand the time behavior
of nano-scale antiferromagnets as ferritin~\cite{lh95}.
\end{subsection}
\end{section}

\begin{section} {A simple conceptual approach of the
translational symmetry breaking of a solid}

For a while we exclude any  calculations and just
rely on very simple and basic concepts of condensed matter
physics and quantum mechanics 
to derive the ``necessary'' structure of the spectra of
ordered condensed matter in finite size samples.
 For the sake of simplicity, we begin  with the problem of the solid
phase. We successively expose the fundamental classical hypothesis
underlying the theory of solids. Quantization of this picture
enlightens the translational symmetry breaking mechanism and finite
size effects give a new light on the absence of solid order in
1-dimensional physics. 

\begin{subsection}{An essential classical hypothesis}
Let us consider a finite sample of solid with $N$ atoms of
individual mass m. The 
Hamiltonian of this piece of solid contains a kinetic energy term
 and an  interaction term $ U({\bf r}_i-{\bf r}_j)$,
 which essentially depends on distances
 between the $N$ atoms, and is translation invariant.
Nevertheless any piece of solid in nature breaks translational
symmetry!

The first step in the description in classical
phase space of the
dynamics of this object with $2dN$ degrees of freedom,
 consists in sorting these variables in two sets:
\begin{itemize}
\item the center of mass variables: ${\bf R}_{c.o.m}$ and
${\bf P}_{c.o.m}$, the dynamics of which is a pure kinetic term
${\cal K}$:
\begin{equation}
{\cal K}= \frac{{\bf P}^{2}}{2 N m}
\label{freemotion}
\end{equation}
\item and the $2d(N -1)$ internal variables, which obey a dynamic
with interactions:
\begin{equation}
{\cal H}_{int}=  \sum_{i \in [1,..,N]} \left[ \frac{{\bf p_i}^{2}}{2  m}
+  U({\bf r}_i-{\bf r}_j)\right]
\end{equation}
\label{phonons}
\end{itemize}

Then invoking the inertia principle, the analysis of the problem
focuses
on the Galilean frame, where the center of mass is at rest. In
this frame, the internal excitations are  analyzed in first
approximation as modes of vibrations: the phonons, which present a
dispersion law linear in  ${\bf k}$ for small wave vectors ${\bf k}$.

In so doing, an {\bf essential dichotomy} is introduced between the
global variable  and its dynamics on one
hand and the internal excitations on the other: this dichotomy is
at the basis of the concept of an ordered phase~\cite{a84}.
A technical asymmetry is also introduced in the treatment of the
dynamics of these two sets of variables: the center of mass
dynamics is described in a classical framework which explicitly breaks
 the translation
invariance of the total Hamiltonian of the solid ${\cal K} +
{\cal H}_{int}$.   On the other hand
the internal excitations  are looked at
in a translationally invariant (eventually quantum) point of
view.
This point of view may seem inconsistent in particular when
looking at a finite sized, eventually small, piece of solid.

Taking as a definition of the solid phase the essential
distinction between the global variable and the internal ones,
 we will show
that the technical asymmetry in the treatment of these variables
 can be easily
overcome, thus explaining both the localization of a piece of
solid in real space, and the influence of space dimensionality 
on the definition of this solid.

\end{subsection}
\begin{subsection}{Quantization of the classical approach, 
finite size spectra, thermodynamic limit and translational symmetry
breaking}

In order not to break artificially the translational symmetry of
the problem we consider a solid with periodic boundary conditions.

If we take for granted that it is legitimate to disconnect the
center of mass dynamics from the internal excitations we may
consider a solid at $T=0$ with no internal excitations: the
vacuum of phonons that we will write $|0>$.

The translationally invariant eigen-states of ${\cal K}$ are the
plane waves with wave-vectors ${\bf k}$ where $
k_{x,y,z}= n_{x,y,z}\frac{2\pi}{L}$,  $L$ is the linear length
of the sample and $n_{x,y,z}$ non zero integers. Their eigen-values
 are of the general form:
\begin{equation}
{ \frac{{\hbar}^2 {\bf k}^2}{2mN}}.
\end{equation}
The total energy of the solid in these states is thus of the
form:
\begin{equation}
E_{0}(k)= \frac{{\hbar}^2 {\bf k}^2}{2mN} + E_g,
\label{comeigen}
\end{equation}
where $E_g$ is a constant measuring the zero point energy of the
internal degrees of freedom. These eigen-states are shown in
Fig.~\ref{solspectrum} connected by the red continuous line noted
$|0>$.

\begin{figure}
\hspace{2.5cm}\resizebox{7cm}{!}{\includegraphics{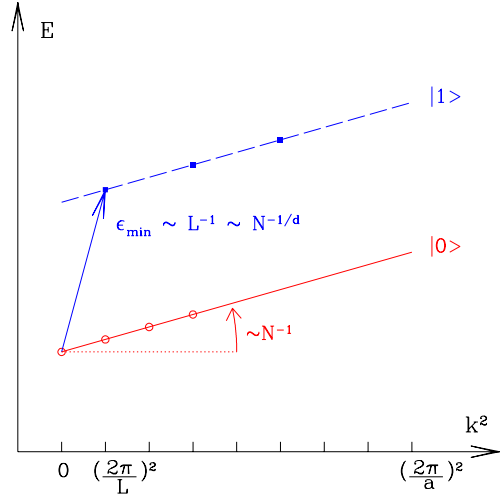}}
    \caption[99]{ Typical spectrum of a finite size 
solid.
The tower of eigen-levels joined by the continuous line and noted
 $\left|0\right>$ is the Anderson tower of states needed to form
a symmetry breaking vacuum of phonons of the solid:
 such a state is non stationary on a finite size sample. 
The second set  $\left|1\right>$ (dashed line)
 is associated with the lowest phonon.
 }
    \label{solspectrum}
\end{figure}
In order to localize the center of mass it is necessary to form a
wave-packet with eigen-states of ${\cal K}$ showing a large
distribution of wave-vectors ${\bf k}$: the largest the ${\bf
k}$-distribution be, the better the localization of the center
of mass. Such a wave-packet is non stationary for a finite size,
but its evolution rate goes to zero as ${\cal O}(1/N)$.
Localization of the center of mass is thus a costless operation in
the thermodynamic limit.

Let us look now to the first excitation of the solid with one
phonon of wave vector ${\bf k}_{min}= 2\pi/L$.
This state can typically be written in a symmetry breaking picture
as:
\begin{equation}
|1> = exp\left( \sum_j i{\bf k}_{min}. {\bf r}_j\right)|0>
\end{equation}
It thus involves a linear superposition of eigenstates of ${\cal K} +
{\cal H}_{int}$ with a distribution of wave vectors displaced by 
${\bf k}_{min}$ with respect to the distribution of the localized
ground-state $|0>$. This second set of excitations is displayed in 
Fig.~\ref{solspectrum} with a dashed line noted $|1>$ joining the 
different eigen-states.
  The softest phonon has an energy proportional to
${k}_{min} \propto L^{-1} \propto N^{-1/d}$ which should be
added to the ground-state energy (\ref{comeigen}) giving
eigen-states with eigen-energies:
\begin{equation}
E_{1}(k)= \frac{{\hbar}^2 {\bf k}^2}{2mN} + E_g + {\hbar} c  k,
\label{phonon}
\end{equation}
where c is the sound velocity.
Due to the structure of equation (\ref{phonon})
the line  joining the different translation invariant states of
this soft phonon is parallel to the ground-state line $|0>$.
 This explains
the supposed-to-be structure of the low lying levels of 
a finite size solid exhibited in Fig.~\ref{solspectrum}.
\end{subsection}
\begin{subsection}{Thermodynamic limit, stability of the solid
and self-consistency of the approach}
The consistency of the semi-classical picture implies that the
localization of the center of mass could be done whatever the
degree of excitations of phonons: looking to the finite size
effects this appears to be the case if the dimension of space if
larger or equal to 2. In these situations, for large enough
sizes there appears two different scales of energy: the Anderson
tower of states of the ground-state  collapses as $N^{-1}$ to
the absolute ground-state whereas
the softest phonon collapses on the ground-state only as $N^{-1/d}$.
In this limit, the dichotomy between the dynamics of the global
variable and the internal variables is totally justified.
On the other hand in 1 dimension it is quantum mechanically
 inconsistent to separate global degrees of freedom from internal
ones:  these two types of variables having dynamics that
cannot be disentangled.
\end{subsection}
\end{section}

\begin{section} {An analogy: $SU(2)$ symmetry breaking in the
N{\'e}el antiferromagnet}
Let us now develop the analogy between the solid states and the
antiferromagnetic ones.
\begin{itemize}
\item The global variables of the solid are ${\bf R}_{c.o.m}$ and
the conjugate variable
${\bf P}_{c.o.m}$. In the collinear antiferromagnetic case the
global variables of position of the magnet are the two Euler
angles $(\theta, \phi)$ allowing to point the direction of the
sublattice magnetization in spin space. Their conjugate variable
is the total spin operator ${\bf S}$.
\item The free motion of the center of mass is governed by the
Hamiltonian ${\cal K}$ (the quadratic form of this kinetic energy
being  related to the homogeneity of space). By
analogy we expect the kinetic energy term describing the free
precession of the sublattice magnetization to be of the form:
${\cal K}_{spin} = {\bf S}^2_{tot}/2van$
\footnote{A three
sublattice N{\'e}el order has a more complicated order parameter:
the three Euler angles are needed to localize the 3 sublattice
magnetizations: and the macroscopic object is no more a rigid
rotator as in the case of the collinear N{\'e}el order but
a (symmetric) top. There is in that last case an extra internal
spin kinetic energy term and as already explained in the
previous section the Hibert space of the problem is larger. See
ref.~\cite{bllp94} for example or the quantum mechanical theory
 of symmetric top molecules.}.
 In such a point of view the constant a is just a
multiplicative term: we know from other sources (fluctuation
dissipation theorem or macroscopic approach of the magnet) that
this is up to a constant the homogeneous  spin susceptibility.
\item The eigen-states describing the free precession of the order
parameter in the vacuum of magnons
 are thus states with total spin $S$ (ranging from $0$ to
$N/2$), and eigen-energies:
\begin{equation}
E_{0}(S(S+1))= \frac{{\hbar}^2 S(S+1)}{2 \chi N} + E_g
\label{ordereigen}
\end{equation}
They form the set $|0>$ of Fig.~\ref{Neelspectra}. By forming a
wave-packet out of this set one can localize the direction of the
sublattice magnetization and break $SU(2)$ symmetry.
\item The discussion of the first excitations above the vacuum of 
magnon completely
parallelizes that of the  phonons excitations (same dispersion
law and same finite size scaling law). The eigen-energies
 of the states embedded in the softest magnon (referred as $|1>$
in Fig.~\ref{Neelspectra}) are thus of the form:
\begin{equation}
E_{1}(S(S+1))= \frac{{\hbar}^2 S(S+1)}{2 \chi N} + E_g + {\hbar}
c_s k_{min}
\end{equation}
where $c_s$ is the spin wave velocity.
\item The possibility of a spin rotational symmetry breaking at
the thermodynamic limit is embodied in the finite size behavior
of the low lying levels of the spectra~(Fig.~\ref{Neelspectra}).
In dimension $d \geq 2$ the eigen-states of the sets $|0>$
 (resp. $|1>$) collapse on
their $S=S_{min}$ component as ${\cal O}(N^{_{-1}})$, more rapidly
than the decrease in energy of the softest magnon which is
${\cal O}(N^{_{-1/2}})$. In dimension 2 and higher, the $SU(2)$
breaking mechanism prevails on the formation of magnon excitations
justifying the classical approach and the dichotomy between
global classical variables and internal excitations.
\item These finite size scalings of the Anderson tower of states
and of the true physical excitations (the magnons) 
give a new light on the Mermin
Wagner theorem which denies the existence of N{\'e}el long range
order in 1 dimensional magnets.
\end{itemize}
\end{section}

\begin{section} {The coherent quantum mechanical description of
the N{\'e}el state}

 At the end of this presentation I hope that Eq.~(\ref{Nwf4}) now
appears  as the natural  quantum
mechanical description of a coherent N{\'e}el state. And by the
fact the usual symmetry breaking approach is justified as soon
as it gives self consistent results (i.e. non zero order parameter).

The technical answer seems beyond doubt.

 The question is now, do coherent states as those described in
 Eq.~\ref{Nwfu}, which I rewrite here 

\begin{equation} 
|Qu.\; N\acute{e}el\; w.f.> = \sum_{S,M_S} {\frac{(-1)^{M_S}}
{\sqrt{ 2S + 1}}}
\left(   \begin{array}{ccc}
         S_A & S_B & S\\
	 S_A & - S_B &  M_S\\
          \end {array}  \right)
\widetilde{ |S,M_S>}_{0}
\label{Nwf5}
\end{equation}
 exist in real life?

 I see no mechanism which can lock the
 difference of phases of the macroscopic number of
 states $\widetilde{|S,M_S>}_{0}$ entering Eq.~(\ref{Nwf5})
 to the correct
values and it seems that many perturbations could destruct such a
coherence, if, by an infinitesimal chance, it existed! 

 So I will plead
that in real life the system may be in any incoherent
superposition of the degenerate $\widetilde{|S,M_S>}_{0}$ which does not build
in spin space a given direction to the sublattice magnetization!

But nobody has to care for it, experiments are not sensitive to
the direction of the sublattice magnetizations but only to
correlations functions: as the square of the staggered
magnetization. This correlation function is identical in all the
states of the Anderson tower in the Ising model; if it survives
to quantum fluctuations introduced by ${\cal V}$,
 we expect it to be nearly identical in  all the 
$\widetilde{|S,M_S>}_{0}$  
states at least for total spin
up to $S \sim \sqrt N$ (above these value of the total spin
there might be some difficulties to disentangle magnons from the
Anderson tower of states of the ground-state). This
has been checked to be true in the Heisenberg model on the
triangular lattice~\cite{blp92}.

As a last remark, the
homogeneous spin susceptibility is always dominated by the
 largest spin
states of the Anderson tower: that is states with total spin
${\cal O}(\sqrt N)$. Don't forget that a state with total spin
$\sqrt N$ has a macroscopic magnetization by site: $m=
S_{tot}/\frac {N}{2} \propto \frac {1}{\sqrt N}$ that is
essentially zero in the thermodynamic limit.  
\end{section}

\begin{section} {Space symmetry breaking of the N{\'e}el state.}
The N{\'e}el state usually breaks some space symmetries of the
lattice.
\begin{itemize}
\item In the square lattice case   (see Fig.~\ref{sqN})
one-step translations are
not a symmetry operation of the ground-state but the point group
is unbroken. This appears in the Anderson tower of states of
Fig.~\ref{sqN}, where the Irreducible Representations (IR) of the
QDJS have alternatively wave-vector ${\bf k} = (0,0)$ or
 ${\bf k} = (\pi,\pi)$~(depending on the parity of the total spin),
 but are  trivial  IR of the point group.
\item On the hexagonal lattice, which is not a Bravais lattice,
the situation is somewhat different (see Fig.~\ref{hexN}):
 the collinear N{\'e}el order
does not break either the translation group, nor $C_3$, the 
group of 3-fold rotations (noted ${\cal R}_{\frac{2\pi}{3}}$).
Only the trivial representation of these two
groups appears in the QDJS (see Fig.~\ref{hexN}). But both the
 inversion group
($C_2$, symmetry operation ${\cal R}_{\pi}$) and 
the reflection with respect to an axis joining
the center of the hexagons ($\sigma$) are broken: these symmetry
breakings appear in the Anderson tower where there is
 in the QDJS an
alternation of even and odd IRs of these two groups.
\end{itemize}
Determination of the space symmetries of each $S$ components of
the Anderson tower can be done exactly using symmetry arguments:
the space symmetries of each
$\widetilde{ |S,M_S>}_{0}$ depend on $S$, on the shape and total
number of spins of the sample~\cite{bllp94,lblp95,le95,f03}.
In the following chapter we will give an example of such
 a determination for
the $J_1-J_2$ model on the triangular lattice.  In a given range
of parameters $1/8<J_1/J_2<1$, there is a competition between
different orders and selection by quantum fluctuations of the
more symmetric one. The study of this example will show the
strength of the symmetry analysis
 and the exact nature of this phenomenon of
``order by disorder''.  

\end{section}
\end{chapter}

\begin{chapter}{``Order by disorder''}
\begin{section}{Some history}
The concept of ``order by disorder'' was introduced in 1980 by
Villain and co-workers\cite{vbcc80} in the study of a 
frustrated Ising model on the square lattice. In this model 
the  next neighbor couplings along all the rows
are ferromagnetic as well as those on the odd columns (named A in
the following).  The couplings
 on the  even columns (named B) are antiferromagnetic. 
It is assumed that 
\begin{equation}
0 \;< \;|J_{AB}|\;< \; J_{BB} \; < \; |J_{AA}|.
\end{equation}
The ground-states of this model have A columns (resp B)
ferromagnetically (resp. antiferromagnetically) ordered. For a
system with a number of sites $ N=0 \; [{\rm mod} \, 4]$, the degeneracy  of
this ground-state is $2^{\sqrt N}$, its entropy per spin $S_0 =
\frac{1}{\sqrt N} Ln 2$ is negligible in the thermodynamic
limit.
At $T=0$ the ground-state has no average magnetization and is
disordered. The picture changes when thermal fluctuations are
introduced: it is readily seen that a B chain sandwiched between
two A chains with parallel spins has lower excitations than a B
chain sandwiched between two A chains with anti-parallel spins.
This gives a larger Boltzmann weight to the ferrimagnetically
ordered system.  Villain and co-workers have been able to show
exactly that the system is indeed ferrimagnetic at low $T$.
They were equally able to show that site dilution
(introducing non magnetic species) was in a certain domain of
composition and temperature able to select the same ordered pattern,
whence the name of ``order by disorder''.

During the nineties several authors have studied a somewhat
less drastic problem in the classical  or quantum
Heisenberg model : it is the selection of a special
kind of long range order among a larger family of  ordered
solutions classically degenerate at 
T=0~\cite{s82,ont85,h89,dm89,jdgb90,cj92,k93}. In the classical
models, the selection of the simplest ordered structure by thermal
fluctuations , is due to a larger  density of low lying
excitations around these solutions, whence an increased
Boltzmann weight
of the corresponding regions and a thermal (entropic) selection
of order.

 This same property of the density of low lying
excitations can also explain a selection of specific spin
configurations when going from the $T=0$ classical approach of the 
Heisenberg model to the semi-classical one.
 Suppose that many classical spin configurations are
degenerate in the classical limit, the existence of a larger density
of excitations around a specific configuration is the signature of a
weaker restoring force toward this configuration (larger well
width in phase space). Insofar as the semi-classical spin-wave
approach is valid, this implies that the zero point quantum energy 
$\sum_{{\bf q} \in BZ^{*}} \frac{\omega_{\bf q}}{2}$ of Eq.~\ref{esw}
 is smaller for this solution, which  will thus be energetically
selected by the ``quantum'' fluctuations.
Both mechanisms (thermal or quantum) lay on the properties of the
low lying excitations around the classically $T=0$ degenerate
solutions.

This selection of order is, in most of the cases,
 less drastic in the continuous spin models, than in the original
problem of Villain. In most of the cases, the degeneracy of the
ground-state is less severe than in the Villain case. In the
Ising domino problem, the degeneracy of the ground-state is
$2^{\sqrt N}$ and the thermal selection emphasizes 4
ground-states among these $2^{\sqrt N}$. In the Heisenberg
problem, as we will see below, in most of the cases the less
ordered solution has a degeneracy of order ${\cal O}(N^{\alpha})$,
with $\alpha$ the number of sublattices, whereas the final order
selected by quantum fluctuations has only a degeneracy 
${\cal O}(N^{\beta})$ with $\beta < \alpha$. From that simple point
 of view one can qualitatively state that the selection of order is
 less drastic than in the Villain problem.
A special mention should be done of the Heisenberg model on the
kagome, checker-board or pyrochlore lattices. In these cases, on
which we will return at the end of these lectures, the degeneracy
of the classical ground-state is exponential in $N$, there is a
residual entropy per spin at $T=0$ and  a
selection (if any) of some partial order is a more difficult
issue.

\end{section}

\begin{section}{"Order by disorder" in the $J_1-J_2$ model
 on the triangular lattice }

The existence of competing interactions is indeed the main 
cause of classical ground-states degeneracy. As a generic example, 
one can consider the so-called $J_1-J_2$ model on a triangular 
lattice with two competing antiferromagnetic interactions.  
This Hamiltonian reads: 
\begin{equation}
{\cal H} = 2 J_1 \sum_{<i,j>} {\bf S}_i.{\bf S}_j +
2 J_2 \sum_{<<i,k>>} {\bf S}_i.{\bf S}_k
\label{eq-Heij2}
\end{equation}
where $J_1$ and $J_2 = \alpha J_1$ are positive and the first 
and second sums run on the first and second 
neighbors, respectively. The classical study of this model has been 
developed by Jolicoeur {\it et al.}~\cite{jdgb90}. They have 
shown that for small $\alpha$ ($\alpha < 1/8$) the ground state
corresponds to a three-sublattice N{\'e}el order with 
magnetizations at $120^o$ from each other, whereas for $1/8 < \alpha < 1$, 
there is a degeneracy between
a two-sublattice N{\'e}el and
a four-sublattice N{\'e}el order (see Fig.~\ref{4-sub&2-sub}). 
Chubukov and Jolicoeur~\cite{cj92} and Korshunov~\cite{k93} have then shown
that quantum fluctuations (evaluated in a spin wave approach)
 could, like thermal ones,
lift this degeneracy of the classical ground states and lead to 
a selection of the collinear state (see Fig.~\ref{4-sub&2-sub})
 \cite{de93}.

\begin{figure}
\hspace{2.5cm}\resizebox{8cm}{!}{\includegraphics{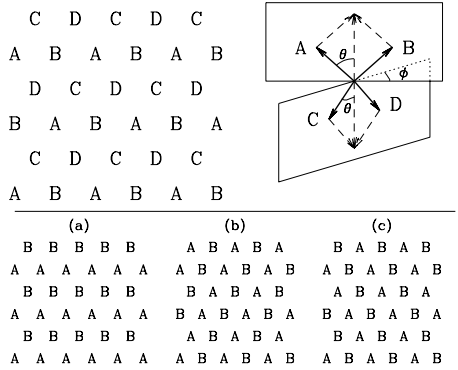}}
\caption[99]{
Top: 4-sublattice classical ground state. 
Spins in the sublattices $A$ and $B$,
as well as spins in $C$ and $D$,
make an angle $2\theta$.
The plane of the spins of $A$ and $B$ makes an angle $\phi$ with the
plane of the spins of $C$ and $D$.
Bottom:  the collinear solutions with the three possible arrangements 
(in this case, classical spins in sublattices A and B are antiparallel).
}
\label{4-sub&2-sub}
\end{figure}

The first study of the exact spectrum of Eq.~(\ref{eq-Heij2}) done by
Jolicoeur {\it et al.} was not incompatible with this conclusion, but was
insufficient to yield it immediately. I will show now how the
study of the degeneracy of the Anderson tower allows a direct
derivation of this phenomenon. This part of the lecture closely follows
the paper by Lecheminant {\t et al.}\cite{lblp95}.

As we have done in section (2.1), let us  first
study the exactly solvable models which display either
four-sublattice order
or collinear order. These models are obtained by extracting from
the Heisenberg Hamiltonian expressed in terms of the Fourier
components of the spin:
\begin{equation}
{\cal H} = 6 J_1 \sum_{\bf k} {\bf S}_{\bf k} . {\bf S}_{-\bf k}
\left[\gamma_{\bf k} + \frac{\alpha}{3}
\left(\cos {\bf k}.\left(2 {\bf u}_1 + {\bf u}_2\right) +
\cos {\bf k}.\left({\bf u}_1 + 2 {\bf u}_2\right) +
\cos {\bf k}.\left({\bf u}_2 - {\bf u}_1\right) \right) \right],
\label{eq-h1j2}
\end{equation}
where $\gamma_{\bf k} = 1/3 \sum_{\mu} \cos{\bf k}.{\bf u}_{\mu}$ 
(${\bf u}_{\mu}$ are three vectors at 120 degrees from each other and 
connecting a given site to first neighbors), 
those which describe either the 4-sublattice structure or the
collinear ones.

\begin{subsection}{Symmetry analysis of the Anderson tower of
the 4-sublattice N{\'e}el order.}
The four ${\bf k}$ vectors which keep the four-sublattice order invariant are
${\bf k} = {\bf 0}$ and the three middles of the Brillouin 
zone boundaries (called in the following $\bf k_I$, $\bf k_H$ and $\bf k_G$). 
It is straightforward
to write the contribution of these Fourier components to ${\cal H}$ 
in the form: 
\begin{equation}
 ^4{\cal H}_{0} = \frac{8}{N}\left(J_1 + J_2 \right) \left( {\bf S}^2
- {\bf S}_{A}^2 -
{\bf S}_{B}^2 - {\bf S}_{C}^2 - {\bf S}_{D}^2 \right),
\label{eq-h0j2}
\end{equation}
where ${\bf S}$ is the total spin operator and ${\bf S}_{\alpha}$ are 
the total spin operator of each sublattice. $ ^4{\cal H}_{0}, {\bf S}^2,
 {\bf S}_{A}^2, {\bf S}_{B}^2, {\bf S}_{C}^2$ and ${\bf S}_{D}^2$ 
form a set of commuting observables. The eigenstates  of $^4{\cal H}_{0}$
have the following energies: 
\begin{equation}
\begin{array}{clc}
 ^4E(S, S_{A}, S_{B}, S_{C}, S_{D}) = & \displaystyle
\frac{8}{N} \left(J_1 + J_2 \right)
\left[S(S+1) -  S_{A}\left(S_{A} + 1 \right) \right. \\
& \left. \displaystyle
- S_{B}\left(S_{B} + 1 \right)
- S_{C}\left(S_{C} + 1 \right) -S_{D}\left(S_{D} + 1 \right) \right]
\label{eq-h01j2}
\end{array}
\end{equation}
where the quantum numbers $S_{A},S_{B},S_{C},S_{D}$ run from 
$0$ to $N/8$ and the total spin results from a coupling of 
four spins $S_{A},S_{B},S_{C},S_{D}$.

 The low lying levels 
of Eq. \ref{eq-h01j2} are obtained for $S_{A}=S_{B}=S_{C}=S_{D}=N/8$:
\begin{equation}
 ^4E_{0}(S) = -\frac{J_1+J_2}{2}\left( N+8 \right) + \frac{8}{N}
\left( J_1 + J_2 \right) S \left( S + 1 \right).
\label{eq-enh02j2}
\end{equation}
These states, which have  maximal sublattice magnetizations
$S_A^2 = S_B^2 = S_C^2 = S_D^2 ={\frac {N}{8}}(\frac {N}{8}+1)$,
are the rotationally invariant projections of the
bare\footnote{We may say Ising-like N{\'e}el state, as these states
can be deduced from  Ising states of  the four sublattices pointing in
the principal directions of a regular tetrahedron.}
N{\'e}el states with four sublattices. Their total energy 
collapses to the absolute ground-state as $N^{-1}$ and form the Anderson
tower of the 4-sublattice N{\'e}el order (noted $\{ ^4\tilde E\}$
in the following).

As we will now show, this multiplicity $\{ ^4\tilde E \}$ can be
entirely and  uniquely described by its
symmetry properties  under spin rotations and transformations of 
the space group of the lattice.

Let us begin by the {\bf SU(2) properties} of  $\{ ^4\tilde E\}$.
These states result from the  coupling of four identical spins
 of length $N/8$. There is $N_S$ different ways to couple these
4 spins: 
 the degeneracy of each $S$ subspace  is thus $(2S+1)N_S$,
 where the $(2S+1)$ factor comes from the magnetic
 degeneracy of each $S$ eigen-state.
$N_S$ is readily evaluated 
by using the decomposition of  the product of four  spin $N/8$ 
representations of
$SU(2)$ (${\cal D}^{N/8}$) 
\begin{equation}
\{ ^4\tilde E\}= {\cal D}^{N/8}\otimes  {\cal D}^{N/8}\otimes  {\cal D}^{N/8}\otimes  {\cal D}^{N/8}
\label{e4}
\end{equation}
in  spin $S$ irreducible representations 
(${\cal D}^{S}$).
One obtains: 
\begin{eqnarray}
N_S &= & \frac{1}{2}\left(-3 S^2 + S\left(N+1\right) +2 +\frac{N}{2}
			  \right)~~~~ {\rm for~~~} S\leq \frac{N}{4}, \\
      & = &\frac{1}{2}
\left(\frac{N}{2}-S+1\right) \left(\frac{N}{2}-S+2\right)
				~~~~ {\rm for~~~} S\geq \frac{N}{4}+1.
\label{eq-mulsj2}
\end{eqnarray}
Note that this degeneracy depends both 
on $S$ and $N$ and not only on the total spin $S$ as 
is the case for a two or three-sublattice problem.

The determination of the {\bf space symmetries} of these eigenstates 
allows a complete specification of $\{ ^4\tilde E\}$.
\begin{itemize}
\item The four-sublattice order is invariant in a two-fold rotation: 
the eigenstates of  $\{ ^4\tilde E\}$ belong to
the trivial representation of $C_2$.
\item $\{ ^4\tilde E\}$ forms a representation  of $S_4$,
 the permutation group of four elements.
 The eigenstates of $\{ ^4\tilde E\}$
could thus be labeled by the irreducible representations (I.R.) 
of  $S_4$ 
(see Table~\ref{table-1}).

\begin{table}
\begin{center}
\begin{tabular}{|c|c c c c c |}
\hline
$S_4$ & $ I $ & $(A,B)(C,D)$ & $(A,B,C)$ & $(A,B)$ & $(A,B,C,D)$\\
${\cal G}$ & $I$ & $t$ & ${\cal R}_{2\pi/3}$ & $\sigma$ & ${\cal R}^{'}_{2\pi/3} \sigma$ \\
  $ N_{el}$ & 1 & 3 & 8 & 6 & 6\\
  \hline
   $\Gamma_1 $ & $ 1 $ & $ 1 $ & $ 1 $ & $ 1 $ & $ 1 $ \\
   $\Gamma_2 $ & $ 1 $ & $ 1 $ & $ 1 $ & $-1 $ & $-1 $ \\
   $\Gamma_3 $ & $ 2 $ & $ 2 $ & $-1 $ & $ 0 $ & $ 0 $ \\
   $\Gamma_4 $ & $ 3 $ & $-1 $ & $ 0 $ & $ 1 $ & $-1 $ \\
   $\Gamma_5 $ & $ 3 $ & $-1 $ & $ 0 $ & $-1 $ & $ 1 $ \\
\hline
\end{tabular}
\end{center}
\caption[99]{Character table of the permutation group $S_4$.
First line indicates classes of permutations. 
Second line gives an element of the space symmetry class 
corresponding to the class of permutation. These space symmetries are:
 the one step translation $t$
  ($A\to C$),  ${\cal R}_{2\pi/3}$ (resp. ${\cal
R}^{'}_{2\pi/3}$)
the three-fold rotation around a site of the
$D$ (resp. $B$) sublattice, and $\sigma$ the axial symmetry keeping invariant $C$ and $D$.
$N_{el}$ is the number of elements of each class.
}
\label{table-1}
\end{table}
\item Each element of the space group maps onto a permutation of  $S_4$ : 
one step translations onto products of transpositions as $(AC)(BD)$,
three-fold rotations onto circular permutations of three sublattices $(ABC)$
and so on. The
complete mapping of the space symmetries of the four-sublattice 
order onto the permutations of $S_4$ is given in
Table~\ref{table-1} together with the character table of $S_4$.
\item Each
irreducible representation of $S_4$ can thus be characterized
in terms of its space symmetry properties. As noted above they are all
invariant in $\cal{R_{\pi}}$.  $\Gamma_1,\Gamma_2,\Gamma_3$  belong
to the trivial IR of the translation group, characterized by
 the wave-vector ${\bf 0}$, whereas $\Gamma_4$ and
$\Gamma_5$
have a wave-vector ${\bf k}_{H}, {\bf k}_{I}$ or ${\bf k}_{G}$.
$\Gamma_1$  and $\Gamma_2$ belong  to the trivial I.R. of
$C_3$, whereas $\Gamma_3$ is the 2 dimensional representation of
this same group. Finally,  $\Gamma_1$ and $\Gamma_4$ are even under 
axial symmetry,
whereas $\Gamma_2$ and $\Gamma_5$ are odd.
\item The number of replicas of $\Gamma_i$ that should appear 
for each $S$ is then computed in the $S,M_S$ subspace
with the help of the trace of the permutations of $S_4$:
\begin{equation}
	\label{eq-ngammai}
	n_{\Gamma_i}^{(S)} = \frac{1}{24} \sum_l {\rm Tr}(R_l|_S) \chi_i(l) N_{\rm
	el}(l)
\label{noc}
\end{equation}
where $R_l$ is an element of the class  $l$ of  $S_4$, $N_{\rm el}(l)$ 
is the number
of elements of the group in this class and
 $\chi_i(l)$ the character of the class $l$ 
in the I.R. $\Gamma_i$ (see Table \ref{table-1}).
The values of the traces for a given total spin $S$ are then found
as:
\begin{equation}
{\rm Tr} \left( R_l\bigg |_{S} \right) =
{\rm Tr} \left( R_l\bigg |_{M_S=S} \right) -
{\rm Tr} \left( R_l\bigg |_{M_S=S+1} \right).
\label{eq-tr2sz4}
\end{equation}
In each $M_S$ subspace of $\{ ^4\tilde E\}$ ,
it is straightforward to find the trace of the elements of  $S_4$:
\begin{eqnarray}
\nonumber
{\rm Tr} \left(I_d\bigg |_{M_S} \right) &=&
	\sum_{t,v,x,y=-N/8}^{N/8} \delta_{t+v+x+y ,M_S} \\
\nonumber
{\rm Tr} 
	\left(\left(A,B\right) \left(C,D\right)\bigg|_{M_S} \right) &=&
	\sum_{t,v=-N/8}^{N/8} \delta_{2t+2v, M_S}\\
\label{eq-holpri}
{\rm Tr} \left(\left(A,B,C\right)\bigg |_{M_S} \right) &=&
	\sum_{t,v=-N/8}^{N/8} \delta_{3t+v,M_S} \\
\nonumber
{\rm Tr} \left(\left(A,B\right)\bigg |_{M_S} \right) &=&
	\sum_{t,v,x=-N/8}^{N/8} \delta_{2t+v+x,M_S} \\
\nonumber
{\rm Tr} \left(\left(A,B,C,D\right)\bigg |_{M_S} \right) &=&
	\sum_{t=-N/8}^{N/8} \delta_{4t,M_S}
\end{eqnarray}
 where $t,v,x,y$, are the $z$-components of the total spin 
of each sublattice (constrained to vary between $N/8$ and $-N/8$) and
$\delta_{i,j}$ denotes the Kronecker symbol.
Using equations (\ref{noc}, \ref{eq-tr2sz4}, \ref{eq-holpri})
one readily obtains the number of occurrences of each $\Gamma_i$
for any  $S$ subset of $\{ ^4\tilde E\}$ (Table~\ref{table-2}).
Note that this result depends on $S$ and on the size of the
sample.

\begin{table}
	\begin{center}
		\begin{tabular}{|c|c c c c c c c c c|}
			\hline
			$N=16$ &&&&&&&&& \\
			$S$ & 0 & 1 & 2 & 3 & 4 & 5 & 6 & 7 & 8 \\
			\hline
			$n_{\Gamma_1}(S)$ &1 & 0 & 2 & 0 & 2 & 1 & 1 & 0 & 1 \\
			$n_{\Gamma_2}(S)$ &0 & 0 & 1 & 0 & 0 & 0 & 0 & 0 & 0 \\
			$n_{\Gamma_3}(S)$ &2 & 0 & 2 & 1 & 2 & 0 & 1 & 0 & 0 \\
			$n_{\Gamma_4}(S)$ &0 & 2 & 2 & 3 & 2 & 2 & 1 & 1 & 0 \\
			$n_{\Gamma_5}(S)$ &0 & 2 & 1 & 2 & 1 & 1 & 0 & 0 & 0 \\
			\hline
		\end{tabular}
\hspace*{-0.5cm}\begin{tabular}{|c|c c c c c c c c c c c c c c c|}
			\hline
			$N=28$ &&&&&&&&&&&&&&&\\
			$S$ & 0 & 1 & 2 & 3 & 4 & 5 & 6 & 7 & 8 & 9 & 10 & 11 & 12 & 13
			& 14\\
			\hline
			$n_{\Gamma_1}(S)+n_{\Gamma_2}(S)$& 2&0&5&1&5&3&4&2&4&1&2&1&1&0&1\\
			$n_{\Gamma_3}(S)$& 3&0&4&2&5&2&5&2&3&1&2&0&1&0&0\\
			$n_{\Gamma_4}(S)+n_{\Gamma_5}(S)$& 0&7&6&11&9&12&9&10&6&6&3&3&1&1&0\\
			\hline
      \end{tabular}
	\end{center}
\caption[99]{Number of occurrences $n_{\Gamma_i}(S)$ of each irreducible 
representation $\Gamma_i$ with respect to the total spin $S$.
For $N=28$,
$n_{\Gamma_1}$ and $n_{\Gamma_2}$ as well as
$n_{\Gamma_4}$ and $n_{\Gamma_5}$ have been added because this sample
does not present any axial symmetry.
}
\label{table-2}
\end{table}
\end{itemize}

This symmetry analysis completes the determination of the QDJS of 
 $\{ ^4\tilde E\}$. These properties of the Anderson tower are
stable under the action of the discarded part of the Heisenberg
Hamiltonian. If the ordering of levels is not destroyed by quantum
 fluctuations, the associated quantum numbers remain good quantum
numbers of the low lying levels $\widetilde{ |S,M_S>}_{0}$ 
of the $J_1-J_2$ model (\ref{eq-Heij2}).  We have thus obtained
the complete determination (all quantum numbers , and all the
degeneracies) of the family of low lying levels describing the
ground-state multiplicity $\{ ^4\tilde E\}$ of the four-sublattice N{\'e}el solutions.
\end{subsection}
\begin{subsection}{Symmetry analysis of the QDJS of the Anderson
tower of states of the 2-sublattice collinear solutions.}

Let us now consider the  collinear solutions (Fig.~\ref{4-sub&2-sub}). 
They are particular
solutions of the 4-sublattice case and  we will rapidly get through 
 the same scheme of analysis, indicating mainly the new points.
The two vectors which keep the two sublattices invariant are
${\bf 0}$ and the middle of one side of the Brillouin 
zone (the vectors $\bf{k_I}$, $\bf{k_H}$ and $\bf{k_G}$ 
correspond respectively to the
collinear solutions $(a)$, $(b)$ and $(c)$ in Fig.~\ref{4-sub&2-sub}). 
Extracting  a specific set of two wave-vectors
from Eq. \ref{eq-h1j2}, we find the following 
contribution to the total Hamiltonian: 
\begin{equation}
 ^2{\cal H}_{0} = \frac{8}{N}\left(J_1 + J_2 \right) \left[ {\bf S}^2
- \frac{1}{2} \left( {\bf S}_{\alpha }^2 + {\bf S}_{\beta}^2 \right) \right].
\label{eq-colj2}
\end{equation}
The corresponding low energy spectrum for $S_{\alpha}=S_{\beta}=N/4$ is: 
\begin{equation}
 ^2E_{0}(S) = -\frac{J_1+J_2}{2}\left( N+8 \right) + \frac{8}{N}
\left( J_1 + J_2 \right) S \left( S + 1 \right)
\label{eq-col1j2}
\end{equation}
and is degenerate with the four-sublattice low energy spectrum
(see Eq. \ref{eq-enh02j2}). 
But here the two-sublattice have maximal spins
$S_{\alpha }=S_{\beta}=N/4$. These new solutions arise from the
three symmetric couplings of the 4-sublattice spins:
$ \bf{S}_{\alpha }= \bf{S}_{A}+ \bf{S}_{B}$ or 
 $\bf{S}_{\alpha }= \bf{S}_{A}+ \bf{S}_{C}$ or
 $\bf{S}_{\alpha}= \bf{S}_{A}+ \bf{S}_{D}$
with the symmetric counterparts for $ \bf{S}_{\beta}$.
These collinear solutions have thus a $Z_3$ degeneracy
(see Fig.~\ref{4-sub&2-sub}).
The representation space is thus the sum of three products
${\cal D}^{N/4}\otimes{\cal D}^{N/4}$. It is not a direct 
sum since ${\cal D}^{N/4}(A,B) \otimes {\cal D}^{N/4}(C,D)$ and
${\cal D}^{N/4}(A,C) \otimes {\cal D}^{N/4}(B,D)$ have in 
common the same (symmetric) irreducible representation 
with a total spin $N/2$. On an $N$-sample, the representation 
space of the ground state of the collinear solution is: 
\begin{equation}
\{^2\tilde E\} = 3 {\cal D}^{S=0} \oplus 3 {\cal D}^{S=1} \oplus ....
\oplus 3{\cal D}^{S=N/2-1} \oplus {\cal D}^{S=N/2}. 
\label{eq-spacetwo}
\end{equation}
The degeneracy is thus $3(2S+1)$ for all $S$ values 
except for $S=N/2$, where it is only $(2S+1)$.

The space group analysis is identical to the analysis done for the
four-sublattice order, but the number of occurrences of each I.
R.  $\Gamma_i$ is now different , since the space 
$\{ ^2\tilde E\}$ is smaller than $\{ ^4\tilde E\}$. 
For each $S$ value there are only
three replicas of ${\cal D}^{S}$ arising from the $Z_3$ symmetry 
( Eq. \ref{eq-spacetwo} and Fig.~\ref{4-sub&2-sub}).
 This allows the direct
computation of the traces of the operations of $S_4$ in each $S$
subset of $\{ ^2\tilde E\}$.
Using the  coupling rules of  two angular momenta
(and in particular the fact that the $S$ eigen-state resulting
from the coupling of two integer spins changes sign as
$(-1)^S$ with the interchange of the two parent spins) one 
obtains (for $S \neq N/2$):
\begin{equation}
\left\{
\begin{array}{l}
\displaystyle {\rm Tr} \left( I_d |_{S}\right) = 3\\ 
\displaystyle {\rm Tr} \left( \left(A,B\right)\left(C,D\right)|_{S}\right) = 
1 + 2\left(-1\right)^{S}\\ 
\displaystyle {\rm Tr} \left( \left(A,B,C\right)|_{S}\right) = 0\\ 
\displaystyle {\rm Tr} \left( \left(A,B\right)|_{S}\right) = 1\\ 
\displaystyle {\rm Tr} \left( \left(A,B,C,D\right)|_{S}\right) = \left(-1\right)^{S} 
\end{array}
\right.
\label{eq-trij24}
\end{equation}
Therefore  the collinear solution is simply 
characterized by $\Gamma_1$ and $\Gamma_3$ 
for even $S$, and $\Gamma_4$ for odd $S$, whatever the sample size.

The symmetries of all states of the tower are now fully
determined both for the 4-sublattice order  $\{ ^4\tilde E\}$
and for the collinear order $\{ ^2\tilde E\}$. If the quantum
Hamiltonian presents one of these kinds of order, the quantum
fluctuations generated by the discarded part of $\cal H$ should
preserve the dynamics and the structure of these low lying subsets.
\end{subsection}
\begin{subsection}{Exact spectra of the $J_1-J_2$ model on small
samples and finite size effects: a direct illustration of the
phenomenon of ``order by disorder''.}
We have determined the low (and high) energy levels of the $J_1-J_2$ Hamiltonian
in each I. R. of $SU(2)$ and of the space group of
the triangular lattice for small periodic samples with $N=12, 16$ and $N=28$.
The spectra are
displayed in Fig.~\ref{fig-N=16} and Fig.~\ref{fig-N=28}.
\begin{figure}
\hspace{2.5cm}\resizebox{8cm}{!}{\includegraphics{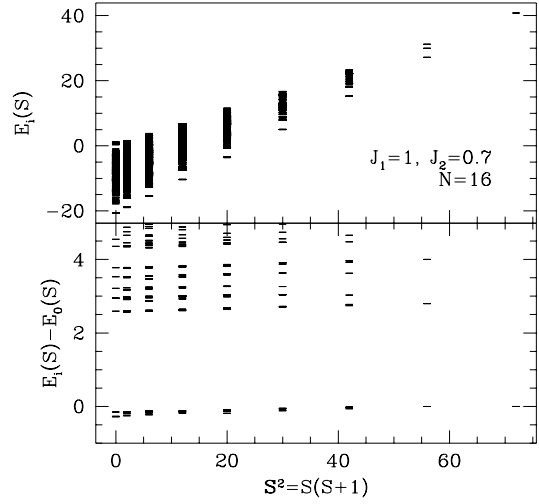}}
\caption[99]{
Top: complete spectrum of the $N=16$ periodic sample with respect to 
${\bf S}^2$.
Bottom: enlargement of the difference between the exact spectrum and
the  energy of the low lying levels of  the model Hamiltonians 
(Eq.\ref{eq-enh02j2} or Eq.\ref{eq-col1j2}).
The  ground-state multiplicity  $\{ ^4\tilde E\}$ 
is  well separated from the magnons.
}
\label{fig-N=16}
\end{figure}
\begin{figure}
\hspace{2.5cm}\resizebox{8cm}{!}{\includegraphics{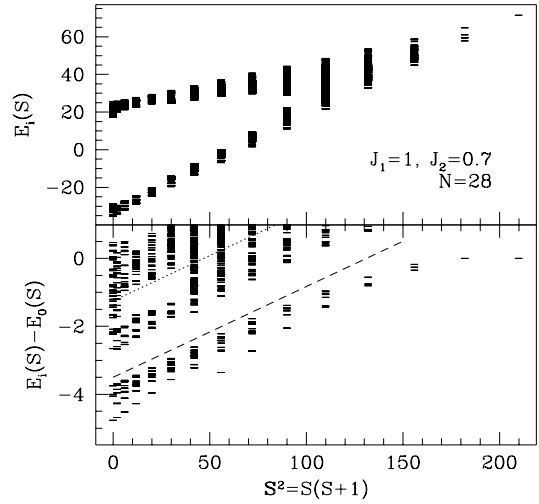}}
\caption[99]{
Partial spectrum of the $N=28$ periodic sample.
(Same legend as for fig.\ref{fig-N=16}). Bottom:
the tower of states of the 4-sublattice order  $\{ ^4\tilde E\}$ lays under the dashed line.
Above appear the first magnons.
Above the dotted line are represented the first excited homogeneous states.
In the magnon multiplicity ($\bf k \neq \bf 0 ,
{\bf k_{H}}, {\bf k_{I}}$ or ${\bf k_{G}}$),
 for $S\leq 5$,
only the lowest 5 states of each I.R.
have been computed.  
}
\label{fig-N=28}
\end{figure}
 We directly see in the upper parts of these figures the set
of QDJS ("Anderson tower of the ground-state") well separated from
 the set of levels
corresponding to the one magnon excitations. We have verified that
this set has the symmetry properties  of the above defined
$\{ ^4\tilde E\}$ subset.
The action of the quantum fluctuations could then be read in 
the  lower parts of the figures. 
As expected, 
the quantum fluctuations lift the degeneracies 
which are present in the exactly solvable model
and stabilize the eigenstates with the lower $S$ values.
Nevertheless the low lying 
energies per site still group around a line of slope  
${\cal O}[(J_1+J_2)/N^2)$.
The  number and space symmetries of these levels 
for each $S$ and $N$ value are exactly those required by 
the above analysis of the four-sublattice N{\'e}el order. 

Moreover, it is already visible on the $N=16$ sample and quite clear 
on the $N=28$ sample that a dichotomy 
appears in this family (see Fig.~\ref{fig-QDJ}).
 The lowest levels of this tower 
of states appear to be $\Gamma_1,\Gamma_3$ or $\Gamma_4$ 
representations depending on the parity of the total 
spin. They precisely build the family  $\{ ^2\tilde E\}$ of isotropic 
projections of the collinear solutions (Eq. \ref{eq-trij24}). 
\begin{figure}
\hspace{2.5cm}\resizebox{8cm}{!}{\includegraphics{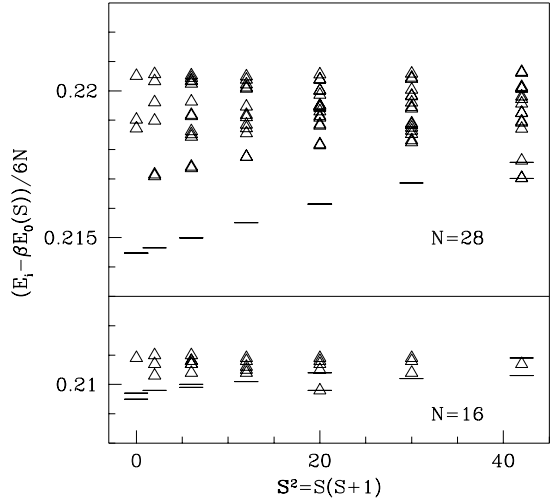}}
\caption[99]{
Enlargement of the $N=16$ and $N=28$ QDJS. A global contribution
$ \beta E_{0}(S)$ is subtracted from the exact spectrum. This
contribution describes the overall dynamics of the order parameter
in this finite sample, $\beta $ measures the renormalization of
this dynamics by quantum fluctuations (see refs.~\cite{bllp94,lblp95}).
The bars represent
eigenstates which belong both to $\{ ^2\tilde E\}$ and $\{ ^4\tilde E\}$.
The triangles indicate states which belong to $\{ ^4\tilde E\}$  but
not to $\{ ^2\tilde E\}$. With increasing sizes,
the subset $\{ ^2\tilde E\}$ is stabilized and separates from the
pure 4-sublattice order.
For $N=28$
the  two  states of $\{ ^2\tilde E\}$ with even $S$ are quasi degenerate and
cannot be distinguished at the scale of the figure.
}
\label{fig-QDJ}
\end{figure}
This strongly suggests that
the 4-sublattice order will  disappear in the thermodynamic limit
and only the collinear order will subsist, as was predicted in
the spin-wave approach~\cite{jdgb90,cj92,k93}.
\end{subsection}
\end{section}
\begin{section} {Concluding remarks}
\begin{itemize}
\item The symmetry and dynamical analysis of the low lying levels 
of a Hamiltonian likely to exhibit ordered solutions 
gives rather straightforward answer to the kind 
of order to be expected. The method is rapid, powerful 
and unbiased. It does not require any {\it a priori} symmetry breaking
choice: if a specific order is selected, one 
should see it directly on the exact spectra. Moreover, as 
it is essentially exact, there are no questions relative 
to the convergence of the expansion as in the spin-wave 
approach. On the other hand, as the sizes amenable to 
computation are limited, there is, in the exact 
approach, a cut-off of the long wavelength fluctuations.
Results so obtained should thus be examined in the
light of a finite size scaling analysis.
This work nevertheless shows that it is not necessary to invoke
quantum fluctuations with very long wave-lengths to select the collinear order.

\item The selection of ``order by disorder'' appears in a
particular clear light. Increasing the sample size, increases the
presence of long wave-length fluctuations. We see on this example
how these long wave-length fluctuations realize a differential
stabilization of the $\{ ^2\tilde E\}$ subset, favoring collinear
order and progressively wiping out 4-sublattice order. This fully
support the spin-wave calculations. This is also a clear
illustration of the previous comment on the ``non drastic''
character of this phenomenon in this peculiar case. Without
fluctuations the system has already some order: the role of the
quantum fluctuations is just to restore a higher degree of symmetry
to the ground-state solution.

\item Going along this route one may
always speculate if in the thermodynamic limit, quantum
fluctuations could not completely restore the symmetries of the
Hamiltonian. Spin-waves calculations, so long they are 
consistent at small sizes with exact diagonalizations and
self-consistent when going to the thermodynamic limit are
credible. 
 This comparison is always  useful and relevant: in the
$J_1-J_2-J_3$ model on the hexagonal lattice~\cite{fsl01,f03},
 there are regions of
 parameter space where finite-size exact diagonalizations give
 N{\'e}el Long Range Order whereas spin-waves in the thermodynamic limit
indicate an absence of sublattice magnetization! We have verified
in each of these cases that at small sizes  the semi-classical
solution in the spin-wave approach was equally 
robust and was only destroyed by very long
wave-length fluctuations.

\item On the other hand, in the situations where we claim an absence of
N{\'e}el Long Range Order and a more exotic phase (see next chapters on
Valence Bond Crystals and Resonating Valence Bond Spin Liquids)
the Anderson tower of states is absent even on the smallest sizes.
In such a case it is quite clear that the spin-wave approach
should be discarded for spin-1/2.
\end{itemize}
\end{section}
\end{chapter}

\begin{chapter}{Valence Bond Crystals}
\begin{section}{Introduction}

In our quest of exotic quantum ground-states, we will now
describe some examples where the semi-classical N{\'e}el order is
not the ground-state of the problem and $SU(2)$ symmetry is not
broken.

 In this chapter we will concentrate on solutions where
there is {\bf long range order in the dimer coverings}: we call these
phases
{\bf Valence Bond Crystals} (in the following noted VBC).

 Such solutions
are well known in 1-dimensional problems as for example in the
A.F. $J_1-J_2$ model: 
\begin{equation}
{\cal H} =  J_1 \sum_{<ij>}  {\bf S}_i. {\bf S}_j +  J_2
\sum_{<<ij>>}  {\bf S}_i. {\bf S}_j
\label{J-1j_2}
\end{equation}
where the first (resp. second) sums run on first (resp. second) neighbors.
In 1-d, for $J_2/J_1 > 0.24$, the ground-state is
dimerized and there is a gap to the first excitations: this is
the simplest case of a VBC.

What is the situation in 2-d?

 In a classical approach, the ground-state of Eq.~(\ref{J-1j_2})
on a square lattice has a soft mode at 
 $(\pi,\pi)$ for  $J_2/J_1 < 0.5$.
At $J_2/J_1 =0.5$, the $(\pi,\pi)$ order is degenerate with 
4-sublattice order and  collinear $(\pi,0)$ or $(0,\pi)$ order.
For $J_2/J_1 > 0.5$, quantum fluctuations select the collinear 
$(\pi,0)$ or $(0,\pi)$ order by the phenomenon of ``order by disorder'' 
(see Fig.~\ref{sqJ1J2}).
\begin{figure}
\hspace{-1.5cm}
\resizebox{!}{4.2cm}{
\includegraphics{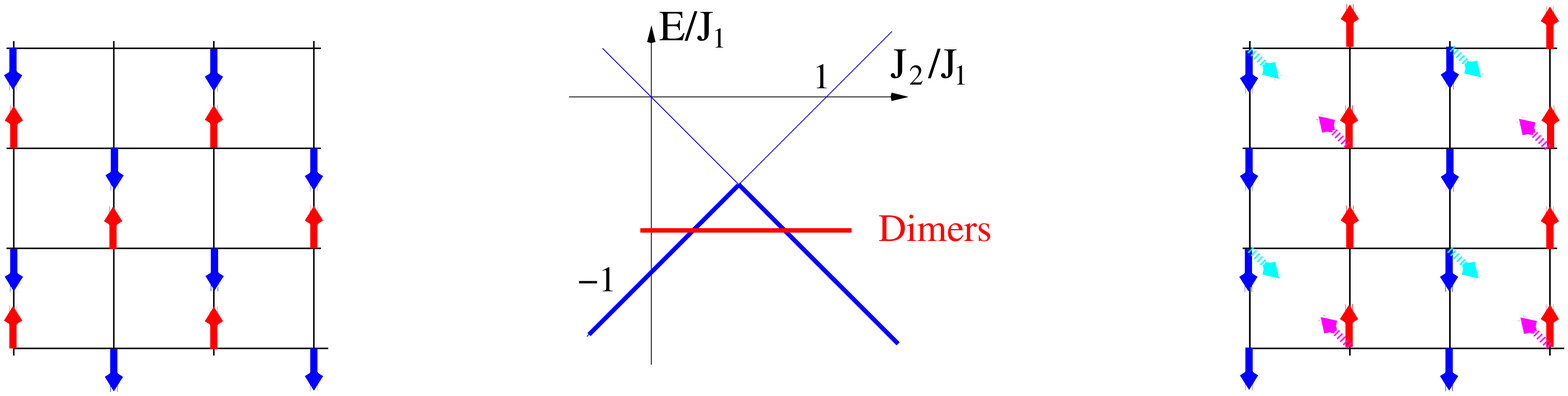}}\\
\hspace*{7.5cm}
Selection of order by
disorder\\
\hspace*{6.5cm}{\& Restoration of symmetry by fluct.}

\vspace*{3cm}
\hspace*{-1cm}
\resizebox{!}{4.cm}{
\includegraphics{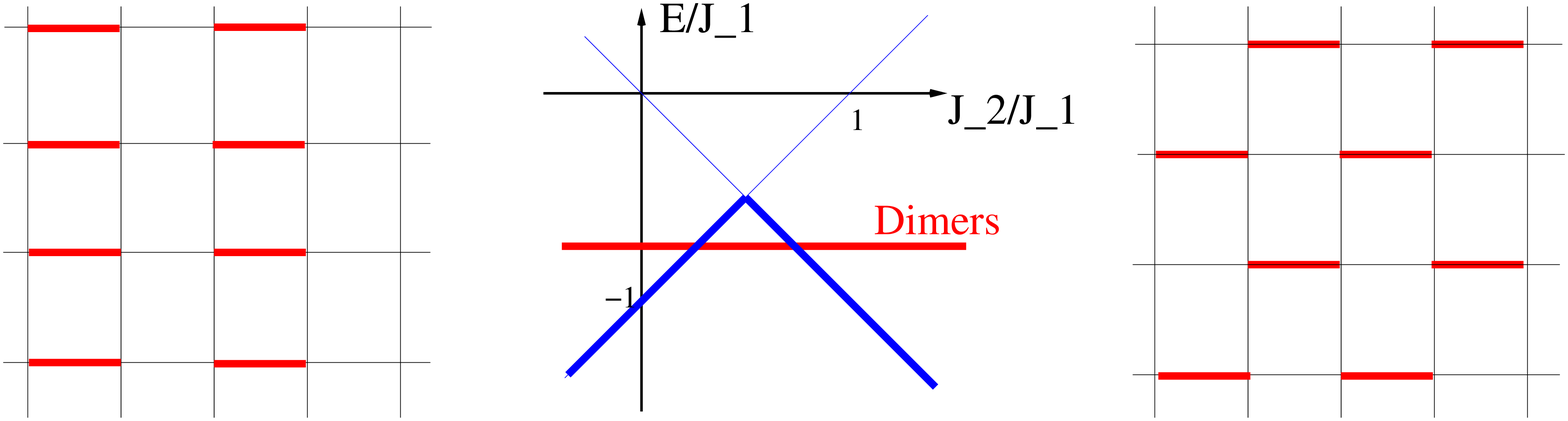}}\\
\vspace*{0.5cm}

{\hspace*{-0.4cm}
 columnar VBC  \hspace{7.5cm} staggered VBC
}
\caption [99]{ Schematization of different variational solutions
of the $J_1-J_2$ model described in the introduction of this
chapter}
\label{sqJ1J2}
\end{figure}

In our naive approach of chapter 1, comparing the energies of
classical N{\'e}el solutions to dimer covering ones, 
we would conclude that dimer covering solutions and VBC are more stable
than any classical N{\'e}el order in a large range of parameters
 around $J_2/J_1 =0.5$ (Fig.~\ref{sqJ1J2}).

In fact ``quantum fluctuations'' stabilize the N{\'e}el states and
the window for an exotic phase is smaller than indicated in
Fig.~\ref{sqJ1J2}. The nature of the quantum phase on the
square lattice at $J_2/J_1 =0.5$ 
is still debated~\cite{sz92,zu96,kosw99,kosw99a,cs00,cbps01}.
A columnar
VBC has been identified in the same model on the honeycomb
lattice for $J_2/J_1 \sim 0.4$ (see ref.~\cite{fsl01} and refs.
therein). For a pedagogical illustration we will
move  to a more clear-cut example:  the
Heisenberg model on the checkerboard lattice~\cite{pc02,fmsl03}
(noted in the following HCKB).
 
\end{section}
\begin{section} {The Heisenberg model on the checker-board
lattice: an example of a Valence Bond Crystal}

The checker-board lattice is made of corner sharing tetrahedrons,
with all bonds equal: this a 2-dimensional slice of a pyrochlore
lattice. The underlying Bravais lattice is a square lattice and
there are two spins per unit cell (Fig.~\ref{checkerboard2}).
\begin{figure}
	\begin{center}
	\resizebox{6cm}{!}{\includegraphics{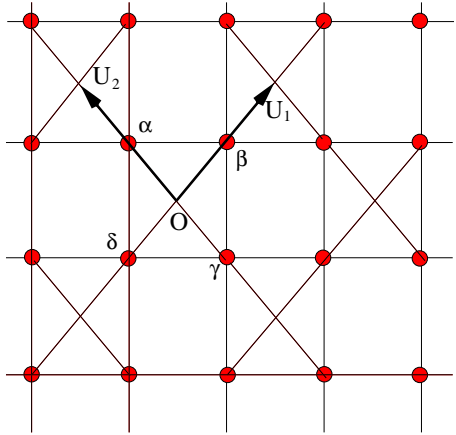}}  \end{center}

	\caption[99]{    The checkerboard lattice: the spins sit
at the vertices shown by bullets, all couplings are identical,
$\bf u_1, u_2$ are the unit vectors of the Bravais lattice.
	}  \label{checkerboard2}
\end{figure}

\begin{subsection} {Classical ground-states}

The Heisenberg Hamiltonian on such a lattice is highly degenerate
in the classical limit. Due to the special form of the lattice
this Hamiltonian can be rewritten  as the sum of
the square of the total spin of corner sharing units $\alpha$
:
\begin{equation}
{\cal H} =  J \sum_{(i,j)\,bonds} {\bf S}_i.{\bf S}_j
\equiv \frac {J}{2} \sum_{\alpha\,units} {\bf S_{\alpha}}^2 -\frac{NJ}{4}.
\label{eq-Heis_checkerboard}
\end{equation}
A classical ground-state is obtained whenever
 $ \forall \alpha \hspace*{0.3cm} {\bf S_{\alpha}}=0$.
Such ground-states have a continuous local
degeneracy and an energy $-(NJ)/4$. This is much higher 
than the dimer covering energy, which is $-(3NJ)/8$. As
we will see below, there is no memory of these classical solutions
in the quantum ground-states and low lying excitations of this
model.
\end{subsection}

\begin{figure}
\begin{center}
\resizebox{8cm}{!}{\includegraphics{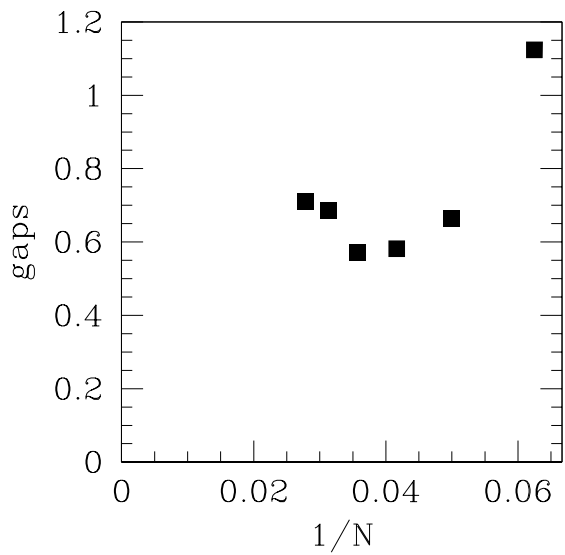}}
\end{center}
\caption[99]{Gap between the absolute ground-state and the first
S=1 excitation of the HCKB model versus sample
sizes.}
\label{gckb}
\end{figure}

\begin{subsection} {The Quantum HCKB model:  Spin Gap}

As we have seen in chapter 2, the first characteristic of the
semi-classical N{\'e}el like solution is the existence of the
Anderson tower of states which collapse to the ground-state as ${\cal
O}(1/N)$ and the absence of spin gap in the thermodynamic limit
(see for example Fig.~\ref{hexN}). 

The first salient feature of the Heisenberg model on the
checker-board lattice is the existence of a large spin gap, which
shows no tendency of going to zero at the thermodynamic
limit~(compare Fig.~\ref{gckb} with Fig.~\ref{hexN}).
This indicates that the ground-state does not break the $SU(2)$
symmetry of the Hamiltonian, and as a corollary we expect that
the spin-spin correlations decrease to zero at large distance
(which seems well verified, see Table IV of ref.~\cite{fmsl03}).
\end{subsection}

\begin{subsection} {Degeneracy of the ground-state and space
symmetry breaking in the thermodynamic limit}

The low lying levels  of the spectra of the  HCKB model in the
singlet space are displayed in Fig.~\ref{S0sector}.

\begin{figure}
\begin{center}
\vspace*{-1cm}
\resizebox{8cm}{!}{\includegraphics{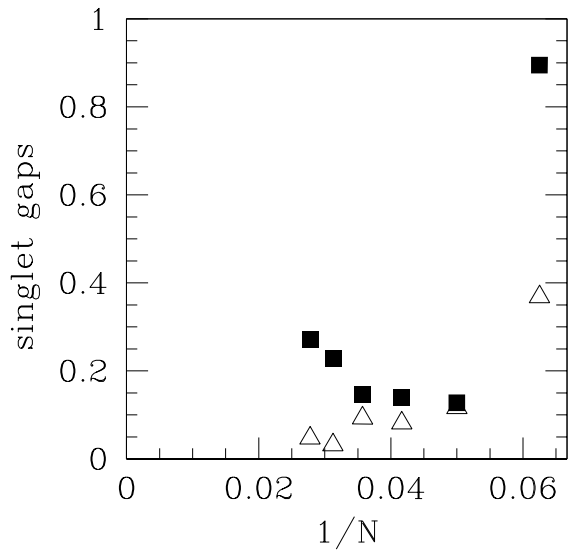}}\end{center}
\caption [99]{ Gaps  to the first (open up
triangles) and second (black squares) level of the singlet sector.
 For the
studied samples these two ``excited'' singlet levels are in the
singlet-triplet gap (See Fig.~\ref{gckb}).}
\label{S0sector}
\end{figure}

In this figure, one reads that the first excited singlet state
very plausibly collapses to the absolute ground-state, whereas a
finite gap to the third S=0 level (perhaps smaller than the spin gap)
 build on with sample size.
 This pleads in favor of a 2-fold  degeneracy of the
absolute ground-state in the thermodynamic limit.

 The absolute ground-state is in the
trivial representation of the lattice symmetry group. Its
wave function is invariant in any translation and
 in any operation of $D_4$: group
of the $\pi /2$ rotations around  point O (or any equivalent point
 of the Bravais lattice)  and axial symmetries with respect to axes 
$\bf u_1$ and $\bf u_2$ (see Fig.~\ref{checkerboard2}).
  The excited
state which collapses on it in the thermodynamic limit
 has a wave vector $(\pi,\pi)$ (its wave function takes a
(-1) factor in one-step translations along  $\bf u_1$ or $\bf u_2$), and it is
odd under $\pi /2$ rotations 
and axial symmetries.
In the thermodynamic limit the 2-fold degenerate ground-state
can thus exhibit a spontaneous symmetry breaking with a doubling
of the unit cell.

 Such a restricted  symmetry breaking does
not allow a columnar or staggered configuration of dimers:
 both of these states have at least a 4-fold
degeneracy (Fig.~\ref{dimVBC}). 
\begin{figure}[h!]
	\begin{center}
	\resizebox{8cm}{!}{\includegraphics{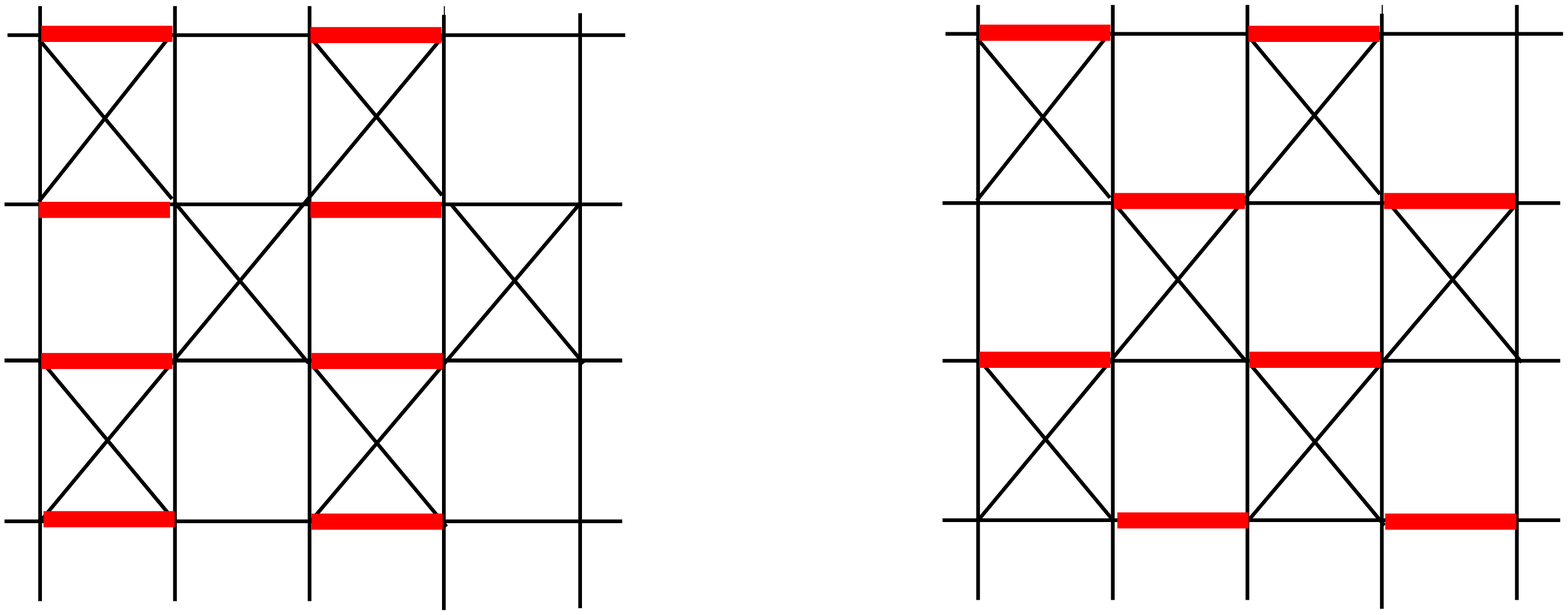}}  \end{center}

	\caption[99]{ Columnar and staggered configuration of
dimers (fat links) on the checkerboard lattice: such symmetry breaking
 configurations are 4-fold degenerate in the thermodynamic limit.}
	  \label{dimVBC}
\end{figure}
 The simplest Valence Bond Crystals  that allow the above-mentioned 
symmetry breaking  are described by pure product wave-functions 
of 4-spin  S=0 plaquettes.

\begin{figure}
	\begin{center}
	\resizebox{8cm}{!}{\includegraphics{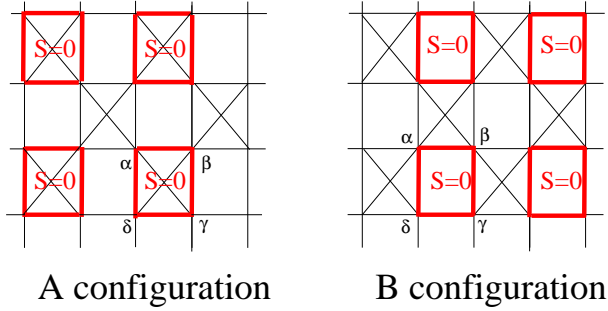}}  \end{center}

	\caption[99]{    S=0 4-spin plaquette  valence-bond   crystals
  on the  checkerboard lattice:  fat links indicate  4 spins
involved in                 a            singlet.
	}  \label{plaqVBC}
\end{figure}

 This family includes eight different
configurations: 
\begin{itemize}
\item
The singlet plaquettes
  may sit either on the squares with crossed links or on
the void squares (A and B configurations of Fig.~\ref{plaqVBC}),
\item  The translation symmetry breaking configurations may be in
two different locations named $A_{1(2)}$ (resp $B_{1(2)}$),
\item An S=0 state on a plaquette of four spins sitting on sites
($\alpha, \beta, \gamma, \delta$) may be realized either by
the symmetric combination of pairs of singlets: 
\begin{equation}
|\psi^{+}> = |\alpha\to\delta>|\gamma\to\beta> + 
|\alpha\to\beta>|\gamma\to\delta>,
\label{splaq}
\end{equation}
 or by the anti-symmetric one:
\begin{equation}
|\psi^{-}> = |\alpha\to\delta>|\gamma\to\beta> -
|\alpha\to\beta>|\gamma\to\delta>.
\label{aplaq}
\end{equation}
where $|\alpha\to\gamma>$ is the singlet state on sites 
$\alpha$ and $\gamma$:
\begin{equation}
|\alpha\to\gamma>  = 
(|\alpha \uparrow, \gamma\downarrow> -
|\alpha\downarrow, \gamma\uparrow>)/\sqrt2. 
\end{equation} 
\end{itemize}

\begin{table}
\begin{center}
\begin{tabular}{|c||c|c|c|}
\hline\hline
Wave-function& ${\cal T}_{{\bf u}_1}$& ${\cal R}_{\pi /2}$&$ {\cal
\sigma}_{{\bf u}_1}$\\
\hline
\hline
$A_{1(2)}^{+}$&$A_{2(1)}^{+}$&$A_{1(2)}^{+}$&$A_{1(2)}^{+}$\\
$A_{1(2)}^{-}$&$A_{2(1)}^{-}$&$(-1)^pA_{1(2)}^{-}$&$(-1)^pA_{1(2)}^{-}$\\
$B_{1(2)}^{+}$&$B_{2(1)}^{+}$&$B_{2(1)}^{+}$&$B_{2(1)}^{+}$\\

$B_{1(2)}^{-}$&$B_{2(1)}^{-}$&$(-1)^pB_{2(1)}^{-}$&$(-1)^pB_{2(1)}^{-}$\\
\hline
$X^{\eta}= A_{1}^{+} +\, \eta \, A_{2}^{+}$&$ \eta\, X^{\eta}$&$ X^{\eta}$
&$X^{\eta}$\\
$Y^{\eta}= A_{1}^{-} +\, \eta \,A_{2}^{-}$&$ \eta\, Y^{\eta}$&
$(-1)^p Y^{\eta}$&$(-1)^p Y^{\eta}$\\
$Z^{\eta}= B_{1}^{+} +\, \eta \, B_{2}^{+}$&$ \eta\, Z^{\eta}$&
$\eta\, Z^{\eta}$&$\eta\, Z^{\eta}$\\
$T^{\eta}= B_{1}^{-} +\, \eta \, B_{2}^{-}$&$ \eta\, T^{\eta}$&
$(-)^p \,\eta\, T^{\eta}$&$(-)^p\, \eta\, T^{\eta}$\\
\hline\hline
\end{tabular}
\end{center} 
\caption[99]{Transformation rules of the product wave-functions
in the elementary operations of the symmetry group
(the space group is defined  with respect to point O
and translations ${\bf u_1, u_2}$).
The wave-functions of the anti-symmetric plaquettes 
have different symmetries depending on the parity p of the number
of plaquettes in the sample.}
\label{symmetryop}
\end{table}

We can thus define eight different product wave-functions
labeled: $|A_{1(2)}^{\epsilon}>$ and $|B_{1(2)}^{\epsilon}>$.
The transformations of these states under the elementary
operations of the  lattice symmetry group are described in
the first four lines of 
Table \ref{symmetryop}.
The symmetric (resp. anti-symmetric) linear combinations of these states
which are irreducible representations of this group are defined
in the four last lines of the same Table.
Comparison of the symmetries of these states for different
samples with those of
the two first levels of the exact spectra indicates a $Z^{+},Z^{-}$ 
symmetry  of the HCKB ground-state doublet. In the 
thermodynamic limit the symmetry
breaking configuration is thus of the B type decorated by the
symmetric 4-spin plaquettes $|\psi^+>$ described in Eq.~\ref{splaq}.

A simple last remark could be done: the
symmetric-plaquette state (Eq.~\ref{splaq})
can be rewritten as the product of two
triplets along the diagonals of the square. This
configuration of spins is not energetically optimal on the squares with
antiferromagnetic crossed links  (A configuration)
but might a priori be favored in
B configuration. Reversely the $\psi^{-}$-plaquette can be rewritten
as the product of two
singlets along the diagonals of the square, and would
eventually be preferred in A configuration.
The variational energy per spin of the product wave-function of
$\psi^{+}$-plaquettes in B configuration  is $E_{var}(B^+)= -0.5$, whereas the
 variational energy per spin of the product wave-function of
$\psi^{-}$-plaquettes in A configuration is $E_{var}(A^-)= -0.375$. The
exact energy per spin is $E_{ex} \sim -0.514 \pm 0.006$. This is a first 
proof that
the real system takes advantage of some fluctuations around the
pure product wave-function $Z^{+}$ to decrease
its energy.

\pagebreak
The study of dimer-dimer correlations (Fig.~\ref{corr} and
Table~\ref{dimer-dimercorr}):
 \begin{equation}
{\cal C}^{4}(1,2;i,j)= 4 \left[< {\bf S}_1.{\bf S}_2\;{\bf S}_i.{\bf S}_j>
 - < {\bf S}_1.{\bf S}_2><{\bf S}_i.{\bf S}_j>\right]
\label{dim-dim-corr}
\end{equation}
 and 8-spin correlation functions~\cite{fmsl03}
shows long range order in the 4-spin
plaquettes, but also the dressing of the pure product state $Z^{+}$ by
quantum fluctuations (see Table~\ref{dimer-dimercorr}).

Are those small size computations relevant for the description of
the thermodynamic limit? The
stronger answer is read in Fig.~\ref{gckb} $\&$ Fig.~\ref{S0sector}:
insofar as the degeneracy of the ground-state and the  gaps to the first
triplet state and the third singlet state 
remain finite in the thermodynamic limit,
the Valence Bond Crystal picture
 (with LRO in plaquettes) will survive to quantum
fluctuations. The gaps results
(Fig.~\ref{gckb} $\&$ Fig.~\ref{S0sector}) 
show that the  studied samples (except the $N= 16$)
have linear sizes
of the order of,
or larger than the spin-spin correlation length.
We thus think that the present qualitative
conclusions  are reliable.
\begin{table}
\begin{center}
\begin{tabular}{|c|c|c||c|c|c|}
\hline\hline
${i,j}$& ex. g.-s.&Z w-f.& ${i,j}$& ex. g.-s.&Z w-f.\\
\hline
{31,32}&.56&.63&{7,13}&.10&.25\\
{7,8}&.43&.42&{19,25}&.10&.25\\
{25,26}&.26&.25&{7,12}&-.10&-.25\\
{13,14}&.26&.25&{31,36}&-.10&-.25\\
{19,20}&.25&.25&{13,18}&-.11&-.25\\
{6,5}&.22&.25&{25,30}&-.11&-.25\\
{6,12}&-.20&-.25&{19,24}&-.11&-.25\\
{25,31}&-.20&-.25&{6,36}&.10&.25\\
{13,19}&-.18&-.25&{12,18}&.11&.25\\
{36,35}&.18&.25&{24,30}&.10&.25\\
{5,11}&-.18&-.25&{35,5}&.10&.25\\
{4,10}&-.18&-.25&{11,17}&.10&.25\\
{12,11}&.17&.25&{29,23}&.10&.25\\
{36,30}&-.15&-.25&{5,4}&-.11&-.25\\
{35,29}&-.15&-.25&{11,10}&-.11&-.25\\
{30,29}&.15&.25&{35,34}&-.11&-.25\\
{17,23}&-.15&-.25&{17,16}&-.11&-.25\\
{18,17}&.15&.25&{29,28}&-.10&-.25\\
{18,24}&-.15&-.25&{23,22}&-.10&-.25\\
{24,23}&.15&.25&{34,4}&.10&.25\\
{28,34}&-.15&-.25&{10,16}&.10&.25\\
{16,22}&-.15&-.25&{28,22}&.10&.25\\
\hline\hline
\end{tabular}
\end{center}
\caption[99]{ Dimer-dimer correlations ${\cal C}^4(1,2;i,j)$ (Eq.
\ref{dim-dim-corr}) in the $N=36$ ground-state. 
 The sites $1,2,i,j$ are described in Fig.~\ref{corr},
the $i,j$ points are enumerated in the first columns.
 This correlation has been measured in the exact ground-state
 wave function (second columns) and  in the variational Z state 
(third columns).
All the values of these correlations between sites of Fig.~\ref{corr}
 can be obtained from this table by a mirror symmetry through the 
bisector of bond $(1,2)$.}
\label{dimer-dimercorr}
\end{table}


\begin{figure}[t!]\vspace*{-1cm}
	\begin{center}
	\resizebox{6cm}{!}{\includegraphics{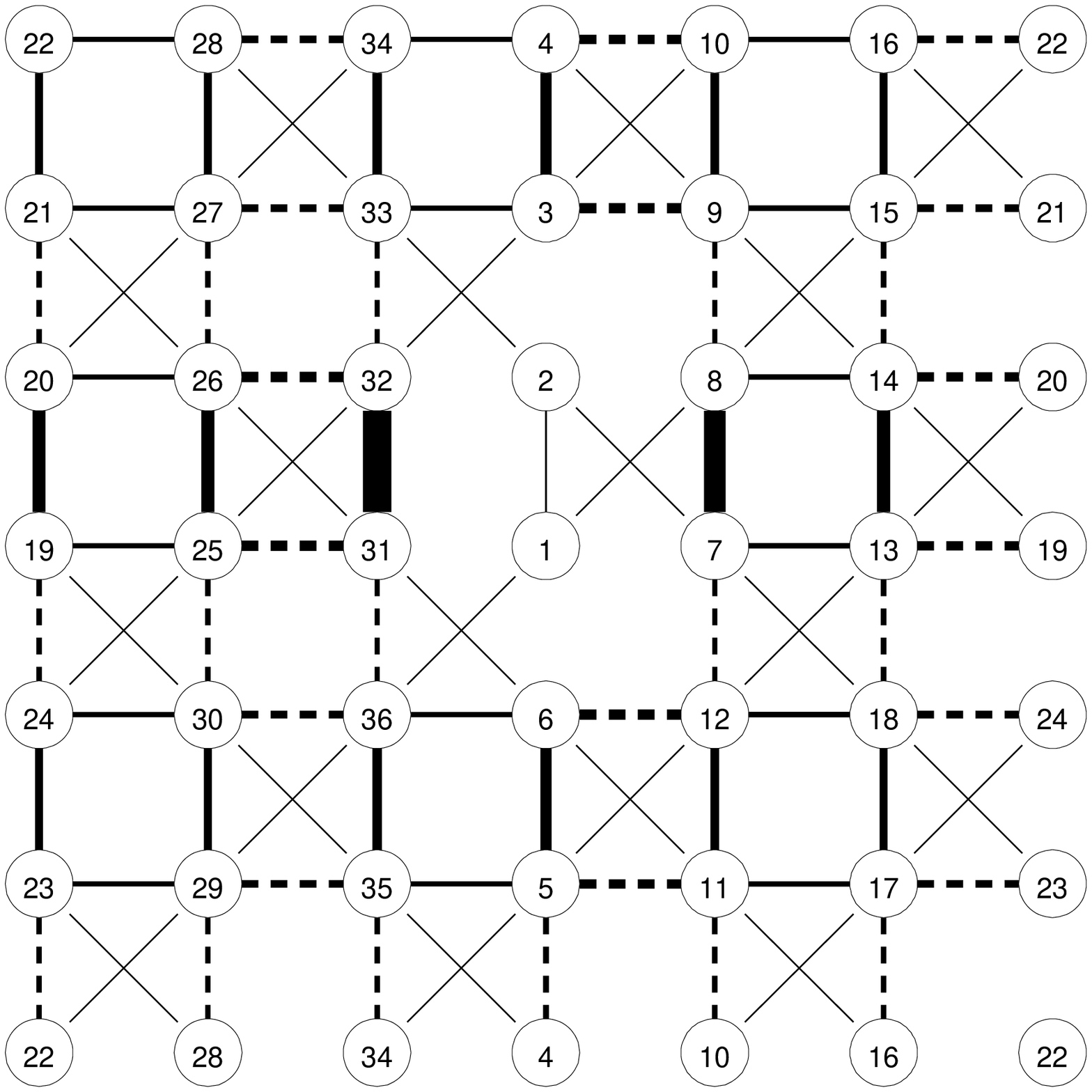}} 
 \end{center}

	\caption[99]{ Dimer-dimer correlations in the exact
ground-state of the 36 sample (Eq.~\ref{dim-dim-corr}).
The reference bond  is the bond $(1,2)$. Positive
(negative)
correlations are drawn as full (dashed) lines. The thickness of the
lines is a measure of the strength of the correlation.
The diagonal lines show the position of the crossed links.
	}  \label{corr}
\end{figure}
\pagebreak

\end{subsection}
\begin{subsection} {Excitations: raw data and qualitative
description of the first excitations}

Looking to Table~\ref{table1} and Fig.~\ref{fig-disp-S1}, it appears
that the triplet excitations are gapped (gap of the order of 0.7)
 and very weakly
dispersive. Singlet excitations too are gapped (4th line of 
Table~\ref{table1} and Fig.~\ref{fig-disp-S0}); they are much more
dispersive than the triplet excitations and less energetic (gap
of the order of 0.25).

\begin{table} [h!]
\begin{center}
\begin{tabular}{|c|c|c|c|c|c|c|}
\hline
  $N$ &  24'& 28 & 32*& 32' & 36 \\
  \hline
$e_0$&-.522&-.520&-.517&-.514&-.520\\
 \hline
  $E^1_{S=1}-E^1_{S=0}$ &0.58& 0.57 & 0.69 & 0.57& 0.71\\
  \hline
  $E^2_{S=0}-E^1_{S=0}$ &0.08& 0.09 & 0.03&0.01& 0.05 \\
  \hline
  $E^3_{S=0}-E^2_{S=0}$ &0.06& 0.05 & 0.18 &0.13& 0.22 \\
\hline
$E^1_{S=1}-E^3_{S=0}$&0.44&0.42&0.47&0.43&0.44\\
\hline
   $n_1$   &51& 82 & 286&135& 110\\
\hline
$ln(n_1)$/N&0.16&0.16&0.18&0.15&0.13\\
\hline
\end{tabular}
\end{center} 
\caption[99]{ Spectrum of the Heisenberg model on the
checker-board lattice. 
Energy per spin in the ground-state $e_0$ and 
energy gaps $E^{n_S}_{S}-E^{n_S'}_{S'}$ between the ${n_S'}$ energy 
level
of the $S'$ spin sector and the ${n_S}$ level of the $S$ sector.
Second line: spin gap. Third line: gap between the absolute
 ground-state and the first singlet excitation . Fourth  line: gap
between the second and third level in the $S=0$ sector.
Fifth line: gap between the third level in the $S=0$ sector
 and the first triplet excitation.
Following lines: 
 $n_1$ is the number of singlet
states in the spin gap (including degeneracies). 
The starred column corresponds to a sample which has
the extra symmetries of the pyrochlore lattice.
}
\label{table1}
\end{table}

\begin{figure}
\begin{center}
\resizebox{8cm}{!}{\includegraphics{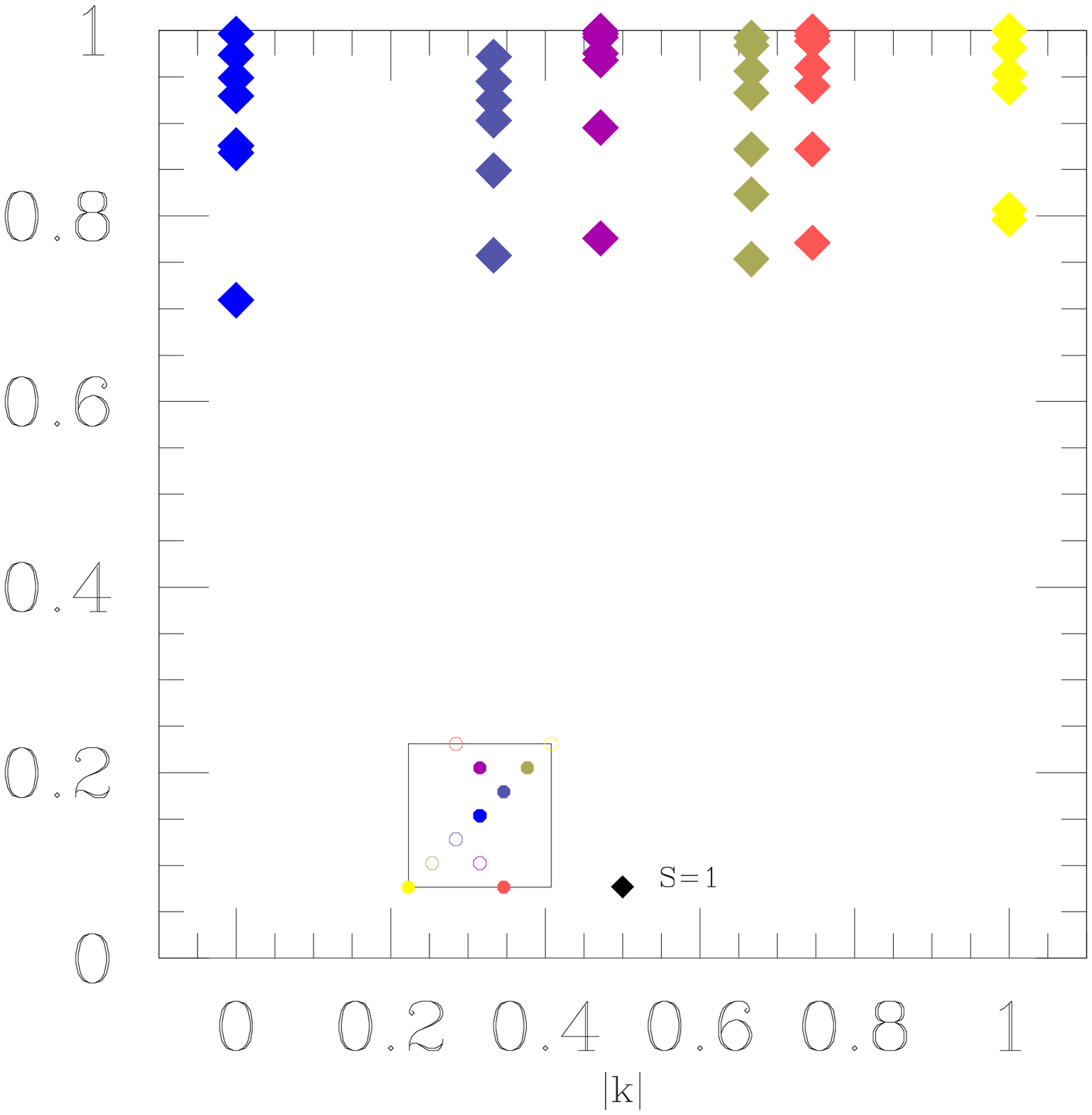}}
\end{center}
\caption{Dispersion relations in the triplet sector  versus
$|{\bf k}|/|{\bf k}_0|$ with ${\bf k}_0= (\pi, \pi)$.} The inset
shows the correspondence between the colors of the symbols and
the wave vectors in the Brillouin zone. Only the triplet
excitations are drawn in this figure.
\label{fig-disp-S1}
\end{figure}

\begin{figure}
\begin{center}
\resizebox{9cm}{!}{\includegraphics{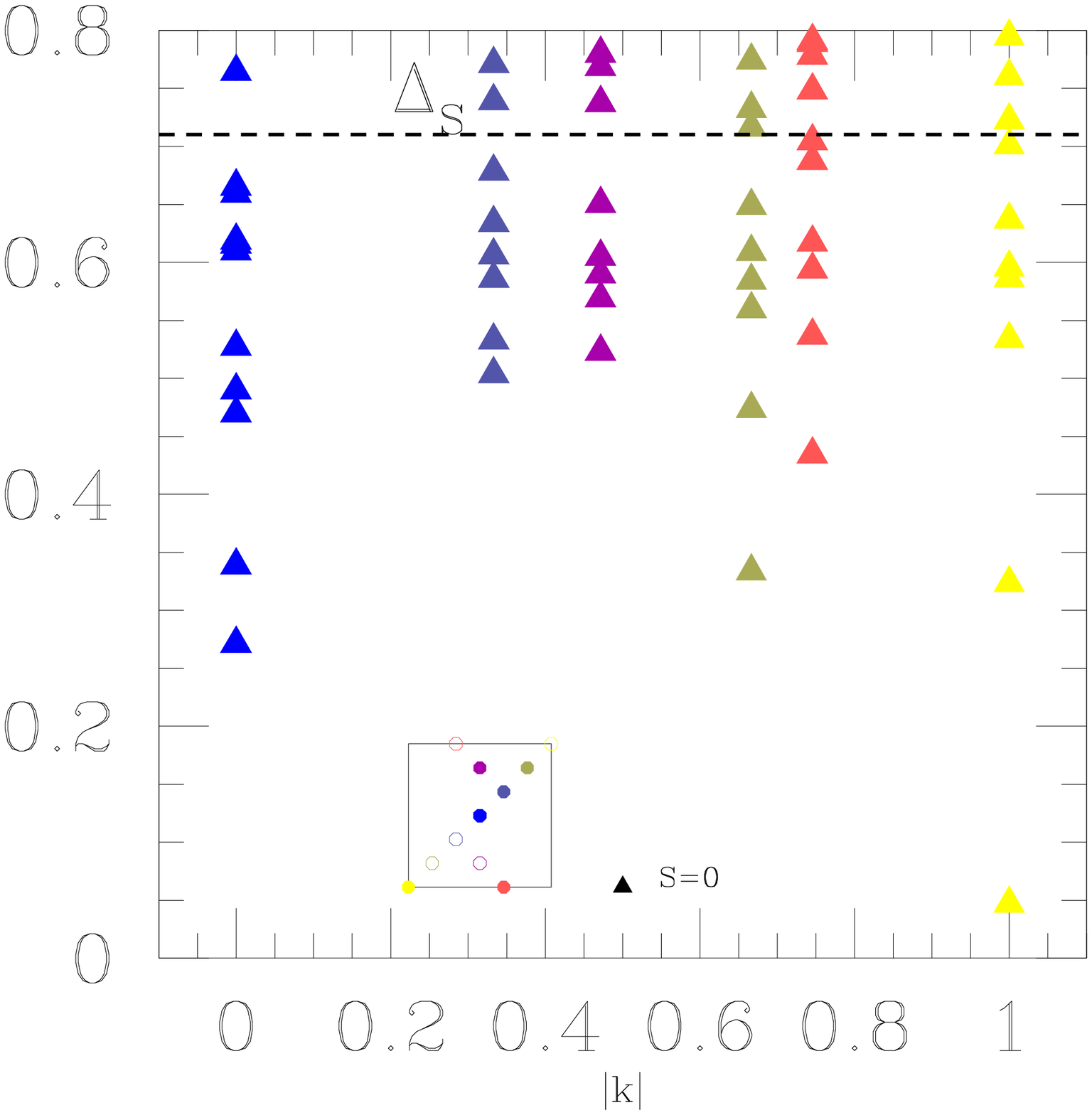}}

\caption{Dispersion relation of the singlet excitations of the $ N=36$
 sample versus $|{\bf k}|/|{\bf k}_0|$ with ${\bf k}_0= (\pi, \pi)$.
The horizontal dashed line indicates the spin-gap.The inset
shows the correspondence between the colors of the symbols and
the wave vectors in the Brillouin zone. Only the singlet 
excitations are drawn  in this figure.
  The first excited level of this figure with ${\bf k}_0= (\pi, \pi)$
(yellow up triangle) is not a true excitation. This level is at
the thermodynamic limit degenerate with the ground-state and
allows the space symmetry breaking of the 4-spin plaquette
order.}
\label{fig-disp-S0}
\end{center}
\end{figure}

There is a very simple variational description of the triplet
 excitations: let us consider the 4-spin plaquettes B of the
ground-state. The S=0 ground-state is formed from the coupling of
two triplets along the diagonals. There are four S=1 states on such
 a plaquette. The lowest S=1 excitation  simply results from the S=1
coupling of the two diagonal triplets. The gap to this variational
excitation is 1. The Bloch waves built on such excitations
are non dispersive. Up to a renormalization of the gap of the
order of $33\%$, this picture appears as a good qualitative
description of the true S=1 excitations of the HCKB model, which
are massive, quasi localized excitations with an energy gap $\sim
0.7$. 

The singlet excitations are more intricate. On a B plaquette the
first S=0 excitation corresponds to the antisymmetric coupling
of dimers $|\psi^->$ described in Eq.~\ref{aplaq}. Its energy gap 
 to the ground-state is equal to 2. This first excitation of the
B plaquettes is more energetic than the object built by a
reorganization of two symmetric $\psi^+$ states on two
neighboring B positions. More precisely the excitation which
promotes the two pairs of spins ($\alpha, \delta)$ and ($\beta,
\gamma$) into triplet states and then couples them in a singlet
states has a gap 1 with respect to the ground-state. To first
order in a strong coupling expansion this excitation is non
dispersive but it can acquire dispersion at higher order. The
exact S=0 excitations are thus certainly a bit more extended and
complex that this first approximation.\footnote{In view of the
strong VBC correlations of the ground-state different authors
developed strong coupling perturbative studies of the
excitations.
The work by Brenig et al.~\cite{bh02} takes as a
departure point the  4-spin S=0 plaquettes on the B
positions and treat the couplings between the B sites as a
perturbation. The limit toward the isotropic point
of this high order perturbation expansion seems to be rather well
behaved but it fails to restore the correct symmetry breaking of
the ground-state in the S=0 sector. This is perhaps not too
surprising in view of the above remarks on the first
excitations in the singlet sector. Berg and
collaborators~\cite{baa03} have
used a more sophisticated method (determination of an effective
hamiltonian by a real space renormalisation method called CORE),
with a different departure point in the $S=0$ sector, they found
domain walls between the two degenerate ground-states as the lowest
excitation in the $S=0$ sector.}

{\bf Remarks on the number of singlet excitations in the singlet-triplet
gap}

The spectra of very small samples of the HCKB model~\cite{pc02} lead to
conclusions on the number of singlets excitations that should be
precised and relativized. The above-mentioned authors, and many
commentators afterward, argued that this large number of singlets
might be reminiscent of the exponential degeneracy in the singlet
 sector of
the Heisenberg model on the kagome lattice. A precise analysis of
the spectra of singlet excitations of the HCKB model shows that
this analogy is unjustified for the following reasons:

{\bf i}) The sample set used to extract this conclusion was a
mixture of true 2-d checker-board samples, quasi 1-dimensional
tubes with a cross section of 4 spins and 3-dimensional
pyrochlore-like samples~\cite{fmsl03}. Table~\ref{table1}
summarizes the results for the restricted family of pure
2-dimensional HCKB samples.
Whereas the low density of singlet states of the
Heisenberg model on the kagome lattice increases as $1.15^N$
with the system size, a similar analysis for the HCKB model gives
the value $0.14^N$.  Changing the ratio of the exponential from a
number larger than one to a number smaller than one changes indeed
completely the picture! In fact the best fit to describe the
number of singlet in the spin-gap is obtained for a power law
fit: $N^{\gamma}$ with ${\gamma}=3.96$. If these low excitations
can be described as modes one would effectively expect some power
law.

{\bf ii}) Last difference between the HCKB model and the
Heisenberg model on the kagome lattice (HK): in the HK model the
continuum of singlets is adjacent to the ground-state whereas, in
the HCKB model, there is a clear-cut gap in the singlet sector
between the ground-state multiplicity and the first singlet
excitations.

This simple analysis of the singlet and triplet excitations of
the HCKB model supports the idea that the excitations are simple
``optical modes'' that could be observed in Raman, RPE, ESR or NMR
spectra. This structure is certainly highly reminiscent of those
of dimerized or spin-Peierls chains, gapped ladders.. A picture
consistent with the strong coupling description of the ground-state.
\end{subsection}
\begin{subsection}{Summary of the generic features of a Valence
Bond Crystal}

I would like to argue that the main features of Valence Bond
Crystals (whatever the dimensionality of space where they are
living) are probably:

\begin{itemize}
\item A spin gap, no $SU(2)$ symmetry breaking and short range
spin-spin correlations,
\item Degeneracy in the thermodynamic limit of the $S=0$ ground-state,
 embedding the spontaneous space symmetry breaking of the phase,
\item Long range order in dimer-dimer and/or larger $S=0$
plaquettes, (that can subsist up to finite temperature?)
\item Gapped excitations, in the $S=0$ sectors as well as in
other $S$ sectors, that can be described as modes, more or
less dispersive. A strong coupling analysis of these modes seems
a priori valid but the examples worked out up to now, on this
model or on the Shastry-Sutherland model~\cite{ss81a,mt00b}, show
that the departure point of the perturbation theory should be
given special consideration.  
\end{itemize}

No experimental evidence exists up to now of a pure Valence Bond
Crystal with spontaneous symmetry breaking.
But a few 2-dimensional systems with a Valence Bond ground-state
have been observed experimentally:
 CaV$_4$O$_9$~\cite{ki94,ttw94,tnyk95,fo96,am96a,am96b,my96,sr96,uksl96,khst96,tku96,khsk97,oyiu97} 
and SrCu$_2$(BO$_3$)$_2$~\cite{k99,ss81a,mu99,who99,ckrz99,mu00,kk00,cms01,mt00b,msku00,ckrz01}. However, in both cases
the ground-state is  non-degenerate  because  the Hamiltonian  has  an
integer spin  in the unit  cell (4 spins  $1/2$)  and the dimerization
does not break any lattice symmetry.
From the theoretical point of view on may argue that the
difference between the theoretical VBC described in this chapter
and these compounds is akin to the difference between the
dimerized phase of the $J_1-J_2$ model on the chain and the spin
Peierls compounds. This is a minor difference and from the
experimental point of view the first excitations of all these
models can be qualitatively described as ``optical modes''. The
detailed characteristics of the two-quasi-particle continua
could be a bit more different.
\end{subsection}
\end{section}
\begin{section}{A simple model of VBC with a critical point: the
hard core quantum dimer model of Rokhsar and Kivelson on the
square lattice}

Looking for a model with a resonating valence bond ground-state,
Rokhsar and Kivelson introduced in 1988 a quantum hard core
model on the square lattice~\cite{rk88}. Their motivation was the
description of systems with strongly coupled real-space Cooper
pairs. At half filling these next-neighbor Cooper pairs can be
seen as next-neighbor dimers. Pauli principle and Coulomb
interaction imply that these dimers are hard core dimers. Insofar
as the spin gap is large enough, it can be speculated that the
manifold of low energy states is spanned by the linearly
independent set of nearest neighbor dimer  coverings\footnote{It
has been shown that such set form a family of non-orthogonal but
linearly independent states~\cite{cck89,mlms02}.}.
The dynamics of the low lying singlet excitations of this model
are described by the Hamiltonian ${\cal H}_{dimer}$:
\begin{equation}
H_{dimer}=\sum_{\rm Plaquette}
\left[
-J\left(
\left|\begin{picture}(13,9)(-2,2)
	\put (0,0) {\line (0,1) {8}}
	\put (8,8) {\line (0,-1) {8}}
	\put (0,0) {\circle*{3}}
	\put (0,8) {\circle*{3}}
	\put (8,0) {\circle*{3}}
	\put (8,8) {\circle*{3}}
	\end{picture}
\right>
\left<\begin{picture}(13,9)(-2,2)
	\put (0,0) {\line (1,0) {8}}
	\put (8,8) {\line (-1,0) {8}}
	\put (0,0) {\circle*{3}}
	\put (0,8) {\circle*{3}}
	\put (8,0) {\circle*{3}}
	\put (8,8) {\circle*{3}}
	\end{picture}
\right|
+h.c.\right)
+V\left(
\left|\begin{picture}(13,9)(-2,2)
	\put (0,0) {\line (0,1) {8}}
	\put (8,8) {\line (0,-1) {8}}
	\put (0,0) {\circle*{3}}
	\put (0,8) {\circle*{3}}
	\put (8,0) {\circle*{3}}
	\put (8,8) {\circle*{3}}
	\end{picture}
\right>
\left<\begin{picture}(13,9)(-2,2)
	\put (0,0) {\line (0,1) {8}}
	\put (8,8) {\line (0,-1) {8}}
	\put (0,0) {\circle*{3}}
	\put (0,8) {\circle*{3}}
	\put (8,0) {\circle*{3}}
	\put (8,8) {\circle*{3}}
	\end{picture}
\right|
+
\left|\begin{picture}(13,9)(-2,2)
	\put (0,0) {\line (1,0) {8}}
	\put (8,8) {\line (-1,0) {8}}
	\put (0,0) {\circle*{3}}
	\put (0,8) {\circle*{3}}
	\put (8,0) {\circle*{3}}
	\put (8,8) {\circle*{3}}
	\end{picture}
\right>
\left<\begin{picture}(13,9)(-2,2)
	\put (0,0) {\line (1,0) {8}}
	\put (8,8) {\line (-1,0) {8}}
	\put (0,0) {\circle*{3}}
	\put (0,8) {\circle*{3}}
	\put (8,0) {\circle*{3}}
	\put (8,8) {\circle*{3}}
	\end{picture}
\right|
\right)
\right]
\label{sqhcd}
\end{equation}
(In their original paper the authors discussed the derivation of
this effective Hamiltonian from a more realistic Hubbard model.)

The first term of Eq.~\ref{sqhcd} describes the spatial flip of
two parallel dimers from horizontal to vertical position and
vice-versa, it could also be seen as a cyclic permutation of the two
dimers around a square: it is a kinetic energy term which favors
resonances between different configurations of parallel dimers
($J$ is always $>0)$.
The second term is a potential energy term likely to be repulsive
in the original electron model. The ground-state for infinitely
large $\frac {|V|}{J}$ is a Valence Bond Crystal, either staggered
(for large $\frac {V}{J}>0$), or columnar (for large $\frac {V}{J}<0$).
See Fig.~\ref{RKModel}.
\begin{figure}
\begin{center}
\resizebox{!}{7.cm}{
\includegraphics{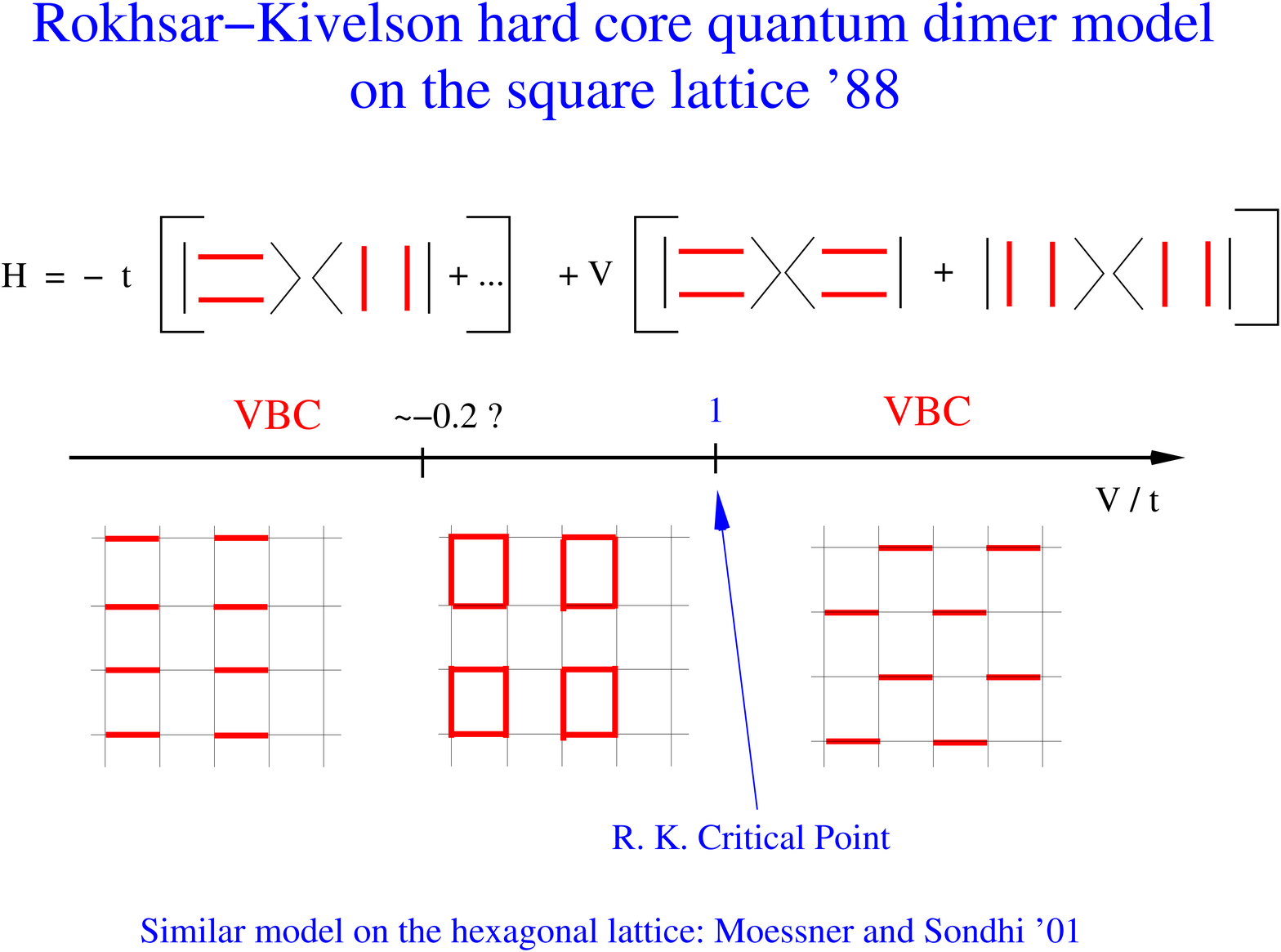}}
\end{center}
\label{RKModel}
\end{figure}

{\bf Topological structure of the Hilbert space of the QHCD model
on the square lattice}

The eigenstates
of ${\cal H}_{dimer}$ can be classified according to their
winding numbers ($\Omega_x, \Omega_y$) across the 2-torus of the
square sample with periodic boundary conditions. There are many
equivalent ways to define these winding numbers. Let us follow
 RK. They draw the transition graph of a dimer
configuration  ${\cal C}$ relative to a reference
configuration ${\cal C}_0$ (which may be the columnar
configuration) as the
superposition of the dimer coverings of the two configurations
${\cal C}$ and  ${\cal C}_0$. The dimers in ${\cal C}$ are
directed from one sublattice to the other and reversely for the
dimers of ${\cal C}_0$. The transition graph thus appears as a
graph of oriented loops. The winding number  $\Omega_x$  (resp.
$\Omega_y$) measures the net number of loops (clockwise minus
counter-clockwise) encircling the torus
in the $x$ (resp $y$ direction).
The Hamiltonian does not couple subspaces with different winding
numbers. These pairs of winding numbers define $N_s$
disconnected topological subspaces (where $N_s$ is the number of
lattice sites).

For $V \geq J \geq 0$,  ${\cal H}_{dimer}$ is positive
semi-definite, the ground-state is unique and nodeless (Frobenius theorem). Moreover 0 is a lower bond
of its energy.

Demons: A lower bond for the ground-state energy is given by a
minimization of the Hamiltonian on each plaquette individually.
If the given plaquette has no parallel dimer (non
flippable plaquette), its energy is zero and if it has parallel
dimers it has a potential energy $V$ and at best a kinetic energy
of $-J$. We can thus write a lower bond energy of the global
system, which is proportional to the number of flippable
plaquettes $n_{flip}$, as $ min\left[0, (V-J)n_{flip}\right]$.

{\bf Phase Diagram of the RK model}

The staggered configuration is  a zero-energy eigenstate of ${\cal H}_{dimer}$.

 At the point $\frac {V}{J} =1$ the model is exactly solvable.

\begin{itemize}
\item
The four staggered configurations are zero-energy eigenstates of ${\cal H}_{dimer}$. As they saturate the low energy bond, they are the ground-states for $V \geq J \geq 0$.
They can be classified
in two different topological classes in which they are the only
representatives. They have a zero energy and any configuration of
 other topological
subspaces has a larger strictly positive energy (at least 
of order ${\cal O}(L)$ in the limit $\frac {V}{J} \to
\infty$ ).
\item
At the point $\frac {V}{J} =1$ the model is exactly solvable. 
There is in each topological subspace a
 ground-state with zero energy. It is
 the equal amplitude superposition of all the
 configurations of that sector.
{\it i}) Simple computation shows that these states are
zero-energy states of ${\cal H}_{dimer}$.
{\it ii}) Since all off-diagonal elements are non-positive the
ground-state is unique and nodeless (Frobenius property, Marshall
theorem).
The equal amplitude states are thus the unique ground-states in
their respective topological sectors. We will  
call these states the RK states. This is the first example in
these lectures of a Resonating Valence Bond wave-function.
\item It has been shown by Kohmoto and Shapir~\cite{ks88}, that
the spin-spin correlations in this state decrease  exponentially.
\item
An important property: any dimer
correlation functions in the RK state can be computed from an
exact
mapping to the classical statistical problem of dimer
coverings first solved by M. E. Fisher and J. Stephenson~\cite{fs63}.
From this work one can conclude that the dimer-dimer correlation
functions at $\frac {V}{J} =1$, decreases algebraically with
distance (as $r^{-2}$). This property implies that the first excitations above
the ground-states are gapless.
\item On the basis of the continuity in
the energy between the staggered phase and the RK states, one may
speculate that the RK point is the quantum critical  end 
of the staggered VBC phase. But  
the excitations of the staggered VBC are non local  and  have energy
of order ${\cal O}(N^{+0.5})$ in the  $\frac {V}{J} \to
 \infty$ limit. To sustain the above point of view one should 
explain how the kinetic term can dress these
excitations so that  they become gapless when $\frac {V}{J} \to
1$. In fact the more probable hypothesis is a first order phase
 transition between
the RK phase and the staggered one. Such a question could
perhaps be answered with Monte Carlo simulations.
\item
The ground-state wave-function at this RK point has a property
which is considered as constitutive of a RVB spin liquid: that
is resonances between all dimer coverings. It must be underlined
here that this resonance phenomenon at the RK point does not bring
any stabilization of the equal amplitude superposition
ground-state when compared to the neighboring staggered VBC
phase. 
\item
RK then argue that for  $\frac {V}{J} <1$, there is, separated
 by a first order phase transition, a new phase which might be
 a ``true'' resonating Valence Bond Spin Liquid~\footnote{i.e. a phase
where the resonances between different dimer coverings are
essential to its stabilization and are so important that there is 
a gap to the first excitations and
 any correlation functions: either
spin-spin,  dimer-dimer or higher order plaquette-plaquette
correlation functions decrease exponentially with distance.}.
The characterization of
this phase is for the moment rather loose: RK argument is
variational and rather week. The first calculation by Sachdev
on a 36 lattice~\cite{s89}, gives evidence for a VBC columnar state for
 $\frac {V}{J} < 0.5$ and not a clear conclusion nearer from
the RK point. Extending the calculations to 64 sites, and 
using various estimators, Leung and
co-workers~\cite{lcr96} estimated that long range columnar order  
probably exists up to the RK point, with the restriction that up
to $\frac {V}{J} \sim -0.2$ the order is very plausibly purely columnar,
whereas in the range $-0.2< \frac {V}{J}< 1$ the order could
reduce to
a 4-spin S=0 plaquette order~\footnote{This conclusion is not
consistent with the degeneracy the authors claim for the ground-state.
The nature of the phase for $\frac {V}{J} <1$
remains an interesting and open  question: 
interesting but technically difficult. The same kind of
difficulty is present in the study of the $J_1-J_2$ model on the
square lattice for $J_2/J_1 \sim 0.5$. At this point of maximum
frustration, N{\'e}el order is destroyed but the exact nature of the
phase is uncertain: columnar order~\cite{sz92,kosw99}, 4-spin
plaquette order~\cite{zu96,cs00} or RVB spin liquid~\cite{cbps01}?
In view of exact spectra for sizes up to N=36, it seems that the
4-spin S=0 plaquette order is the less plausible (because the
${\bf k}= (\pi,\pi)$ states necessary for the 2-fold symmetry
breaking of this state is very high in the spectrum). 
We expect a 4-fold symmetry breaking in the columnar state as well as in the RVB
state~(\cite{mlms02} and refs. therein). The gaps from the
ground-state to the plausible candidates for these 4-fold symmetry
breakings are still very large in the N=36 sample. We are thus
lead to conclude that the N=36 sample is too small to give
informative issue on the dilemma: columnar state or RVB state.
This strongly weakens  the variational argument of
ref.~\cite{cbps01}.}.
It seems  nevertheless widely
 admitted~\cite{f91,ms01}  that this model has
crystalline order everywhere except at the RK critical point.
\end{itemize}

In view of these results for the QHCD model on the square
lattice, of most studies on the $J_1-J_2$ $SU(2)$ model, and of the
$SU(N)$ studies on the same lattice~\cite{am88,rs89}, one may be
tempted to conclude that VBC is the paradigm of the quantum
ground-state on square and possibly bipartite lattices. This
might  be an escapable assumption~\cite{cbps01,fsl01}, but the
fact is that the triangular based lattices (see next chapters)
seem much more favorable to 
Resonating Valence Bond Spin Liquids.
\end{section}
\end{chapter}

\begin{chapter}{Resonating Valence Bond Spin Liquid (Type I)}

The Resonating Valence Bond Spin Liquid is a quantum concept
introduced in 1973 by P. W. Anderson~\cite{a73}, following the
line of thought of Linus Pauling for molecules. When the
semi-classical N{\'e}el
states or simple dimer covering solutions are very far to satisfy
each individual bond, Anderson speculated that the macroscopic
system could take advantage of the quantum resonances between
the exponential number of  dimer coverings to lower its
ground-state energy. Such states have no long range order whence
the name of Spin Liquid, quantum resonances between the
exponential number of equivalent dimer coverings are essential:
it is a Resonating Valence Bond Spin Liquid (abbreviated as RVB
Spin Liquid or RVBSL in the following).

\begin{section}{Introduction: short range versus long range
Resonating Valence Bond wave-functions}
Resonating Valence Bond wave-functions encompass a large class of
wave-functions beyond the equal amplitude superposition of 
next neighbor dimer
coverings that we encounter at the RK point in the last chapter.

It is easy to verify that the whole set of dimer coverings (without any
restriction on the length of the dimers) 
 is an
overcomplete basis of the $S=0$ subspace of the spin system
(compare the numbers of these coverings to the size of the $S=0$
subspace for a N site lattice)~\footnote{For
large enough sizes the next neighbor coverings form a linearly
independent family.}.

Let us suppose that we have designed a family ${\cal E}$ of linearly
independent dimer coverings ${\cal C}_i$, a general RVB wave-function
 will be written as:
\begin{equation}
|RVB> = \sum_{{\cal C}_i \in {\cal E}} A({\cal C}_i) |{\cal C}_i>
\label{genRVB}
\end{equation}
where $|{\cal C}_i>$ are products of dimer wave-functions (with a
sign conventionally fixed, respecting  the lattice topology).

In  variational calculations, one generally use restricted forms
of Eq.~\ref{genRVB}, where the amplitude $A({\cal C}_i) $  of
 a given configuration ${\cal C}_i$ 
is written as the product of amplitudes $h(k,l)$ for each  dimer
$(k,l)$ present in ${\cal C}_i$.

Two situations have been studied:

{\bf i}) either long range RVB wave functions where the function $h(k,l)$
 depends algebraically on the distance $r_{kl}$   between 
 sites $k$ and $l$  (at least for large distances):
\begin{equation}
h(k,l) =\frac {Cst}{r_{kl}^{\sigma}}
\end{equation}
Liang, Dou\c{c}ot et Anderson~\cite{lda88}
 have shown that such wave functions
have N{\'e}el long range order in the Heisenberg model
on the square lattice if $ \sigma < 5 $ and no N{\'e}el long range order
for $ \sigma > 5$~\footnote{They equally show that the
difference in energy of those different wave-functions are
extremely tiny}. Capriotti and co-workers~\cite{cbps01}
 have used a $p$-wave BCS
wave-function for the $J_1-J_2$ model on the square lattice,
which has no long range order in
dimers.

{\bf i}) or the short range Valence Bond w.-f. where the amplitudes 
$h(k,l)$ are  not necessarily strictly restricted to next
neighbors but decrease at least exponentially with distance
 (most of the following is
concerned with that kind of wave functions). By construction such
functions cannot describe N{\'e}el long range order, as
 N{\'e}el order has long
range correlations between spins on the same sublattices. As we
have seen in the previous chapter, this family encompasses the
quantum critical behavior of the QHCD model on the square
lattice. We can equally describe in this basis the Valence Bond
Crystals, which are characterized by  dominant amplitudes
associated to the simple symmetry breaking configurations.
Many properties of these wave-functions have been studied
theoretically~(\cite{s88,s88a,rc89,cck89,mlms02} and references therein),
 we will see some of them in the following.

In this chapter we will first describe with some length the properties
of the QHCD model on the triangular lattice, to compare to the
solution of the same model on the square lattice. We will then
move to the Multi-Spin Exchange Hamiltonian on the same lattice,
which is the first $SU(2)$ model exhibiting a ``true'' resonating
Valence Bond Spin Liquid. A special attention will be given to
the topological degeneracy of the ground-state, and to the
existence of deconfined spin-1/2 ``spinons'' excitations, which
is the most important experimental signature of a RVB Spin Liquid
state. We will close the chapter by a small bibliography on gauge
theory approaches that have been dealing with the same physical
problem. 

\end{section}
\begin{section}{The Quantum Hard Core Dimer  model on the triangular
lattice}

The QHCD model on the triangular lattice has been studied by Moessner
and Sondhi in 2001~\cite{ms01}, when they realized that the
dimer-dimer correlation function on this lattice was not
algebraically decreasing as on the square lattice but 
{\it exponentially 
decreasing with distance}. The model on the triangular lattice
 comprises the same ingredients as on the square lattice: a potential
 energy term between parallel pairs of dimers and a kinetic energy term
 which does a cyclic permutation of parallel dimers on 4-spin plaquettes
 (involving two triangular units).
\begin{equation}
H_{dimer}=\sum_{\rm Plaquettes}
\left[
-J\left(
\left|\begin{picture}(19,11)(-2,2)
	\put (0,0) {\line (1,2) {5}}
	\put (10,0) {\line (1,2) {5}}
	\put (0,0) {\circle*{3}}
	\put (5,10) {\circle*{3}}
	\put (10,0) {\circle*{3}}
	\put (15,10) {\circle*{3}}
	\end{picture}
\right>
\left<\begin{picture}(19,11)(-2,2)
	\put (0,0) {\line (1,0) {10}}
	\put (5,10) {\line (1,0) {10}}
	\put (0,0) {\circle*{3}}
	\put (10,0) {\circle*{3}}
	\put (5,10) {\circle*{3}}
	\put (15,10) {\circle*{3}}
	\end{picture}
\right|
+h.c.\right)
+V\left(
\left|\begin{picture}(19,11)(-2,2)
	\put (0,0) {\line (1,2) {5}}
	\put (10,0) {\line (1,2) {5}}
	\put (0,0) {\circle*{3}}
	\put (5,10) {\circle*{3}}
	\put (10,0) {\circle*{3}}
	\put (15,10) {\circle*{3}}
	\end{picture}
\right>
\left<\begin{picture}(19,11)(-2,2)
	\put (0,0) {\line (1,2) {5}}
	\put (10,0) {\line (1,2) {5}}
	\put (0,0) {\circle*{3}}
	\put (5,10) {\circle*{3}}
	\put (10,0) {\circle*{3}}
	\put (15,10) {\circle*{3}}
	\end{picture}
\right|
+
\left|\begin{picture}(19,11)(-2,2)
	\put (0,0) {\line (1,0) {10}}
	\put (5,10) {\line (1,0) {10}}
	\put (0,0) {\circle*{3}}
	\put (10,0) {\circle*{3}}
	\put (5,10) {\circle*{3}}
	\put (15,10) {\circle*{3}}
	\end{picture}
\right>
\left<\begin{picture}(19,11)(-2,2)
	\put (0,0) {\line (1,0) {10}}
	\put (5,10) {\line (1,0) {10}}
	\put (0,0) {\circle*{3}}
	\put (10,0) {\circle*{3}}
	\put (5,10) {\circle*{3}}
	\put (15,10) {\circle*{3}}
	\end{picture}
\right|
\right)
\right]
\label{thcd}
\end{equation}
The sum over plaquettes runs on the three kinds of plaquettes with
orientations at 60 degrees from each other.

J can be assumed to be positive~\footnote{This was not obvious a
priori and is important in the following as it insures that the
Hamiltonian is positive semi-definite for $ V \geq J \geq 0$}:
 it enforces resonance effects, V
can be positive (repulsion between dimers) or attractive. The
conditions of validity are the same as those of the model on the
square lattice: it is supposed that the spin gap is large enough
so that the first excitations are in the  singlet sector. Insofar
as the spin gap is large, the spin-spin correlations are short
range which is consistent with the restriction to the subspace
of nearest-neighbor Valence Bonds.

The properties of the lattice affect the properties of the QHCD
model on two central points:
\begin{itemize}
\item In the triangular case
 due to the higher entanglement of the lattice
with the two-dimer terms, there is only 4 different topological
sectors classified according to the parity of the winding numbers:
(even, even), (even, odd), (odd, even), (odd, odd). (The
dimer-flip term can change the winding numbers, not their
parities).
\item At temperature much larger than $J$ and $V$, the square
lattice problem has algebraically decreasing dimer-dimer correlations,
whereas on the triangular lattice these correlations decrease
exponentially~\cite{ms01}.
\end{itemize}

As in the square lattice case, at the point $J=V$ the model
is exactly solvable.
The ground-states here too are the equal amplitude superpositions
of all dimer coverings in each topological sector.

Demons: A lower bond for the ground-state energy is given by a
minimization of the Hamiltonian on each plaquette individually.
If the given plaquette has no parallel dimer (non
flippable plaquette), its energy is zero and if it has parallel
dimers its  potential energy is $V$ and  its kinetic energy is
 $\leq -J$. We can thus write a lower bond for the energy of the global
system, which is proportional to the number of flippable
plaquettes $n_{flip}$, as $ min\left[0, (V-J)n_{flip}\right]$.
 At $ V=J$ the equal
amplitude wave-functions in each topological sub-sector
saturate this lower bond. As ${\cal H}_{dimer}$ is positive semi-definite at this point, the equal amplitude wave-functions
 are thus  the unique ground-states of the problem.

Contrary to the case of the square lattice, the degeneracy of this
RVB subspace is only 4 on the triangular lattice
 (whereas it is of order N in the square lattice case).
 As in the square
lattice case, these RVB states  are degenerate with the 6 staggered configurations, which
are ground-states for any $V/J \geq 1$ (see Fig.~\ref{MSModel}).

The sum over all configurations of the equal
amplitude wave-functions is equivalent to the classical dimer
problem (up to the question of the staggered phase which has a
negligible statistical weight in the problem):
 thus the dimer-dimer correlations decrease exponentially with
distance at the point $J=V$. It's the description of a true RVB Spin
Liquid phase with exponentially decreasing spin-spin  and dimer-dimer 
correlations (we thus expect a gap in the singlet sector),
and translational
invariance of the ground-state (all four topological
ground-states, with equal amplitude wave-functions are in the
same ${\bf k}= (0,0)$ sector of the impulsion~\cite{mlms02}).

Monte-Carlo simulations~\cite{ms01} have shown that this phase
extents at least in the range $ 2/3 < V/J \leq 1$. It terminates
at $V/J = 1$, with a first order transition to the staggered
phase (seen in the Monte Carlo simulations as hysteretic
behaviors). The dimer-dimer correlation function is very short
range in all the above-mentioned range of parameter, and very
weakly dependent on temperature, which is suggestive of a gap in
the spectrum.

We will not comment on the rest of the phase diagram, as it is
not relevant to our main point here. 
 It is described in Fig.~\ref{MSModel}.
(For more details, see the original paper~\cite{ms01}).
\begin{figure}
\begin{center}
\resizebox{13.5cm}{!}{
\includegraphics{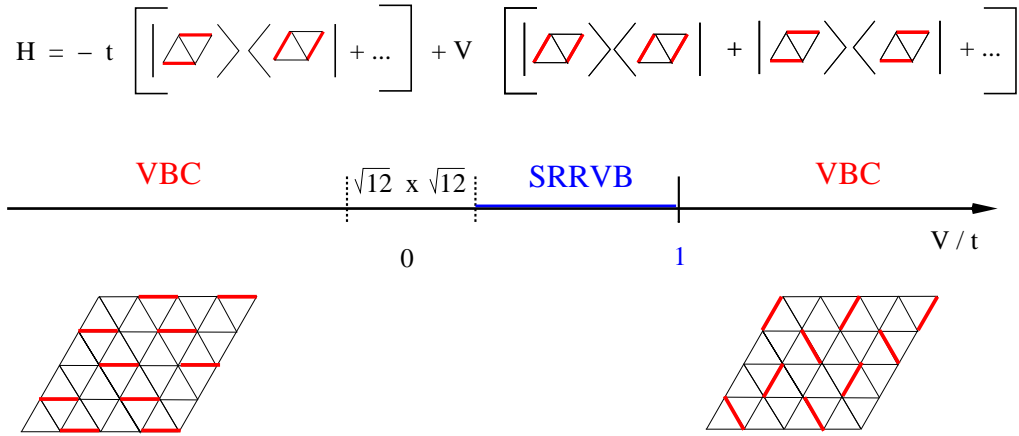}}
\end{center}
\caption[99]{The phase diagram of the Quantum Hard Core Dimer
problem on the triangular lattice}
\label{MSModel}
\end{figure}

{\bf Spinons:}
As noted above the RVB phase has a gap to collective excitations,
which is equally true of VBC. The major difference insofar
between VBC and RVB Spin Liquids is the existence in this new
quantum phase of deconfined spin-1/2 excitations: the spinons.
If you break a Valence Bond in a VBC phase  and try to separate
the two single spins from each other the energy of the system 
increases as the length of the string of misaligned dimers which
appears between the two single spins (take as an example
the staggered or the columnar phase of the QHCD).
This creates an elastic
restoring force which binds the two spin-1/2 together: in such a
Valence Bond Crystals excitations have always an integer spin
($\Delta S = 0$ or $1$). We suspect  that in the RVB Spin Liquid state,
 where the  correlations between local operators
are short range and any disordered configuration as probable as
an other, the restoring force between two single spins beyond a
certain distance will be negligible and the spin-1/2
(``spinons'') will be deconfined. A simple verification can
 be done on the equal amplitude states of the Quantum Hard Core Dimer model
 whatever the lattice: spinons do not interact beyond one lattice step.
One expects  this property to extend in all the RVBSL phase at
T=0.
At high temperature the classical square lattice is known to be
logarithmically confining~\cite{fs63}. Moessner and Sondhi have checked
that the triangular lattice is not confining.
 
The existence of deconfined spin-1/2 excitations, and thus
of a continuum of excitations just above the gap is
the main experimental signature of the 2-dimensional Valence
Bond Spin Liquid. It was recently claimed that this continuum of
excitations has been observed in Cs$_2$CuCl$_4$ which is supposed to be a
two dimensional magnet~\cite{cttt01}.

Two more complex evidences of RVB Spin Liquids phases had been
obtained
 before the discovery of this simple toy model: the first in 1992
in a large $N$, $Sp(N)$ analysis of Sachdev~\cite{s92}, the
second in a more realistic $SU(2)$ spin model by Misguich and
coworkers~\cite{mblw98,mlbw99}. This last work will be the object of the
next section.
\end{section}
\begin{section}{The MSE model or Ring Exchange model on the
triangular lattice}

The multiple-spin  exchange model  (called  MSE in the  following) was
first introduced by Thouless~\cite{t65}  to describe 
the nuclear magnetism  of
three-dimensional         solid    He$^3$~\cite{rhd83}      and     by
Herring~\cite{h66} 
for  the Wigner crystal.   It is an effective
Hamiltonian which governs the spin degrees of  freedom in a crystal of
fermions.  The Hamiltonian is a sum of permutations which exchange the
spin variables  along rings of  neighboring sites.  It  is now largely
believed that MSE interactions on the triangular lattice also describe
the    magnetism     of  solid   He$^3$    mono-layers    adsorbed  on
graphite~\cite{rbbcg98,mblw98,collin01} and that  it  could be a  good
description   of     the  two   dimensional     Wigner    crystal   of
electrons~\cite{bcc01}. In He$^3$, exchange terms including
up to  6 spins are present~\cite{rbbcg98}.
Recent discussions equally concern the strength and importance of 
 the 4-spin exchange term in
La$_2$CuO$_4$~\cite{chapfmcf01,kk03,lst03,d02}.

 Here  we will only focus on
2- and  4-spin interactions which   constitute the minimal  MSE  model
where  a   short-range   RVB  ground-state is  predicted   from  exact
diagonalizations~\cite{mlbw99}. The Hamiltonian, which is also
called by some authors the Ring Exchange model, reads:

\begin{equation}
H=J_2 \sum_{
\begin{picture}(17,10)(-2,-2)
	\put (0,0) {\line (1,0) {12}}
	\put (0,0) {\circle*{5}}
	\put (12,0) {\circle*{5}}
\end{picture}
} P_{i j}
+J_4 \sum_{
\begin{picture}(26,15)(-2,-2)
        \put (0,0) {\line (1,0) {12}}
        \put (6,10) {\line (1,0) {12}}
        \put (0,0) {\line (3,5) {6}}
        \put (12,0) {\line (3,5) {6}}
        \put (6,10) {\circle*{5}}
        \put (18,10) {\circle*{5}}
        \put (0,0) {\circle*{5}}
        \put (12,0) {\circle*{5}}
\end{picture}
} \left( P_{i j k l}+P_{l k j i} \right)
\label{J2J4}
\end{equation}

The  first   sum runs  over  all pairs  of  nearest  neighbors  on the
triangular  lattice and $P_{i j}$ exchanges  the spins between the two
sites $i$ and $j$. 
 The second sum runs over  all the 4-sites plaquettes and $P_{i
j k  l}$ is  a cyclic permutation  around the  plaquette.  The  2-spin
exchange  is equivalent to the  Heisenberg  interaction since  $
P_{ij} = 2 \vec{S}_{i}\cdot\vec{S}_{j}  +1/2 $, but the  four-spin term
contains terms involving 2 and 4 spins  and  makes  the model  a highly
frustrated one.

The general phase diagram of this model is given in
Fig.~\ref{MSEphs}.
\begin{figure}
\begin{center}
\resizebox{!}{10cm}{
\includegraphics{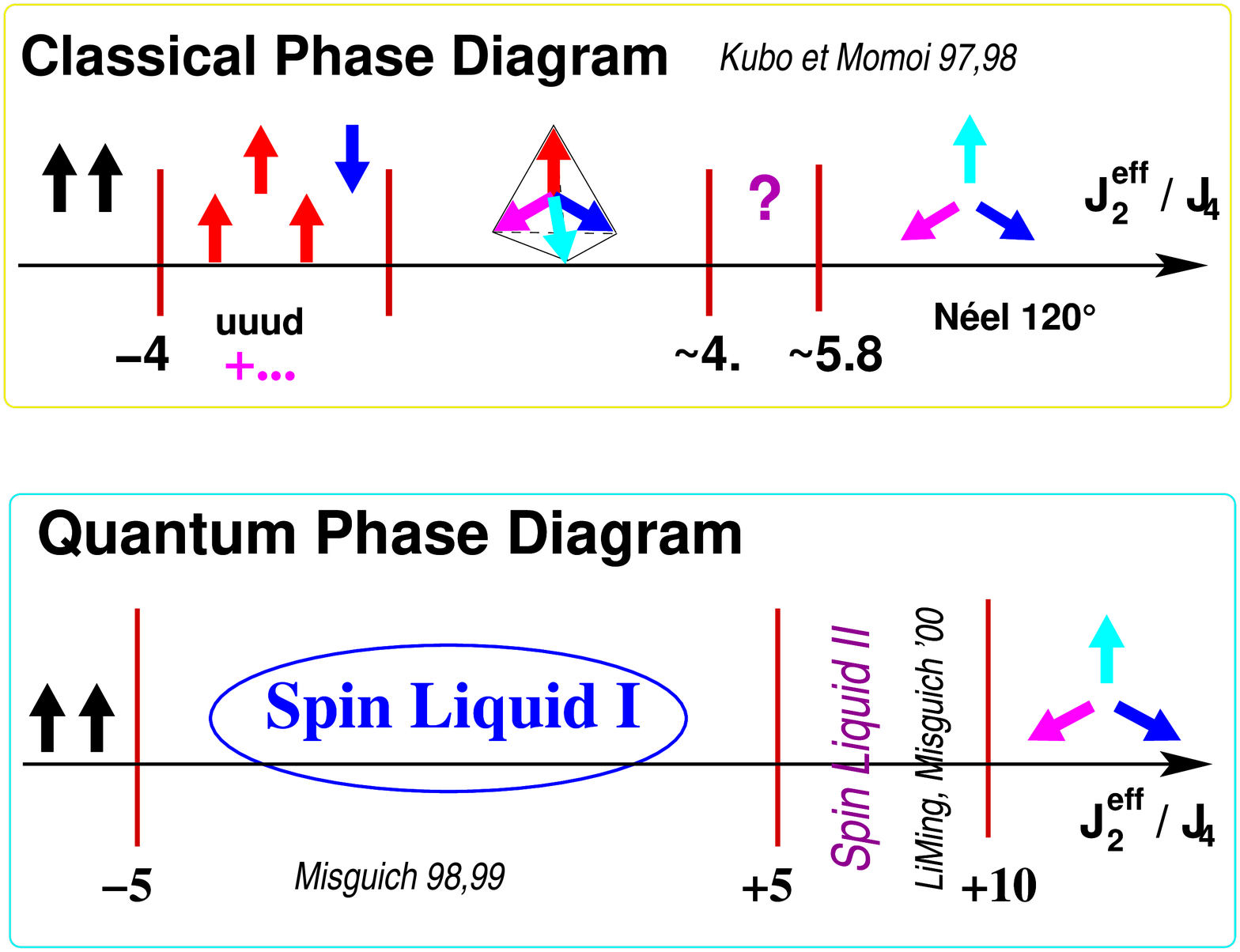}}
\end{center}
\caption [99]{The phase diagram of the ring exchange model
(Eq.~\ref{J2J4}) on the triangular lattice~\cite{lmsl00,mlbw99}.}
\label{MSEphs}
\end{figure}

We will now focus on the phase described as ``Spin Liquid I'' in
Fig.~\ref{MSEphs} and more precisely on the point 
 $J_2<0$ and $J_2/J_4\simeq -2$  which has been studied extensively
 by   means of exact
diagonalizations up to  $N=36$ sites~\cite{mlbw99} (this is a  good 
qualitative description of the low-density solid He$^3$ films).

Finite size effects on the spin gap and energy per spin are
displayed in Fig.~\ref{Egap}.

\begin{figure}
\begin{center}
\resizebox{!}{12cm}{
\includegraphics{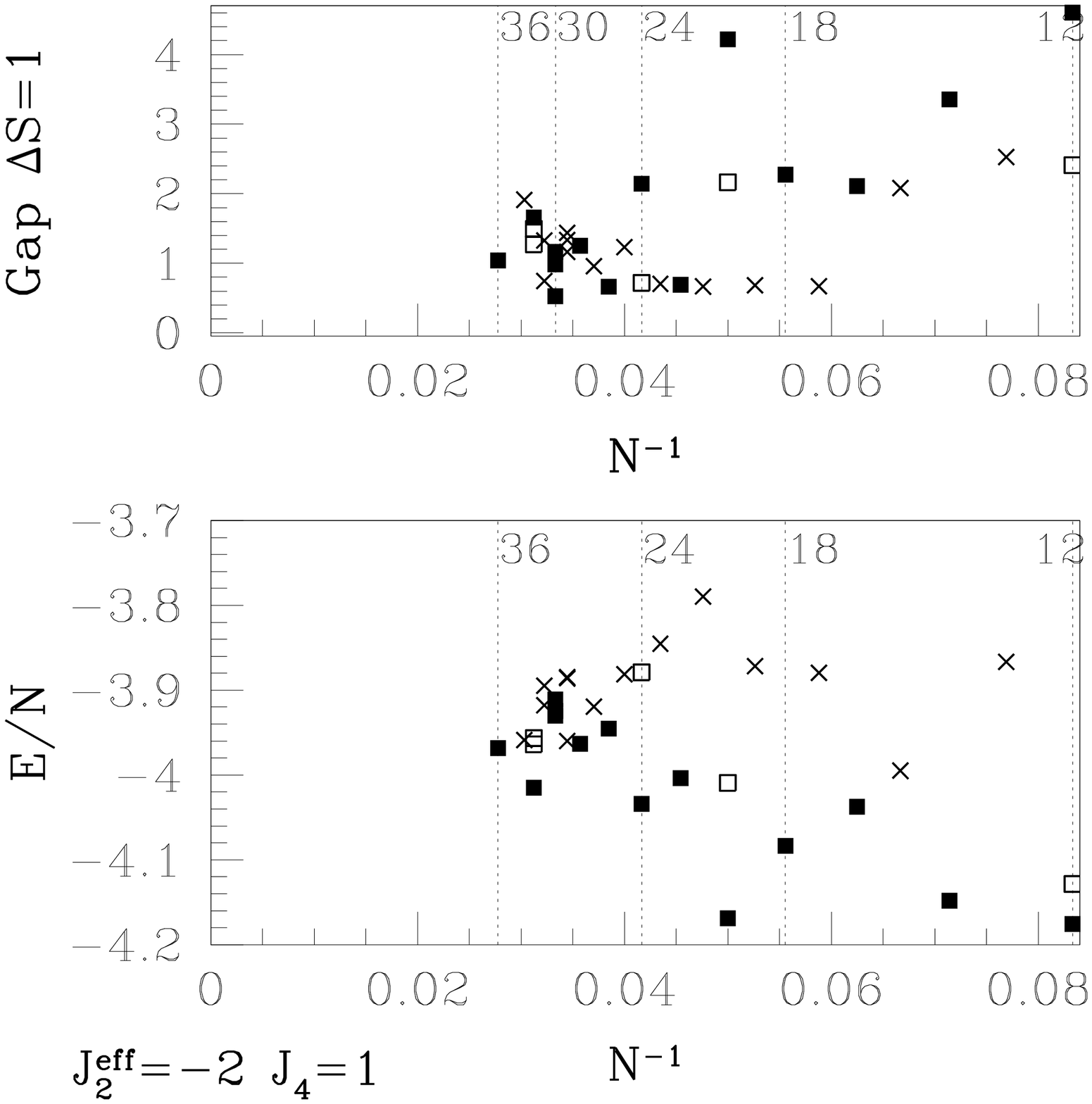}}
\end{center}
\caption [99]{Finite size effects on the spin gap (top graph)
 and energy per
spin (bottom graph) in the Ring Exchange model (Eq.~\ref{J2J4}) for $J_2=-2$, $J_4=1$.
Ref.~\cite{mlbw99} and unpublished results. Samples with an odd
number of sites are indicated by crosses. The full squares are for
even samples with the full symmetry of the infinite lattice and
open squares for even samples with lower spatial symmetries. Study
of the energy per site (bottom figure) is specially interesting.
Shape effects are still important for sizes as large as 32 but is
appears clearly that the energy of the most frustrated samples
(odd number of sites: crosses) is converging to the same limit
as the energy of the unfrustrated ones. This is a good indication that the
larger sizes that we have considered should allow significant
qualitative conclusions.}  
\label{Egap}
\end{figure}

   These data point to a spin-gapped phase
 with a  short correlation length
(of the order of a few lattice steps) and a spin gap of order 1. 

Three properties should be emphasized:
\begin{itemize}
\item No sign of a VBC could be found. All correlations functions 
spin-spin, dimer-dimer, 4- and 6-spin plaquette-plaquette seem
short range~\cite{mlbw99,m99} and consequently all
susceptibilities associated to local observables are zero in the
ground-state of the MSE spin liquid (see ref.~\cite{mlms02}).
\item The ground-state displays at the thermodynamic limit a
degeneracy that has been shown to be purely of topological
origin~\cite{mlms02}.

\item  The system probably supports unconfined spinons
\footnote
{On the basis of too small samples we
had concluded in our 1999 paper that spinons were probably confined
(which was a bit unpleasant and contradictory with the existence of
a topological degeneracy and the absence of any order in a local
order parameter.) Extending the
calculations to larger sizes up to $N=33$, we have now results
that clearly point  to deconfined spinons for large enough
distances.}.
\end{itemize}

\begin{subsection}{Topological degeneracy:}
This subject is fully developed in
ref.~\cite{mlms02} and illustrated in Fig.~\ref{topdeg}. In this
figure the first graph (top left)
 shows the low lying singlet states for the
N=36 sample. On this graph one sees that these singlet levels
appear as multiplets (the black symbol is one time degenerate and
the red symbol has a 3-fold degeneracy). With the system size the
"red" levels collapse to the black ones exponentially fast, with
a characteristic length which is of the order of $0.6$ lattice
step (finite size scaling in the bottom right figure): this is
the degeneracy, that we argue to be of topological origin.

\begin{figure}

\hspace*{-2cm}
\resizebox{!}{6.5cm}{
	\includegraphics[19pt,264pt][580pt,704pt]{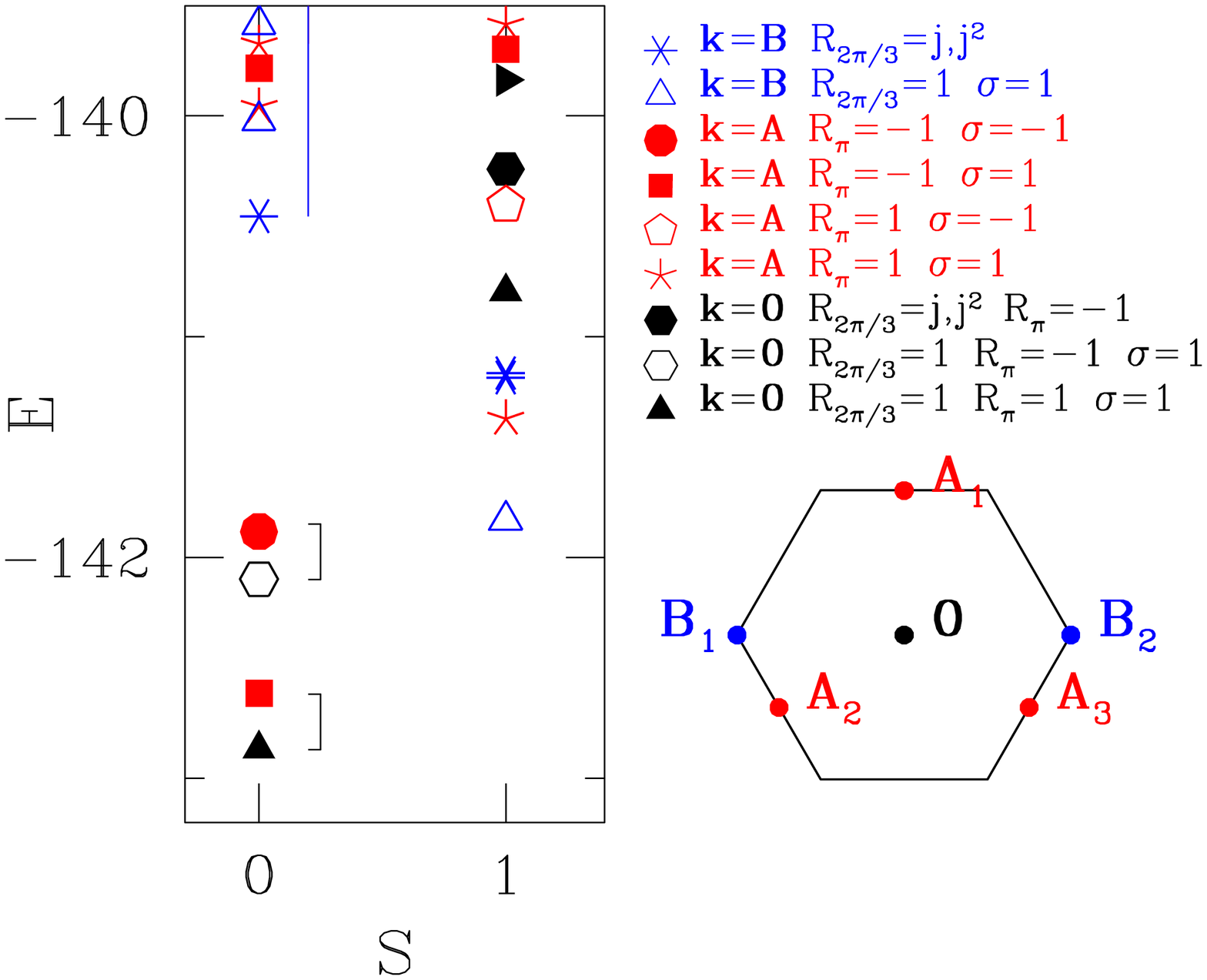}}

\vspace*{-6.7cm}
\hspace*{7cm}
\resizebox{!}{6.5cm}{	\includegraphics{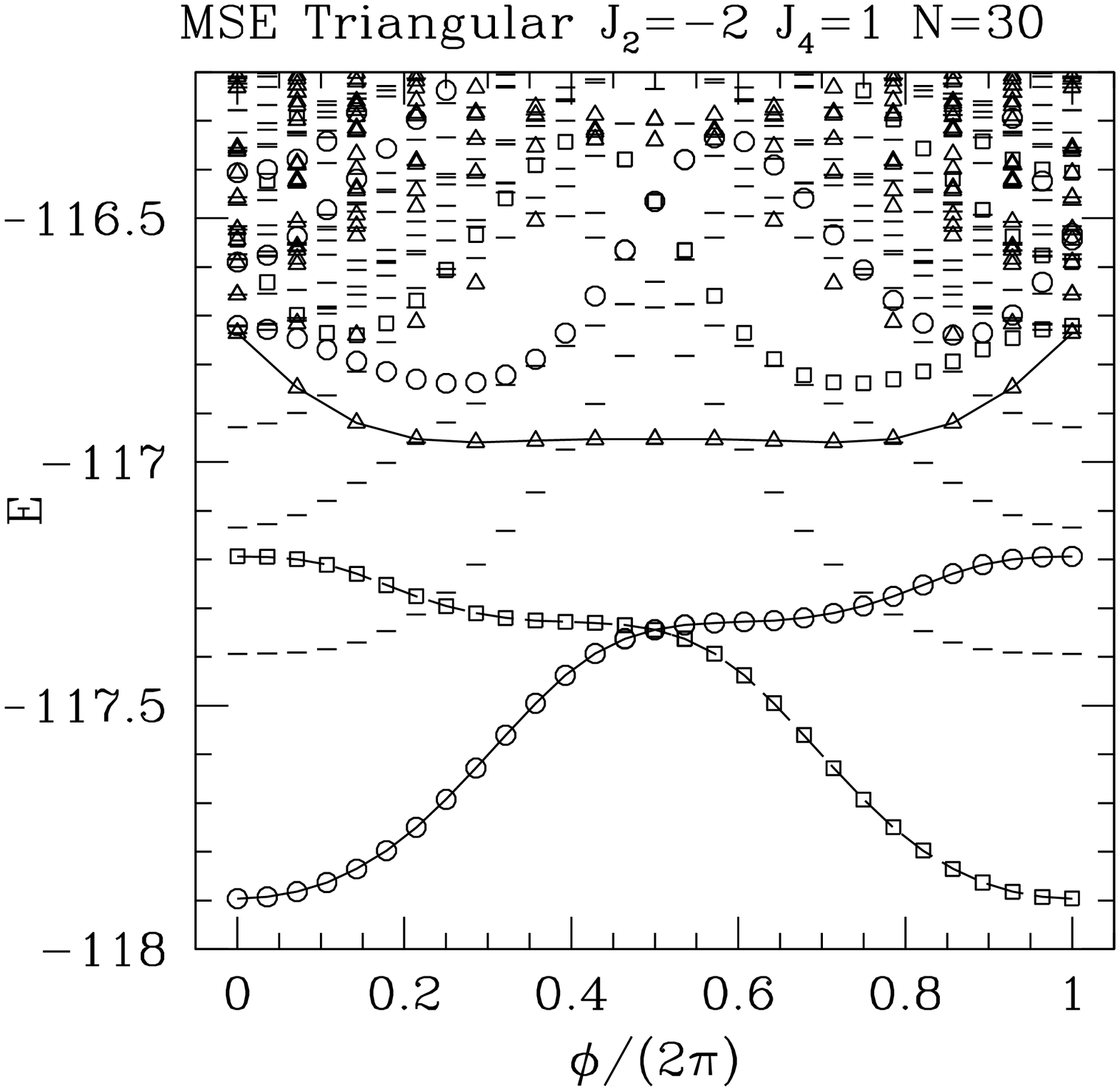}}

\vspace*{2.5cm}

\hspace*{-1.5cm}\resizebox{!}{4.5cm}{ \includegraphics*[15pt,315pt][592pt,706pt]{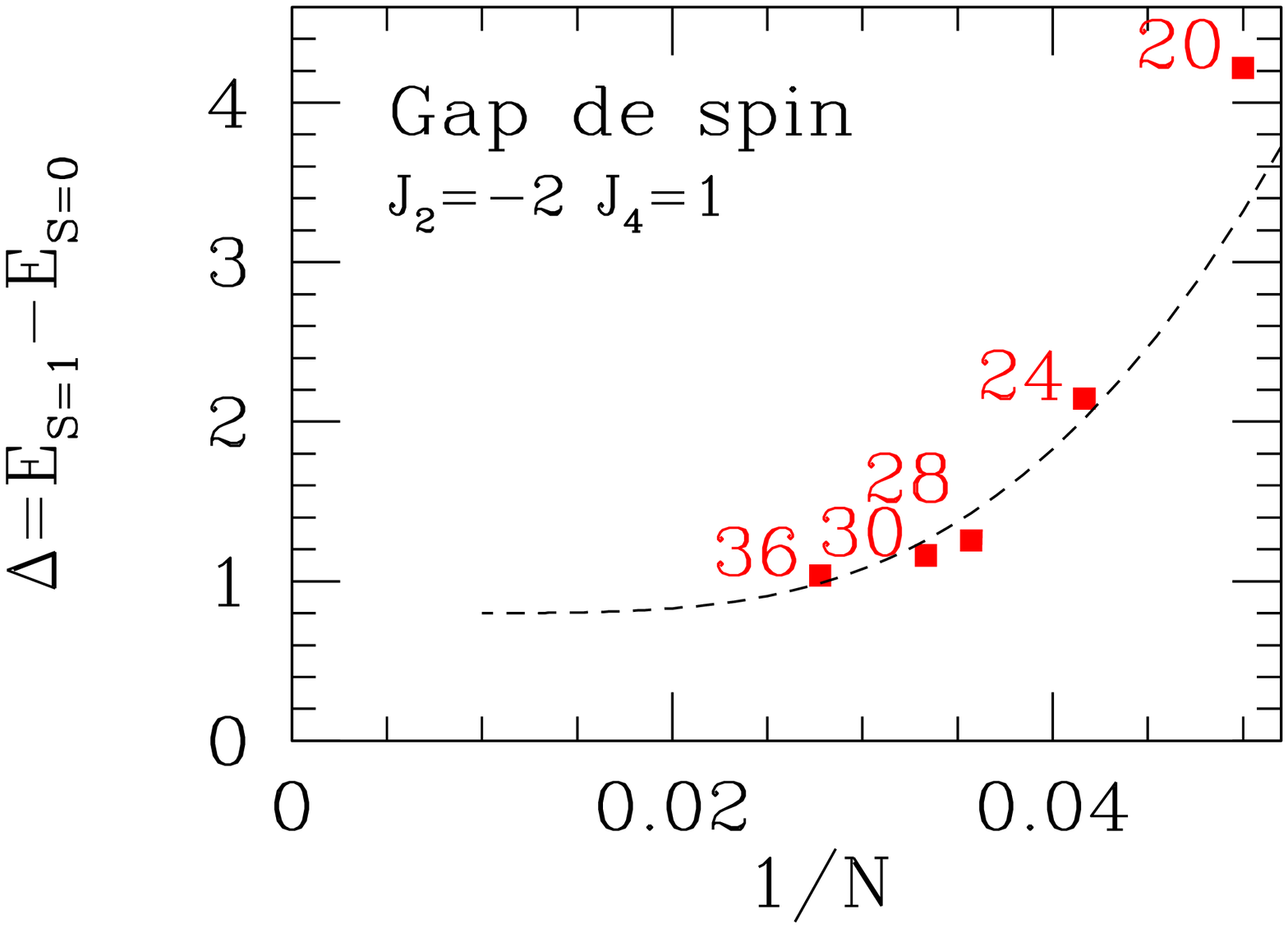}}
 
\vspace*{-4.7cm}
\hspace*{6.5cm}\resizebox{!}{4.5cm}{
\includegraphics{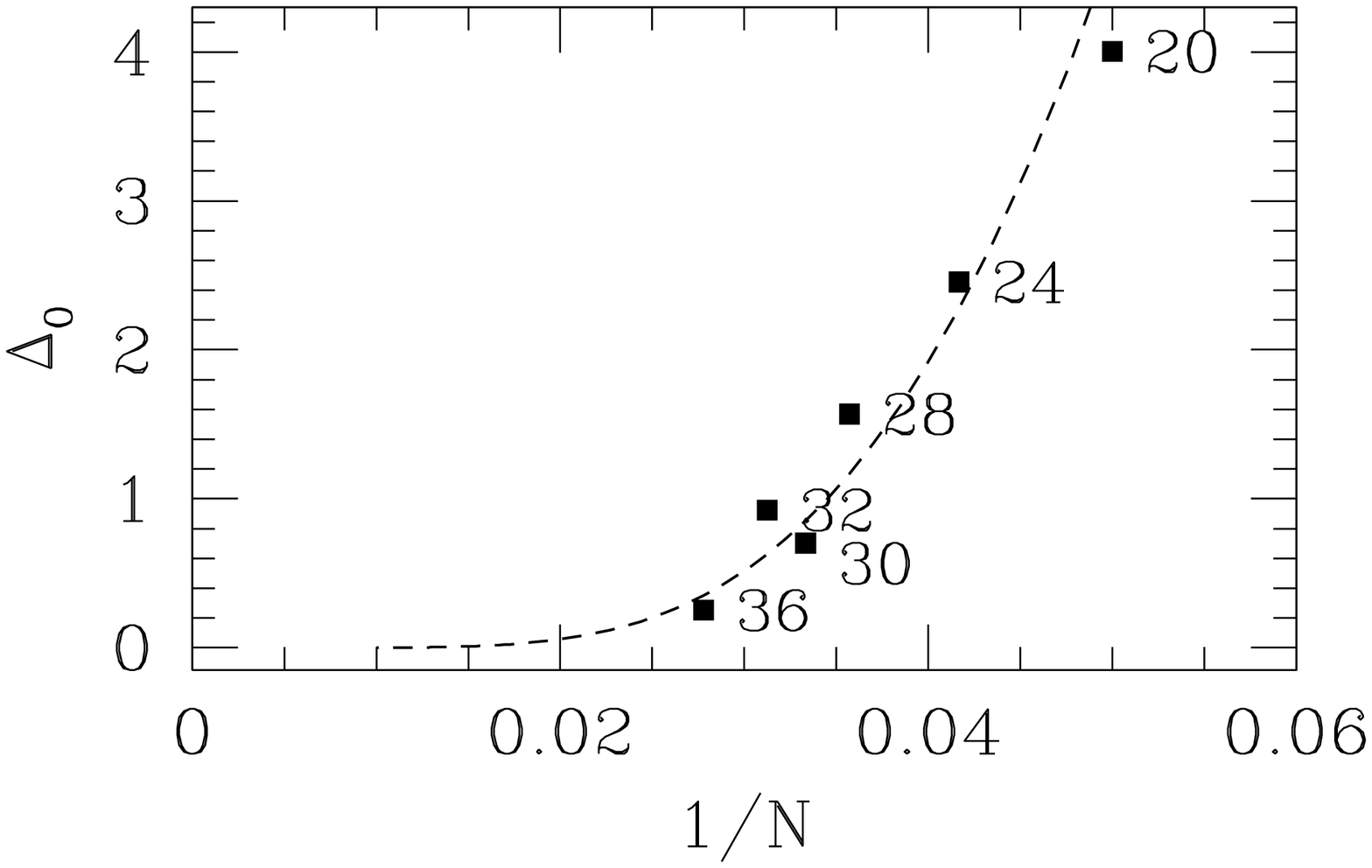}}\\
\hspace*{8.5cm}\vspace*{-0cm} $exp( -L/\xi)$ , $\xi =0.6$.

\caption [99]{ Topological degeneracy in the Ring Exchange model (Eq.~\ref{J2J4}) for $J_2=-2$, $J_4=1$.
Ref.~\cite{mlbw99} and unpublished results. The figure at the top
left shows the low lying levels in the singlet and triplet sectors for
the sample of size $N=36$. Notice that the singlet gap is already
much smaller than the triplet gap. The figure at the bottom left
gives the finite size scaling of the spin gap. The figure at the
bottom right shows how the gap in the singlet sector is closing
as a function of the size. It is exponentially 
decreasing with the linear size
of the lattice, and the correlation length is about 0.6  lattice step.
Such a law  correctly describes the finite size
scaling of the spin gap of unfrustrated samples (bottom left
graph),
 with an estimate of the spin gap $\sim 0.8 \pm 0.1$. The figure in the
top right shows the effect of an adiabatic twist of the boundary
conditions  in exchanging the two quasi-degenerate
topological levels.
}
\label{topdeg}
\end{figure}

In short the topological degeneracy can be understood using two
arguments: {\it i)} the wave-functions describing the MSE spin
liquid can be classified as short range RVB wave functions (all
the correlations functions in local observables are short range),
{\it ii)} the Hamiltonian is a local operator insensitive to a
global property as the parity of the winding number. In other
words for  large enough sizes, it is possible to locally optimize
the energy and the result does not depend on the topological
sector where it is done. As there is 4 topological sectors for a
triangular lattice on a two torus, we have there the origin of
the 4-fold degeneracy.

On lattices with an odd number of rows it is possible to
transform one topological sector in another by a $2\pi$ twist of the
boundary conditions along the even direction (Fig~\ref{topdeg}
top-right graph). 
( This is equivalent to  the introduction of a quantum of 
fictitious flux through the torus: detailed
explanations can be found in ref.\cite{mlms02}). In such an
operation the global spectrum of the Hamiltonian is unchanged but
eigen-states are not mapped on themselves:
 the momentum of the states is translated by
 $(\pi,0)$ in
such an operation. Oshikawa~\cite{o00} concluded that such systems
should exhibit a doubling of the Brillouin zone and the spontaneous
symmetry breaking of a Valence Bond Crystal. This is an incorrect
speculation as we will explain in  section 5.5.
A simple counter-example is given by the HCQD model:
 the ground-state in the different
topological subspaces for (even, even) samples are all in the
${\bf k}= (0,0)$ sector of the momentum. There is thus an alternation
 of the quantum numbers of the degenerate ground-state multiplicity
 when going from (even, even) to (even, odd) samples 
{\it whatever the system size}:
 this indeed is inconsistent with a VBC  spontaneous
symmetry breaking in the thermodynamic limit. In the same line, in the MSE
model, the evolution with the system size of the quantum numbers
associated to the point group equally shows that this topological
degeneracy is by no way associated to some symmetry breaking of
the lattice~\cite{mlms02}.

A last comment  relative to the possibility of using
this degeneracy to produce {\bf quantum bits protected from
decoherence effects}~\cite{k97,ifi02}. At first sight the idea is
attractive: due to the absence of long range order in any local
variable the local susceptibilities vanish in the thermodynamic
limit~(see ref.~\cite{mlms02}) and one expects such quantum bits
to be insensitive to any local cause of decoherence. But the
charm of this property is to be paid by highly non trivial, if
not impossible, writing and reading of the state of the quantum
bit (which would imply  manipulation of  gauge fields ..).
\end{subsection}
\begin{subsection} { Unconfined spinons?}
To estimate the confinement energy of two separated single spins
(see Fig.\ref{spinons}),
we use the following arguments:

{\it i)} A sample of $2N$ spins has a ground-state energy which is
essentially:
\begin{equation}
E ( 2N, S=0) = 2 N e_\infty  + {\cal O}(\frac{1}{\sqrt N}),
\label{eq.a}
\end{equation}
the correction term to the $N \to \infty$ limit is at
most ${\cal O}(\frac{1}{\sqrt N})$ if there is long range order in
spin-spin correlations and is more plausibly ${\cal O}(\frac{exp(- L/ \xi)}{\sqrt N})$ in the present case.

\begin{figure}
\vspace*{-10cm}
\hspace*{2cm}
\resizebox{!}{12cm}{
\includegraphics{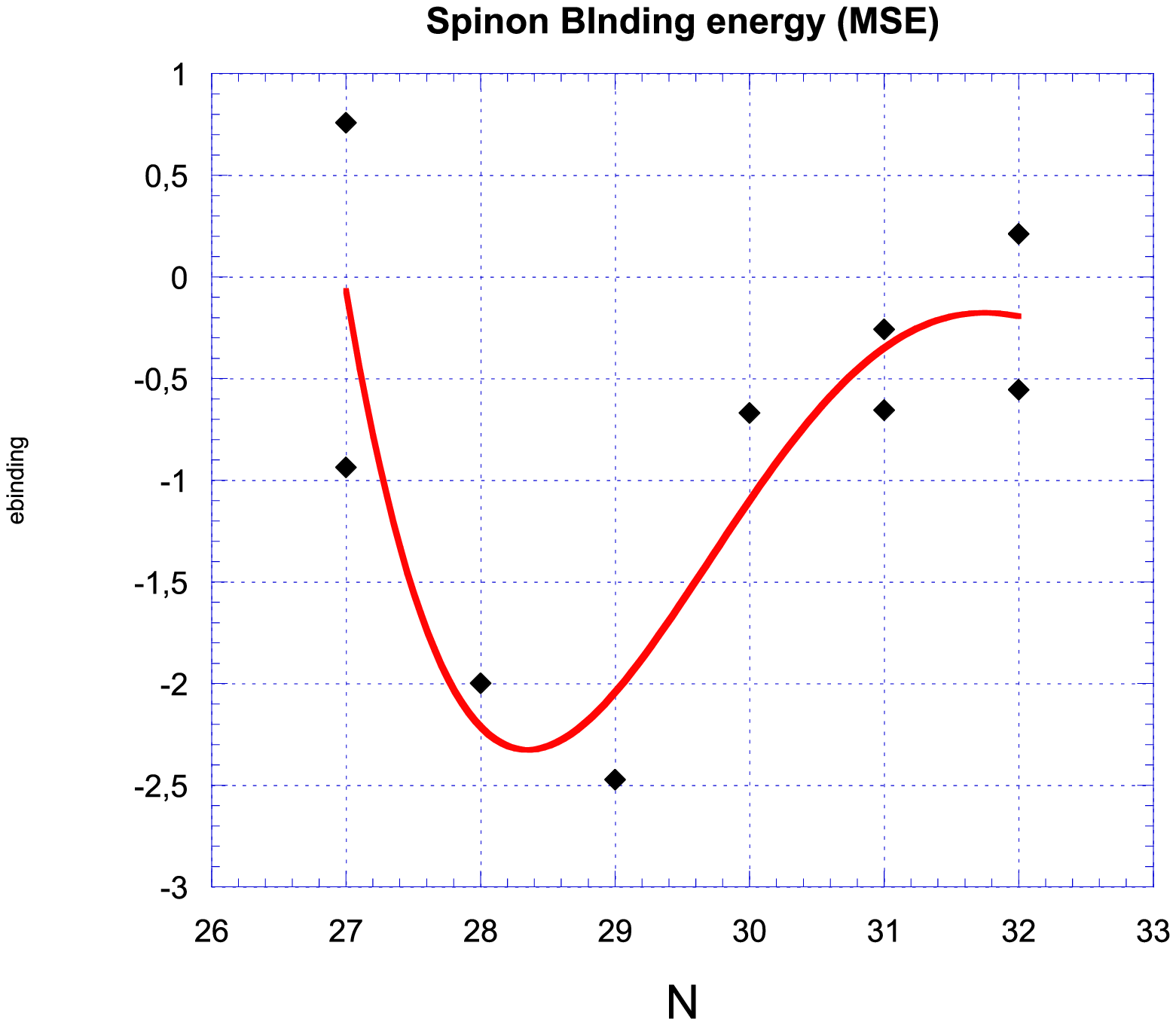}}
\vspace*{2.6cm}
\caption [99]{Spinons binding energy in the Ring Exchange model
(Eq.~\ref{J2J4}) for $J_2=-2$, $J_4=1$. One sample only has been
used for the sizes 28, 29, 30; this precludes an estimation of the
uncertainty due to the sample form for intermediate results.
Notwithstanding this, the displayed results do not infirm the
conclusion of a zero energy binding for large enough sizes. The
red curve is an imperfect and arbitrary polynomial fit to indicate
 a general tendency
(unpublished results, work in progress).}
\label{spinons}
\end{figure}
{\it ii)} A sample withe $2 N + 1$ sites can accommodate N valence
bonds + one single spin, we can write its ground-state energy
as:
\begin{equation}
E ( 2N +1, S=1/2) = (2 N+1)\; e_\infty  + \Delta_{spinon} + {\cal O}(\frac{1}{\sqrt N} \cdots)
\label{eq.b}
\end{equation}
where $\Delta_{spinon}$ measures the energy gap for the creation
of one spinon.

{\it iii)} If we now look to the first $\Delta S =1$ excitation of an
even sample, we expect (in the hypothesis where spinons are the first
excitations of this system) that:
\begin{equation}
E ( 2N,  1st\; exc.\;lev.) = 2 N e_\infty  +  2 \Delta_{spinon} + E_{binding} + {\cal O}(\frac{1}{\sqrt N} \cdots)
\label{eq.c}
\end{equation}
where $ E_{binding}$ is the binding energy of two spinons. If $ 
E_{binding}$ is negative in the thermodynamic limit then the
spinons will be confined, the first excitations will be integer
spin excitations 
  as expected in a Valence Bond Cristal.  If $ E_{binding}
\to 0$ with $\Delta_{spinon} \to \Delta_0 \neq 0 $
with increasing size, we then expect the spinons to be unconfined
and the first excitations of the model are fractionalized~\cite{k01,ns01,ps02,s01,bfg02,dnkks02,msf02}

Using these three equations
(Eqs.~\ref{eq.a},~\ref{eq.b},~\ref{eq.c}) for consecutive sizes
one obtains an estimate of the binding energy for each sample
size. The results are shown in Fig.~\ref{spinons}. They give a
positive
indication in favor of unbound spinons in the MSE model.
\end{subsection}\end{section}

\begin{section}{RVB Spin Liquids in other spin models}
We suspect that RVB Spin Liquids could be observed in other spin
models. As already discussed in the previous chapter, 
the case of the $J_1-J_2$ model on the 
square lattice is still debated, the correlation length being probably
 larger than the largest sizes actually available. The $J_1-J_2$ model
on the hexagonal lattice may have 
spin liquid phases for $ J_2/J_1 \sim 0.3$ and around the point
with ferromagnetic $ J_2$ and $ J_2/J_1 = - 0.25$~\cite{fsl01}.
Here too, and contrary to the MSE model the range of
parameters where a spin liquid phase might appear is relatively
small, the local S=0 objects  probably  extend over a few
lattice cells, and as
a consequence the gaps are rather small and the shape effects a
bit chaotic for the available sizes.

An explanation of the robustness of  the short-range RVB  phase in the MSE
model can be guessed  from the analogy between multiple-spin  interactions
and QHCD models.  From the analysis of QHCD  models we understand that
RVB phases are possible when VBC are energetically unstable.  Columnar
VBC are   stabilized  by   strong    parallel dimer  attraction     and
staggered~\cite{ms01} VBC appear   when  the repulsion  between  these
parallel    dimers  is strong.    In    between,  an   RVB  phase  can
arise~\footnote{In  Ref.~\cite{ns01}   an   RVB state  is   selected by
introducing defects in the   lattice   in order to destabilize     the
competing VBC  states.}.    From this   point  of  view,  increasing the role of the kinetic term of the model is essential: it is exactly
 the role of the    four-spin
ring exchange term of the MSE model (Eq.~\ref{J2J4})\cite{m99}.

The role of the triangular lattice (or of a preferred triangular
sublattice as in the case of the $J_1-J_2$ model on the hexagonal
lattice) should probably also be emphasized: J.C. Domenge's
 preliminary work on the MSE model on the square lattice points to
a smaller extent of the Spin Liquid phases~\cite{d02}.
\end{section}
\begin{section}{Short range RVB: topological degeneracy and
absence of symmetry breaking}

In this section we develop two important properties of type I 
short range RVB Spin Liquids,
 in a more general point of view than in the preceding sections
 and independently of a
specific Hamiltonian, with the only restriction that it is a
short range operator and that the ground-state and first excited
levels can be described
with resonant Valence Bond superpositions of short range dimers
(not exclusively first neighbors).

(For simplicity  this section includes 
some parts of the paper by Misguich 
{\it et al.}\cite{mlms02} but
 some demonstrations and examples, which should not
alter the general understanding are omitted 
and in some cases the reader is referred to the original work to
complete the picture.)

\begin{subsection}{
Topological degeneracy of the ground-state
multiplicity of a type I RVB with half integer spin in the unit
cell}

\subsubsection{Definition of the topological sectors}
\label{deftopsec}

\begin{figure}
        \begin{center}
        \resizebox{3cm}{!}{\includegraphics{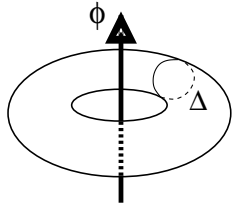}}
        \end{center} \caption{2-torus with one cut $\Delta$.}
        \label{torus} 
\end{figure}

Let us  draw a cut $\Delta$ encircling the torus created by periodic
boundary   conditions  (see Fig.~\ref{torus}).   This hyper-surface of
dimension $d-1$ cuts bonds of the lattice but there is no site sitting
on it. The position  of the cut is  arbitrary. 
 The   family   of
nearest-neighbor dimer coverings  can be decomposed into two subspaces
${\cal E}^\pm_\Delta$ depending on the    parity $\Pi_\Delta$ of   the
number  of    dimers    crossing the cut    $\Delta$~\footnote{
The definition of Rokhsar and Kivelson~\cite{rk88} that we used
 before is
equivalent to this one but less practical in the present context.}.
 By considering a set of
$d$ cuts $\Delta_{i=1,\dots,d}$ encircling  the torus in all  possible
directions one obtains $2^d$ families of dimer covering.

Any movement of dimers  can be represented  as  a set of closed  loops
around  which  dimers are   shifted in a  cyclic way.    A {\em local}
operator  will only generate contractible  loops which will cross each
cut a {\em  even} number of times.   The number of dimers crossing the
cut can therefore only be changed by an  even integer and the parities
$\Pi_{\Delta_i}$ are unchanged.
        
This  property remains true as  long as one  works in a subspace where
the  dimer lengths are  smaller than  the linear  system size, that is
when  the topological sector are  well  defined (if a  dimer length is
half the   linear of the system   one cannot decide  by which  side it
goes).  On the other hand, we checked on the triangular (resp. Kagome)
lattice that these 4 sectors are  the only topological sectors: local
3- (resp 4-) dimer moves can be used to transform a configuration of a
given sector into any other configuration of that sector.
          
These subspaces are   orthogonal  in the thermodynamic
limit~\cite{b89a}.  The graph of the scalar product $<c^+|c^->$ of two
dimer configurations belonging to different subspaces ${\cal E}^+$ and
${\cal E}^-$  has at least one long  loop encircling the  torus in the
$L_x$ direction.  When $L_x$ goes to infinity this contribution to the
scalar product is smaller than  $2^{-L_x/2}$.  Consider two normalized
vectors  $\left|\Psi^+\right>$  and  $\left|\Psi^-\right>$ belonging to  two
different sectors:
\begin{equation}
	\left|\Psi^\pm\right>=\sum_{c^\pm\in\mathcal{E}^\pm}\Psi^\pm(c^\pm) \left|c^\pm\right>
\end{equation}
Because of the exponential number  of dimer coverings in each  subspace
it is not obvious that $\left|\Psi^+\right>$  and  $\left|\Psi^-\right>$ are
orthogonal in the thermodynamic limit. But it is nevertheless the
case~\cite{mlms02}, and: 
\begin{equation}
\left<\Psi^+|\Psi^-\right> < {\cal O}(2^{-L/2}).
\label{ineqB}
\end{equation}

In  the following, unless explicitly   mentioned, we consider the  2D
case for simplicity but most of the topological arguments about dimer
covering immediately extend to higher dimensions.

\subsubsection{Two-fold degeneracy in even$\times$odd samples}

In the special  case of tori  with an odd number  of rows (and  an odd
number  of spin-$\frac{1}{2}$ per crystallographic  unit cell),  one step
translation along the $x$ axis (called  ${\cal T}_x$ in the following)
maps  ${\cal E}^+$ on  ${\cal  E}^-$ and  reversely. Some  point-group
symmetry can  also do this job. A  $\pi$ rotation about a lattice site
nearby the  cut (called $\Rpi$ in  the following) has the same effect.
If the cut is chosen  parallel to a symmetry  axis  of the sample,  a
reflection   with  respect to   this axis   (called  $\Sigma_y$ in the
following)  will equally  map   ${\cal E}^+$   on  ${\cal  E}^-$   and
reversely.

All  these symmetry operations isolate a  single column $C$ of lattice
sites  between $\Delta$ and  its transform  $\Delta'$.   In that  case
columns have an odd number of  sites and an odd  number of dimers must
connect  some sites  inside  $C$ with   sites  outside $C$.  Therefore
$\Pi_{\Delta}$ differs from   $\Pi_{\Delta'}$ and  the two   subspaces
${\cal E}^+$ and ${\cal E}^-$ are exchanged.

For  a  large enough system  these  two sectors  are 1) orthogonal, 2)
uncoupled by any   local  Hamiltonian and  3) exchanged   by  symmetry
operations   (even$\times$odd).  This is  enough to  insures that they
have the  same  spectrum, irrespectively of the   physics of the model,
provided it can be described in the short-range  dimer space. In fact,
quantum  numbers of these doublets  of degenerate  states are fixed by
symmetry.

We decompose an eigenstate $|\psi_0>$ on the two topological subspaces
defined relatively to the cut $\Delta$:
\begin{equation}
        \left| \psi_0\right> =\left|\psi_0^+\right>
        +\left|\psi_0^-\right>
\end{equation}
where  $\left|\psi_0^{\pm}\right>$   belong respectively   to the sets
${\cal E}_\Delta^{\pm}$.  $\left|  \psi _{0}\right>$, as an eigenstate
of  the Hamiltonian with periodic boundary  conditions,  belongs to an
irreducible representation of  the translation group. In the following
we  will also  assume  an   $\Rpi$  and $\Sigma_y$  invariance of  the
Hamiltonian and that,   for simplicity, $|\psi_0>$  transforms under a
one-dimensional representation under $\Rpi$ and $\Sigma_y$.

\begin{eqnarray}
        {\cal T}_x \left| \psi_{0}\right> =
        e^{i{\bf k}_0\cdot{\bf u}}\left|\psi_{0}\right>\nonumber\\
        \Rpi \left| \psi_{0}\right> = \rho^{\pi}_0 \left| \psi_{0}\right>\nonumber\\
        \Sigma^{y} \left| \psi_{0}\right> = \sigma^y_0 \left| \psi_{0}\right>
\end{eqnarray}
In       the  thermodynamic     limit  $\left|\psi_0^+\right>$     and
$\left|\psi_0^-\right>$ are orthogonal and   ${\cal T}_x$, $\Rpi$  and
$\Sigma^y$ map ${\cal E}^+$ on ${\cal E}^-$ and reversely:
\begin{eqnarray}
        {\cal T}_x \left| \psi_0^\pm\right> =
                e^{i{\bf k}_0\cdot{\bf u}}\left| \psi_0^\mp\right>\nonumber\\
 \Rpi \left| \psi_0^\pm\right> =
                \rho^\pi_0 \left| \psi_{0}^\mp\right>\nonumber\\
 \Sigma^y \left| \psi_0^\pm\right> =
                \sigma^y_0 \left| \psi_{0}^\mp\right>.
        \label{nq}
\end{eqnarray} 
Let us now build the variational state:

\begin{equation}
\left|\psi_{1,\Delta}\right> =
        \left|\psi_{0}^+\right> - \; \left|\psi_{0}^-\right>.
        \label{eq:psi1Delta}
\end{equation}

Eqs.~(\ref{nq}) imply:
\begin{eqnarray}
        {\cal T}_x \left| \psi_{1,\Delta}\right>
                = - e^{i{\bf k}_0\cdot{\bf u}}\left| \psi_{1,\Delta}\right>\nonumber\\
        \Rpi \left| \psi_{1,\Delta}\right>
                = -\rho^{\pi}_0 \left| \psi_{1,\Delta}\right>\nonumber\\
        \Sigma^{y}\left| \psi_{1,\Delta}\right>
                = -\sigma^y_0 \left| \psi_{1,\Delta}\right>
\end{eqnarray}
$\left| \psi_{1,\Delta}\right>$ has thus a wave-vector ${\bf k}_1$, a
rotation quantum number $\rho_1^\pi$ and a reflection quantum number
$\sigma^y_1$ related to the quantum numbers of $\left|\psi_0\right>$
by the relations:
\begin{eqnarray}
        {\bf k}_1 = {\bf k}_0 +(\pi,0)\nonumber\\
        \rho^{\pi}_1= - \rho^{\pi}_0\nonumber\\
        \sigma^y_1 = - \sigma^y_0.
        \label{symmetryqn}
\end{eqnarray}
It  is  thus  a  state  orthogonal to   the   ground-state (even  on a
finite-size  system where the topological  sectors are not rigorously
orthogonal).

Since  any   local  Hamiltonian has  exponentially   vanishing  matrix
elements between different sectors we have
\begin{equation}
	\left<\psi_0^+\right|
	{\cal H}_0
	\left|\psi_0^-\right> \to 0
\end{equation}
and    $|\psi_{1,\Delta}>$  is   thus degenerate    with the  absolute
ground-state,        their      symmetries    being    related      by
relations~(\ref{symmetryqn}).

\begin{subsubsection}{Degeneracies in even$\times$even samples}
\label{sss:exact_deg_e_e}

As remarked  by  Bonesteel~\cite{b89a} on a  square  lattice a $\pi/2$
rotation  exchanges sector $(+,-)$  and  sector $(-,+)$ but  sectors
$(-,-)$ and $(+,+)$ remain inequivalent.   A similar phenomenon occurs
on the triangular lattice  where $2\pi/3$ rotations permute cyclically
3 of the  4 sectors. As a result there are equally in even$\times$even
 samples
exact degeneracies due to these mappings~\cite{mlms02}. Readers
interested in the details of these degeneracies are referred to
the original paper~\cite{mlms02}.
\end{subsubsection}

\begin{subsubsection}{4-fold degeneracy in RVBSL phases}
\label{sec:4fdeg}

The numerical   data  on the type I Spin Liquid phase of the
 MSE  and QHCD models~\cite{mlms02}  suggest that  the  ground-state
degeneracy is  4  in  such a phase whatever may be
the   shape of the   sample.

There is no global mathematical proof of this property but
the following physical arguments make it extremely plausible.

Let us make the following assumptions:
 
a) The ground-state can be described
in  a  short-ranged  dimer  basis.    b)   All $n$-dimer   correlations
($n=2, 3, \cdots$)  are  short-range and  the  corresponding correlation
lengths are bounded.  c) The  Hamiltonian is local.

 From hypotheses a)
and c) it is clear that for a large enough system the four topological
sectors are not  mixed  in the  ground-state  and the spectrum  can be
computed separately in   each sector.  As seen before, we
  do  not  have any  symmetry
operation which connects all {\em four} sectors
(these operations connect only  two or three of the topological
subspaces, depending on the geometry of the sample)
 and we need a physical
argument to explain  why  energies should be  the same in each
sector  (in the thermodynamic  limit)?

   Because of their  topological
nature, it is  not  possible  to  determine  to which sector   a dimer
configuration belongs  by looking only at  a {\em finite area} of the
 system.
In other words,   any  dimer configuration defined  over  a  large but
finite part of the system can be equally  realized in all sectors. The
Hilbert space available   to the system   is the same  over any finite
region of the system.  In the absence of  any form of long-range order
the  system  can  therefore optimize   all  its correlations  with  an
arbitrary high accuracy equally well in each sector. At this point we
 can only conclude that the four sector will have the same
energy {\em density} and we cannot exclude the existence of a gap 
between
the different topological sectors.

 However the numerical results 
obtained
in the QHCD and MSE models indicate that it is not the case and that 
the
four ground-states have asymptotically the same {\em total} energy. 
 We think that this should be true for a general short-range RVBSL.

\end{subsubsection}

\begin{subsubsection} {Miscellaneous remarks on the RVB ground-state degeneracy}
\label{remarks}

--- {\em   Dimers     and twist  operator}.   The    variational state
$\left|\psi_{1,\Delta}\right>$   can   be    deduced  simply      from
$\left|\psi_0\right>$ by changing the  sign of the dimers crossing the
cut $\Delta$. Such an  operation can also been seen  as a $2\pi$ twist
of the spins of column 0.
 From the physical point
of   view   the   reason   why    $\left|\psi_{1,\Delta}\right>$   has
asymptotically the same energy as $\left|\psi_0\right>$ becomes clear:
in the absence of stiffness ($\sim$ absence of sensitivity to a
boundary twist in the thermodynamic limit)
and  of long-range spin-spin correlations,
the  perturbation  induced by  the boundary  condition 
cannot propagate and does not change  the energy of the initial state:
its only  effect is  to change  the  relative phases of the  different
topological  components of the   wave function,  and consequently  the
momentum  and space symmetry quantum  numbers of  the initial state of
the even-odd samples.

---  {\em Fractionalization    and topological  degeneracy.}  To   our
knowledge  all  present theoretical  descriptions  of   fractionalized
excitations       in        2D        magnets      or          related
problems~\cite{wen91,s01,sp02,k01,msf02}  (we should   also   mention
topological  properties of  Laughlin's   wave function for  fractional
quantum Hall effect~\cite{h85,wn90})     imply       topological
ground-state degeneracies.  In  such pictures, the physical  operation
which transforms a ground-state into another is the virtual creation of
a pair  of spinons (by  dimer breaking)  followed by  its annihilation
after the  circulation of  one of them   around the torus.   In such a
process a $\pi$  phase-shift   is introduced between   the topological
sectors (as in  the above recipe).  For samples  with an odd number of
rows  this operation connects eigenstates    with different ${\bf  k}$
vectors    (and    space   quantum     numbers)   as    described   in
Eqs.~(\ref{symmetryqn}). 

--- {\em Numerical studies  with a different topology.} An interesting
check  of  the pure topological  nature  of  this degeneracy  could be
obtained by studying the  problem no more on  a torus but on a surface
with  a different   genus.    On a  sphere   we expect  an  absence of
degeneracy.  Unfortunately
if  a lattice  can be represented   on an infinite  plane,  both the
number of links  $L$ and plaquettes $P$ depend  linearly on the number
of sites  $N$ and Euler's relation  $P-L+N=2-G$ constrains  the genius
$G$ to  be 2~!
The torus    is the only  possible   topology  if we require    a full
translation invariance in both directions.    In a recent work,  Ioffe
{\it et al.}~\cite{ifi02}  have studied the  absence  of sensitivity to
disorder as an evidence for topological phenomenon in the liquid phase
of  the QHCD model on  the  triangular lattice.   They also used  {\em
open} boundary conditions to modify the topology of the system and argue in
that case that the low-energy spectrum is free of edge states which
 could
hide the actual ground-state degeneracy.

--- {\em Example of RVB  phase with 2 spins  in the unit cell.} A spin
liquid  state, seeming very similar  to the state  observed in the MSE
model  on the triangular lattice, has  been observed  in the $J_1-J_2$
model on the hexagonal lattice~\cite{fsl01} for $J_1=-1,\;J_2=0.3$. No
quasi-degeneracy of the ground-state  has been  noticed. It should  be
remarked that in  this  system  there are  2 spins   $\frac{1}{2}$ per
crystallographic unit cell and no  degeneracy is expected on the basis
of the topological arguments.
\end{subsubsection}
\end{subsection}
\begin{subsection}{Symmetry breaking in gapped phases}
\label{sec:no_sym_break}

From the mathematical point of  view, ground-state wave functions that
break one-step translations or  space  group symmetries can  be  built
from linear  combinations of the degenerate  ground-states of the {\it
even-odd} samples.  In a completely  equivalent way, ground-state wave
functions    that  break rotation     symmetry    can  be   built   in
even$\times$even samples.  One could thus  superficially conclude that
spontaneous symmetry breaking is possible in RVBSL, we will show below
that this assumption is false.

There are many  features which show that this degeneracy property is a
subtle one, both from the mathematical and physical viewpoints.

The  possible alternation of the spatial  properties  of the low-lying
excitations with the  parity of the number of  rows of the sample
(as observed in the QHCD model on the triangular lattice) is a
first   difficulty.  The degeneracy  of  the  RVBSL  is in fact  quite
different from that  appearing in  a VBC.   We do  not expect the  VBC
ground-state degeneracy  to depend on the  genus of the sample, as the
RVBSL does.

From the  physical point of  view  also the  two situations  are quite
different. 
An infinitesimal  symmetry breaking perturbation
is able to select one  symmetry breaking ground-state  of the VBC, but
as we will show below this is impossible in the RVBSL.

Let us call ${\cal A}$ the extensive non-diagonal observable
appearing in the VBC in the thermodynamic limit. On a columnar
VBC modulated in the {\bf u} direction, this observable is: 
\begin{equation}
	{\cal A}=
	\sum_{j=1}^{N}  e^{{i{\bf K}_1} \cdot {\bf r}_j}
	P_{S=0}( {\bf r}_j, {\bf r}_j + {\bf u})   
\end{equation}
 where $ P_{S=0}( {\bf r}_j, {\bf r}_j + {\bf u})$ is the
projector on the singlet state of two neighboring spins.
${\cal A}$ connects eigen-states with wave-vector ${\bf k}_0$ 
to states with wave-vector
${\bf k}_0 + {\bf K}_1$ . 

On a  finite size  sample,   with periodic  boundary conditions,   the
expectation value of ${\cal A}$ is zero in any eigenstate,
but  $<{\cal A}^2>$   could be non zero.  If
the order parameter ${\cal P}$ defined by:
\begin{equation}
{\cal P}^2= <\psi_{g.s.}|{\cal A}^{\dagger}{\cal A}|\psi_{g.s.}> /N^2
\end{equation}
 does  not  vanish  in the thermodynamic
limit, the system has columnar dimer long range order with wave vector
${{\bf K}_1}$.

Let us now consider a perturbation of the Hamiltonian:
\begin{equation}
H_{\delta} = H_0 - \left(\delta {\cal A} +
h. c.\right).
\end{equation}
At T=0, the intensive linear response on the observable ${\cal A}$ is
measured by the susceptibility:
\begin{equation}
\chi = \frac{2}{N} <\psi_{g.s.}|{\cal A}^{\dagger}
\frac {1 }{  H_0 - E_{g.s} }{\cal A} |\psi_{g.s.}>
\label{suscep}
\end{equation}

This  susceptibility is bounded from below\cite{ssgpt99}:
\begin{equation}
\frac {4 {\cal P}^4 N^2}{f}< \chi 
\label{ineq}
\end{equation}
where $f$ is the oscillator strength:
\begin{equation}
f = \frac{1}{N} \left<\psi_{g.s.}| \left[{\cal A},\left[H_0 ,{\cal A}\right]\right]|\psi_{g.s.}\right>
\end{equation}
The demonstration uses the properties of the spectral
decomposition associated to the operator ${\cal A}$:
\begin{equation}
S(\omega) = \frac {1 }{N}  \sum_{n \neq 0}\left|<\psi_{g.s.}|{\cal A}|n> \right|^2 \delta (\omega -\omega_n)
\end{equation}
where $\omega_n =E_n -E_{g.s}$
\begin{equation}
{\cal P}^2= \frac {1 }{N} \int S(\omega) d\omega
\end{equation}
Using the Cauchy Schwartz inequality one obtains:
\begin{equation}
{\cal P}^4 \leq \frac {1 }{N^2} \int \omega S(\omega)d\omega \int \omega^{-1} S(\omega)d\omega
\end{equation}
where  
\begin{equation}
\int \omega S(\omega)d\omega = f/2
\end{equation}
\begin{equation}
\int \omega^{-1} S(\omega)d\omega = \chi /2
\end{equation}
which proves inequality~(\ref{ineq}).
For a short range Hamiltonian the oscillator strength $f$ is
$ {\cal O}(1)$ and  inequality (\ref{ineq}) implies that  
the T=0 susceptibility associated to a finite order parameter
diverges at least as the square of the sample
size: any infinitesimal  symmetry
breaking perturbation will select a symmetry
breaking state.

We will now show that for a RVBSL, where all the correlations
functions are short-ranged with  correlation lengths bounded by $\xi$,
the susceptibilities of the medium remain finite in the
thermodynamic limit. To do so we distinguish in Eq.~\ref{suscep}
the contributions from the quasi-degenerate states of the
topological multiplet (called $|\alpha_i>$)
 from the contribution of the other states of the spectrum,
above the physical gap $\Delta$. We thus obtain the following upper
bound for the susceptibility:
\begin{eqnarray}
\chi\; &=& \chi_{\rm top}\; + \chi_{\Delta} \\
\chi_{\rm top}\; &=&\frac{2}{N} <\psi_{g.s.}|{\cal A}^{\dagger}\frac{ 
|\alpha_1><\alpha_1| }{E_{\alpha_1} - E_{g.s.}}{\cal A} |\psi_{g.s.}>
\label{chitop}
\\
\chi_{\Delta} &\leq& \frac{2}{N \Delta } \left[ <\psi_{g.s.}|{\cal A}^{\dagger}{\cal A} \right.
|\psi_{g.s.}> \nonumber \\
 &-&<\psi_{g.s.}|{\cal A}^{\dagger}|\alpha_1><\alpha_1|{\cal
A}|\psi_{g.s.}>]
\label{chidelta}
\end{eqnarray}
where   $|\alpha_1>$ stands for   the    state(s) of the   topological
multiplet connected to the absolute ground-state by ${\cal A}$.  Using
the local properties  of ${\cal A}$, ${\cal  A}^{\dagger}|\alpha_1>$ is
in the same topological sector as $|\alpha_1>$ and $<\psi_{g.s.}|{\cal
A}^{\dagger}  |\alpha_1>$ is at  most of ${\cal O}(N \times 2^{-L/2})$
(see  paragraph~\ref{deftopsec}).   As  $E_{\alpha_1}  - E_{g.s.}$  is
supposed to   decrease  as  $exp(-L/\xi)$  (see  Fig.~\ref{topdeg})
$\chi_{\rm top}$  goes to a constant when the
size of  the   sample  goes to   infinity,  provided $\xi$   is  small
enough~\footnote{Strictly  speaking $\xi^{-1}$ should   be $\leq\log 2$ but
Ineq.~\ref{ineqB}    is a dramatic
overestimate of the  scalar product in the case  of an RVBSL~\cite{mlms02}. The reason
is    that   if  two   dimer     coverings  $c^+$   and $c^-$ maximize
$\left<c^+|c^-\right>$  they only differ along  a  single large ($\sim
L$) loop. They have different  local  correlations along the loop  and
their energy difference is of order $L$  and it is very unlikely that
their weights in the states $\psi^+(c^+)$ and $\psi^-(c^-)$ are
{\em both} of order one.}.
In a  system with exponentially  decreasing correlations, ${\cal P}^2$
decreases as $1/N$ and $\chi_{\Delta}$ is trivially constant at the
thermodynamic limit.

In such a phase  an  infinitesimal field  cannot    induce a symmetry
breaking and   there  could not  be any {\it spontaneous}
symmetry breaking.

\end{subsection}
\end{section}
\begin{section}{Other approaches of type I RVB Spin Liquids}

As already said, Sachdev in  a large $N$, $Sp(N)$ analysis of the
Heisenberg model on the triangular lattice equally found in the
extreme quantum case an RVB ground-state and
deconfined spinons~\cite{s92}.
 
An interesting mapping of the dimer models on Ising model in
transverse field has been studied by many authors~(See 
refs.~\cite{ms01b,msf02} and references therein). This approach
is one of the ways to give evidence of the parentage between this
RVB spin liquid with its fractionalized excitations and the
deconfining Ising Gauge theories~\cite{k79,lfs01,sf01b,msf02,bfg02}
\end{section}

\begin{section} {Summary of the properties of   type I RVB Spin liquids}

To conclude let us summarize the properties of a type I RVB Spin liquid:
\begin{itemize}
\item It is a phase which does not break either SU(2) symmetry
nor any spatial symmetries of the lattice. Its ground-state is
{\bf unique} up to a topological degeneracy which exists only in
systems with an odd number of spin-1/2 in the unit cell, 
living on a 2-torus (more generally the degeneracy is $2^g$,
 with $g$ the genus of the torus). In that sense it 
is awkward to call such a phase a disordered phase! None of the
classical ideas associated to  disorder are relevant to understand the
properties of this RVB Spin Liquid phase. If we have to
compare it to a liquid phase it is more the superfluid phase of $^4He$
that we should have in mind! 
\item All correlation functions in local
observables have only short range order, and consequently
 the susceptibility associated to any local observable is
zero a T=0.
\item This phase has a gap for all excitations, either in the
singlet or the triplet sectors and it supports fractionalized
excitations (the ``spinons'').
 The first excitations in the singlet sector
correspond in the gauge theory language to the bosons of the
gauge field that Senthil and M. P. A. Fisher call "visons"~\cite{lfs01,sf01b}. Due to the properties of the excitations we expect them 
to form continua in the spin sectors. The neutron experiment of
Coldea and co-workers on  Cs$_2$CuCl$_4 $ is perhaps the first
 experimental proof of
such a state~\cite{cttt01}.
\item An RVB spin liquid state is
expected in presence of competing and frustrating interactions.
The bandwidth in which this phenomenon can be observed is
strongly reduced with respect to the original couplings and often
only a small fraction of the original couplings: the case of the
MSE system at $J1= -2$ and $J_4 =1$ (the point where all the results
 given here have been calculated) is in some sense rather exceptional,
 as at this point  the spin gap
is  almost as large as the cyclic 4-spin exchange.
\end{itemize}
\end{section}

\end{chapter}
\begin{chapter}{Resonating Valence Bond Spin Liquid (Type II)}

Type I RVBSL have a gap in the singlet sector. In the Ising gauge
theory approach this gap is essential for the consistency of
the theory~\cite{sf01b}. The Ising gauge field  quasi particles called
 visons by Senthil and co-workers are vortices of the gauge field, they carry a $Z_2$ gauge flux, no spin and
 have long range interaction with spinons.
If the spectrum of these particles
 has a gap then the spinons are
unconfined and the phase is ``fractionalized''. If they
condense, the long range interaction between them and the spinons
frustrate the  motions of the latter which remained confined. The
gap in the singlet sector (above the topological degeneracy)
 is a crucial ingredient of Type I RVBSL: as we have learned in
the previous chapter it is a property of the
MSE Hamiltonian near the ferromagnetic phase.

As displayed in
Fig.~\ref{MSEphs}, the MSE Hamiltonian near the
antiferromagnetic three sublattice N{\'e}el order displays a second
Spin Liquid phase, with no long range order in any observable,
but no gap in the singlet sector~\cite{lmsl00}. Such a behavior
is not a standard Spin Liquid behavior. Could it be
a quantum critical behavior similar to that
 of the Rokhsar Kivelson model at $ V=J$?
Keeping in mind the restrictions inherent on small size
approach of such phenomena, we nevertheless think that the
observed phenomenon
is distinct from the critical RK behavior on the square lattice
 (see below).
In view of the similarities between the phase seen in the MSE
model and the phase observed in the Heisenberg model on the
kagome lattice, I took the step to describe the physics that can
be observed at that point as something different: a
RVB Spin liquid of type II.

 In fact as most of the  studies done in many 
 groups have been devoted to the Heisenberg model
 on the kagome lattice (noted HK model in the following), it is on 
 this last  model that we  will  center  this last chapter.

\begin{section}{Miscellaneous models on the kagome lattice}
There
has been a large number of studies devoted to different
antiferromagnetic models on
the kagome lattice, which is a 2-dimensional lattice of corner
sharing triangles~(Fig.~\ref{kagome}).

\begin{figure}
\hspace{2cm}\resizebox{7cm}{!}{\includegraphics*{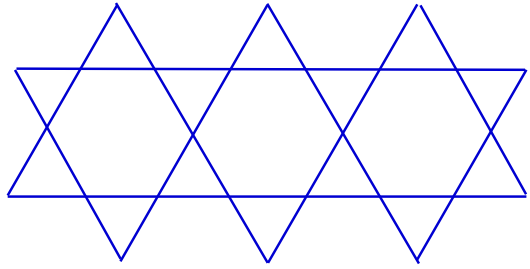}}
    \caption[99]{ The kagome lattice: a lattice of corner
sharing triangles. In Japanese kagome is a
common word which designs both a basket and the special canwork
for baskets. This is equally a trademark for a popular tomato sauce. 
 }
    \label{kagome}
\end{figure}

The next neighbor
 Ising model on such a lattice is disordered, its entropy is very
large $S_{kag}^{Ising}= 0.502$, more than half the independent
spin value, much larger than the triangular lattice value
 $S_{tri}^{Ising}= 0.323$  and of the order of Pauling
approximation for independent triangles
$S_{Pauling}=0.501$~\cite{p38}. This
suggests that the correlations in this system are very weak: the
model remains disordered at all temperatures~\cite{kn53,hr92}. 

Moessner and Sondhi have studied this Ising model in a transverse
magnetic field (the simplest way to include in the model some
quantum fluctuations): the model fails to order for any
transverse field, at any temperature~\cite{msc00,ms01b}.

The n.n. classical Heisenberg model on the kagome lattice has also an
extensive T=0 entropy: this is a property easily understood,
  shared with the same model on different lattices with
corner sharing units as the checker-board lattice or the true
three dimensional pyrochlore lattice. On all these lattices the
n.n. Heisenberg Hamiltonian can be rewritten as the sum of the
square of the total spin $S_{\alpha}$ of individual units $\alpha$ (a
tetrahedron in the 2-d and 3-d pyrochlore cases and a triangle for
the kagome lattice), which share only one vertex. Classical
ground-states are obtained whenever
 $\forall \alpha\; {\bf S}_{\alpha}=0$.
This condition fixes the relative position of the three classical
spins of a triangle at $120$ degrees from each other in a plane.
But it does not fix the relative orientation of the plane of a
triad with respect to the planes of triads on corner sharing
triangles: the model has thus an infinite local degeneracy
and an extensive entropy~\cite{chs92,hr92}. This state too is reluctant
to order: nevertheless  thermal fluctuations select coplanar
configurations~\cite{chs92,hr92,rcc93}, without long range order in
the plane. The order parameter of such a phase is the direction of
the local helicity (sometimes called vectorial chirality) and defined
by its components as:
\begin{equation}
 {\zeta}^{\gamma} = \sum_{on\; a \;triangle} \epsilon^{\alpha\beta\gamma} S^{\alpha}_iS^{\beta}_j
\end{equation}
where $\epsilon^{\alpha\beta\gamma}$ is the antisymmetric
tensor. This kind of order is by analogy to liquid crystals
sometimes called a nematic order. The existence of such an order
parameter is  probably not without relation with
 the instability of the classical
Heisenberg model on the kagome lattice to Dzyaloshinsky-Moriya
interactions\cite{e02,ecl02}.
\end{section}

\begin{section}{The next-neighbor spin-1/2 Heisenberg model on the kagome lattice: an extreme play-ground for ``quantum fluctuations''}

The n.n. spin-1/2 quantum Heisenberg model on the kagome lattice has
equally been the object of many
studies~\cite{e89,ce92,s92,le93,ze95,lblps97,web98,m98,smlbpwe00,mm01}.
From these studies one can remember the following facts:

\begin{subsection}{Ground-state energy per spin}
The Heisenberg model on the kagome lattice has an extremely low
energy per bond ($<{\bf S}_i.{\bf S}_j>_{per\, bond}= -.437)$
$ \sim 87\%$
 of the energy per bond of independent
triangles. On this lattice the energy per bond of the spin-1/2
system is much lower than the classical energy $\frac
{E_{qu.}}{E_{cl.}} \sim 1.74$, a ratio much larger than in any
 other 2-dimensional magnet,  that can only be compared to the value
 obtained for the Bethe chain (1.77) 
(see Table.~\ref{ener-_param}).  The kagome lattice
is the 2-dimensional lattice which offers the larger stabilization
 due to
quantum fluctuations. 
\end{subsection}

\begin{subsection}{Correlations}

Spin-spin correlations~\cite{le93}, dimer-dimer correlations
(Fig.~\ref{dim-dimkagMSE}), chirality-chirality correlations~\cite{ce92}
are short range, which is consistent with  the previous point.
\begin{figure}[h!]
\vspace*{-6cm}\hspace*{3.5cm}
\resizebox{12cm}{!}{\includegraphics{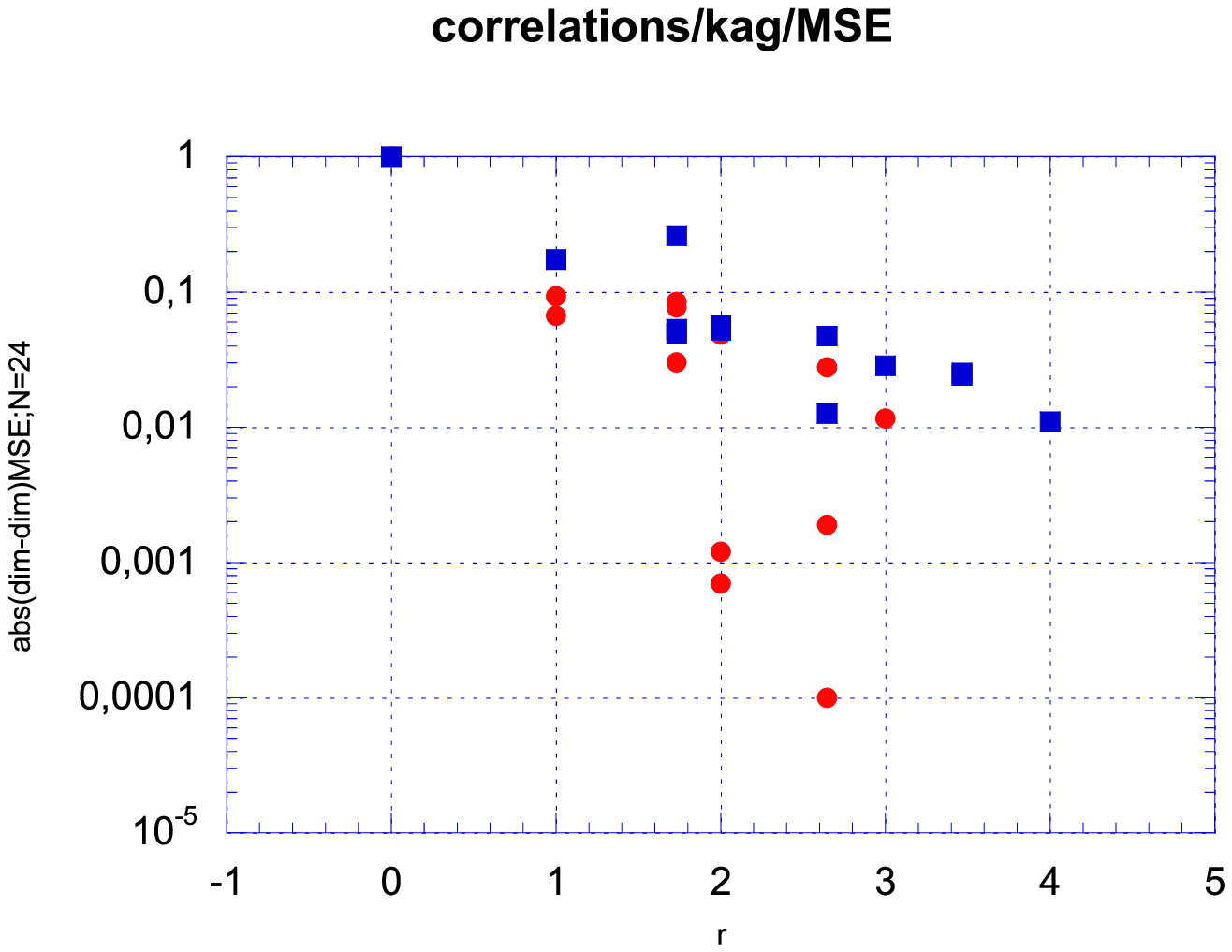}}
\vspace*{3.5cm}
\caption[99]{Dimer-dimer correlations in the ground-state of the
kagome Heisenberg model (blue squares) and in the MSE model 
(red bullets) versus distance. Although the decrease of these
correlations is weaker in the KH model than in the MSE model, it
is nevertheless roughly exponential in these first two decades, as
the spin-spin correlations are~\cite{ce92,le93}}
\label{dim-dimkagMSE}
\end{figure}
\end{subsection}

\begin{subsection}{Spin-gap and absence of gap in the singlet
subspace}
There is probably a spin-gap of the order of 1/20th of the
coupling constant~\cite{web98}. In view of the smallness of this
spin-gap with regards to the available sizes caution is
necessary. The above conclusion was drawn from the raw data of
exact spectra of samples with up to 36 spins (see Fig.~\ref{spinon_kag}): more precisely from the
measurement of $E_0(S=S_{min} +1) -E_0(S=S_{min})$, where
$E_0(S)$ is the lowest energy in the $S$ sector, and $S_{min}=$ 0
or 1/2 depending on the parity of the number of spins of the
sample. The size effects on these results are an order of
magnitude smaller than in an ordered N{\'e}el antiferromagnet.
Nevertheless they are still not negligible for these  sizes. We
have thus tried an indirect measurement of the spin-gap along the
following line.

The lowest exact eigenstates in each $S$ sector of a sample
of N sites define the energy per spin of the sample at $T=0$ as a
function of its magnetization $m = S/(N/2)$. For low value of the
magnetization one can fit this energy per spin to an analytic law
of the form~\footnote{This analytic form cannot extend beyond
$m=1/3$, where an angular point appears with a discontinuity of the
first derivative signaling a magnetization
plateau~\cite{lm02,h01}}:
\begin{equation}
e(m) = e(0) + a m + b m^2/2 +{\cal O}(m^3)
\label{eofm}
\end{equation}
The a and b coefficients depend on N. Their physical
significance is clear: a measures half the spin gap and
\begin{equation}
b=\frac{\partial^2 e}{\partial m^2} = \chi^{-1}
\end{equation}
where $\chi$ is the homogeneous susceptibility of the medium for
fields larger than the critical field $H_c$ ($H_c$ in convenient units
is equal to the spin gap).

This indirect determination of the spin gap leads to a strong
renormalization of the data far small sizes (explaining the large
error bars for small sizes data in Fig.~\ref{gapkago}).
 On the other hand the $N$
dependence of the renormalized data is now much weaker as can be
seen by a comparison of the raw
 results of~\cite{web98} and Fig.~\ref{spinon_kag}, with those of
Fig.~\ref{gapkago}. A linear extrapolation versus $1/N$
(which should give a lower bound of the spin-gap) leads to the
value $0.06$ for the spin gap (consistent extrapolations give 
$e_{\infty}=-0.4365$ and $\chi_{\infty}=0.34$). All these
determinations are in agreement with the results obtained from the
raw data. Nevertheless is should be underlined that even for N=36
at the smallest non zero magnetization the linear term of
equation~(\ref{eofm}), is only $90\%$ of the quadratic term:
this shows the limit of confidence in our assumption on the
existence of a spin-gap~\footnote{This spin gap result is in my point
 of view the less reliable result among all those described in these
lectures. In view of the  spin-spin
correlations which seems exponentially decreasing and not
critical~\cite{le93}, and of the fact that results between 27 and 36 may
signal a cross-over behavior in the finite size effects, we will
not question this point further for the moment. It would
nevertheless be
useful to develop a very precise analysis of finite size effects
between 27 and 36 (as we have done for the MSE model see Fig.~\ref{Egap})
to try to confirm this conclusion.}.
 
\begin{figure}
\vspace*{-10cm}
\hspace{3.5cm}
\resizebox{!}{10cm}{
\includegraphics{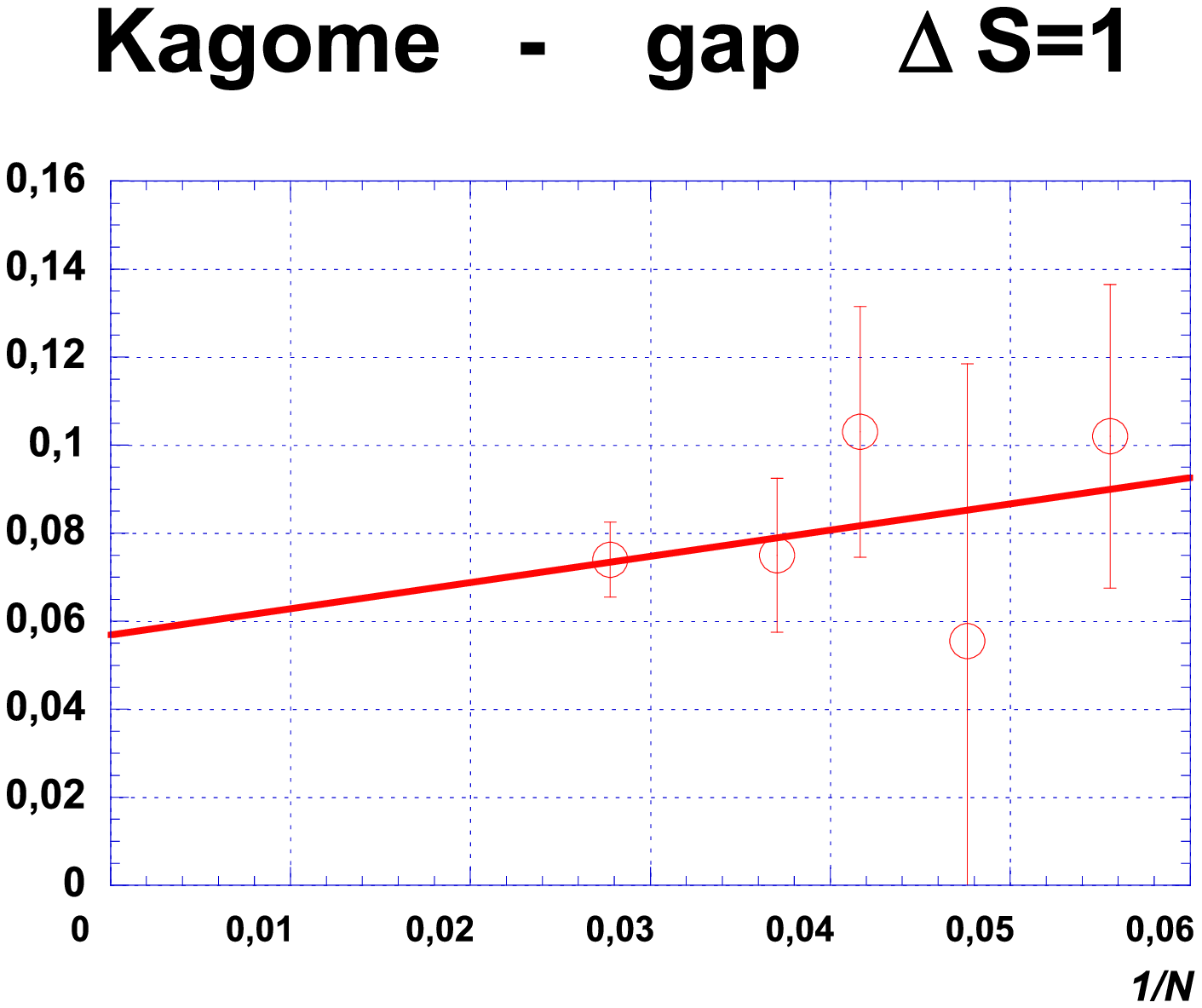}
}
\vspace*{2.5cm}
\caption[99]{Finite size scaling of the spin gap in the spin-1/2
Heisenberg model ${\cal H} = \sum_{<i,j>} {\bf S}_i.{\bf S}_j$ 
on the kagome lattice. Data are obtained by the indirect
procedure described in the text.}
\label{gapkago}
\end{figure}
\end{subsection}

\begin{subsection}{An exceptional density of low lying excitations
 in the singlet sector}

Whatever the ultimate fate of the spin gap a still larger
surprise emerges from the exact spectra:  the absence of
gap in the singlet sector and the anomalous density of low energy
states adjacent to the ground-state.
We have measured the number of singlet levels in the spin-gap
(taken as a natural energy band-width of the problem): this
number increases exponentially fast with $N$ as $1.15^N$
(see Fig.~\ref{numberostate}).

\begin{figure}
\vspace*{-3cm}\hspace*{1.3cm}
\resizebox{10cm}{!}{\includegraphics{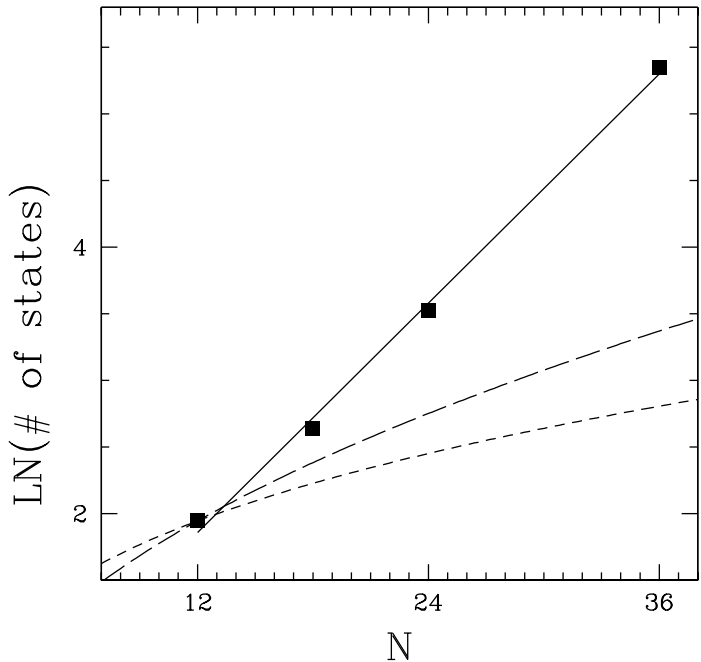}}
\caption[99]{Logarithm of the number of singlet states in the spin
gap versus sample size (black squares). The short dashed and long
dashed curves display the theoretical law (Eq.~\ref{law}) (short dashes: $p=1$, long dashes: $p=2$).
}
\label{numberostate}
\end{figure}

The first immediate consequence of this property is the existence
of a $T=0$ residual entropy in this model. This came as a shock
for many scientists who had the idea that the quantum dynamics
``should'' lift the degeneracy of the incipient Ising model or
dimer models (see next section). This does not seem to be the
case!!

Some remarks are necessary to fully appreciate this property.

The total number of states of a sample of $N$ spins 1/2 is  $2^N$.
These states are stretched on an energy scale of the order of
$NJ$ where $J$ is the coupling constant of the Hamiltonian. This
implies that on most of the spectrum the density of states
increases exponentially with $N$. If we specialize to the $S=0$
sector as we will do below, the picture is not very different:
the number of states is
 $ C^N_{\frac{N}{2}}-C^N_{\frac{N}{2}-1} \sim {\cal O}(\frac{2^N}{N})$
 and here too, in most of the spectrum the density is exponentially
increasing with N.

BUT in all  the phases that we have studied up to now, the nature
of the ground-state and of the low lying excitations leads to a
different behavior at the edges of the band. Typically the
ground-state degeneracy is ${\cal O}(1)$ in VBC and in type I
RVBSL and it is ${\cal O}(N^{\alpha})$ in N{\'e}el ordered states
with $\alpha$ sublattices. In all these situations the low lying
excitations can be described as modes or quasi-particles. In the
corresponding energy range one typically counts $N^{\beta}$
levels associated with one quasi particle excitations.
 This always leads to
density of states increasing as a power law as a function of $N$.
Inclusion of multi-particle excitations can be done in an average
way:
let us suppose that single particle excitations have a dispersion
law:
\begin{equation}
\omega({\bf k}) = k^p.
\label{dispersionlaw}
\end{equation}
In a d-dimensional space the internal energy of such a system
increases with temperature  as $T^{(p+d)/p}$, the specific heat as
$T^{d/p}$ as the entropy ${\cal S}$. For a system with $N$
spins the average energy range $W$ excited at the temperature $T$
is of the form: $W= C^{st} N T^{(p+d)/p}$. This implies that the entropy
depends on $N$ as: 
${\cal S} \propto N \left({\frac{W}{N}}\right)^{\frac{d}{p+d}}$
and thus the average number of excited levels ${\cal N}$
increases as:
\begin{equation}
ln ({\cal N}) \propto N^{\frac{p}{p+d}}
\label{law}
\end{equation}
\footnote{I thank S. Kivelson for giving me this idea for
the computation of  the multi-particle  density of
states. The errors, if any, are mine.}.

As an example, let us consider the  quantum
critical phase of the Rokhsar Kivelson model on the
square lattice at the quantum critical point (Chapter 4 page 60):
 in that case the critical correlations decrease
as $r^{-2}$, the dispersion law is linear in $k$, the logarithm of
 the number of states
increases as $N^{1/3}$ (short dashes of Fig.~\ref{numberostate}),
much slower than the exact results. 

Even in including many particle excitations one would thus expect
a number of levels increasing much more slowly than in the exact
results! Except if we accept that there is infinitely soft low
energy modes ($p \rightarrow \infty$), then we  recover the
correct density of low lying levels and a residual entropy.
It is still unclear  if we can do a connection between
the ``zero mode'' of the classical model at T=0 and this picture.
 And we cannot completely indulge ourselves in saying that quantum 
fluctuations are unable to lift the classical degeneracy
  as quantum fluctuations seems to open a spin gap!

\begin{figure}[b!]
\vspace*{-7cm}
\hspace*{3cm}
\resizebox{12cm}{!}{\includegraphics{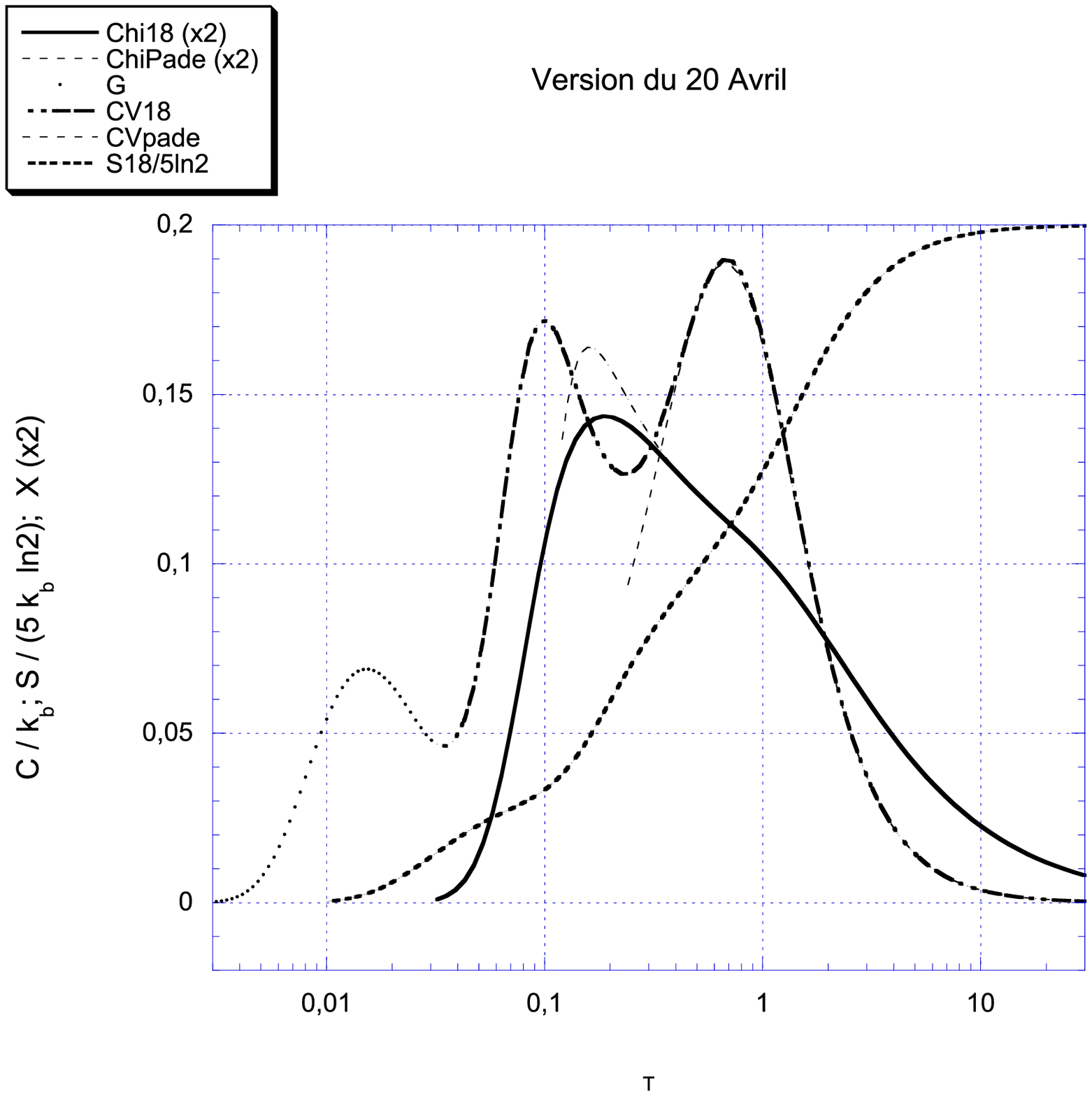}}
\vspace*{0.5cm}
\caption[99]{
Specific heat (dot-dashed curve), entropy (dotted curve)
 and spin-susceptibility (full line) of the Heisenberg model on
the kagome lattice (exact diagonalization on a N=18
sample, and Padd{\'e} approximants to high temperature
series~\cite{ey94}). The third bump in the specific heat at very
low temperature is an artifact of the small size and the
associated discretization in the singlet sector. We think that
the intermediate bump is a real feature of the specific heat
which subsists up to the thermodynamic limit~\cite{smlbpwe00}.
 Note the entropy of
the singlets in a range of temperature below the spin gap where
the spin excitations are negligible as the spin-susceptibility. }
\label{susc_cv_kago}
\end{figure}
A physical consequence of this exceptional density of low lying
singlets can be observed in the specific heat (Fig.~\ref{susc_cv_kago}): at low temperature
the specific heat of this spin system is unusually insensitive to
large magnetic fields. This is easily understood if we suppose
that in this energy range the excitations are essentially
singlets~\cite{smlbpwe00}. This result is to be compared to the
experimental results of Ramirez {\it et al.}~\cite{rhw00} on
$SrCrGaO$ where the specific heat around $5 K$ has an extremely
low sensitivity to magnetic fields up to $10\; Tesla$, whereas the
homogeneous susceptibility in this range of temperature is
probably very low if we notice that it turns down around $50
K$~\cite{mklmch00}.

\end{subsection}
\begin{subsection}{Anomalous density of states in other spin
sectors}
This anomalous density of low lying states has equally been
observed in the spin 1/2 sector (where the law could be fitted to
$N 1.15^N$), in the spin 1 sector as well as in other sectors with
larger total spin. It should be noticed that such a density of
states implies the absence of an intrinsic energy scale for the
low lying excitations: a phenomenon that has been observed in
inelastic neutron scattering~(ref.~\cite{mmpfa00} and refs.
therein) and theoretically in the imaginary part of the dynamic
susceptibility calculated within the dynamical mean field
theory~\cite{gsf01}. A very high spin susceptibility just above
the spin gap is not excluded in spin-1/2 compounds~\cite{ls02}.
\end{subsection}
\begin{subsection}{Unconfined Spinons}
\begin{figure}[h!]
\vspace*{-7cm}\hspace*{3cm}
\resizebox{13cm}{!}{\includegraphics{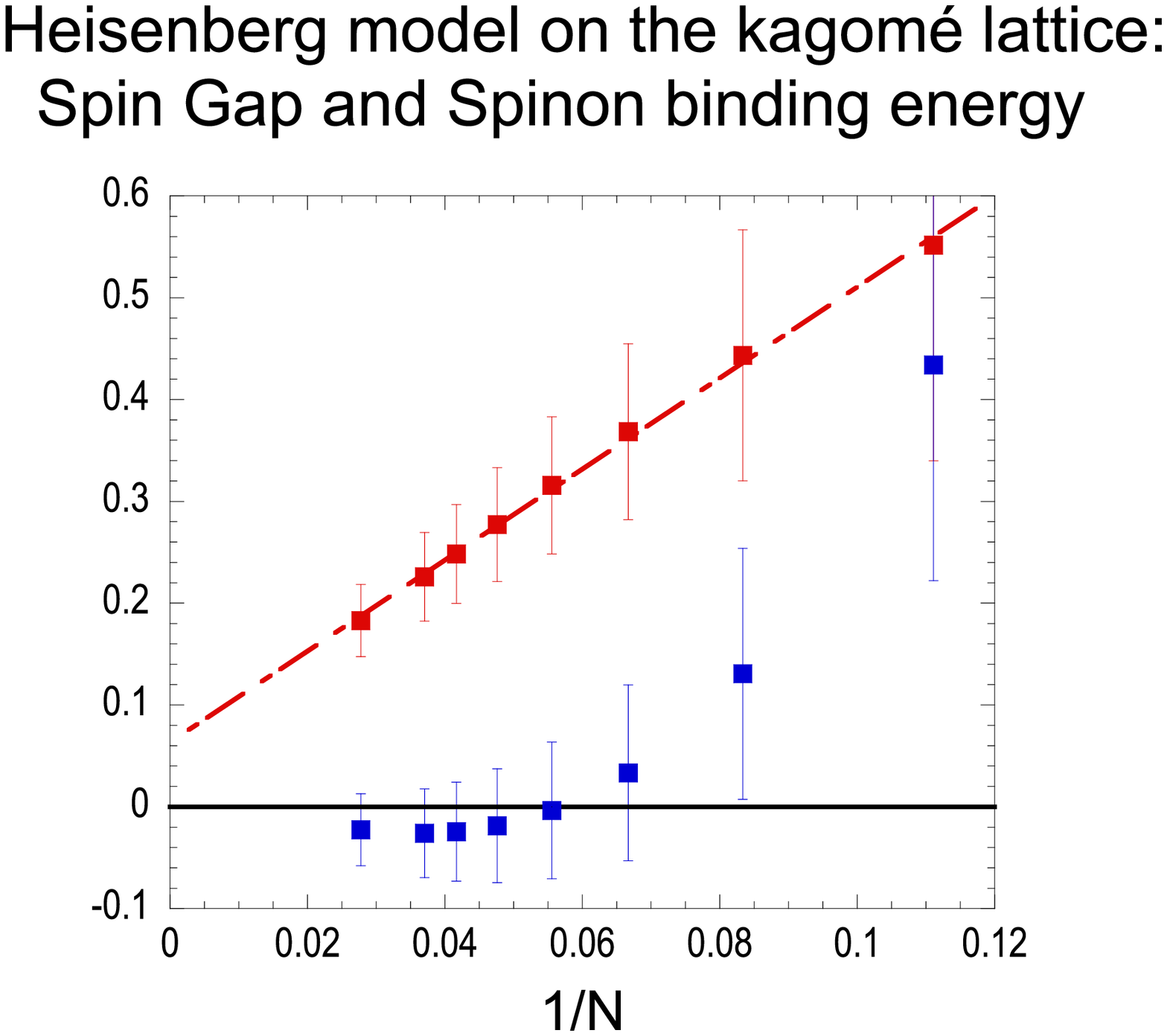}}
\vspace*{1.5cm}
\caption[99]{ Finite size scaling on the spin gap: red
squares (raw data) and
the spinon binding energy: blue squares (unpublished results).}
\label{spinon_kag}
\end{figure} 
The spinon binding energy (Fig.~\ref{spinon_kag})
 has been computed in the HK model along
the same line as it has been done in Chapter 5 page 75 for the
MSE model. The only difference lays in smaller and smoother shape
effects in the HK case, which allows to do analytic fit of the
energy per spin $e(S_{min}, N)$ versus $1/N^{3/2}$ for even and
odd samples separately. Eq.~\ref{eq.b}, is then written, with
reference to the interpolated value of   the energy of even
samples for odd value of N: this takes into account the finite
size effects with more subtlety than the $(2N+1)e_{\infty}$ term
of equation~\ref{eq.b}. All these features explain the smoother
behavior of the spinon binding energy shown in
Fig.~\ref{spinon_kag}.
From this data one can rather safely conclude that spinons are
probably unconfined in the HK model.

The global picture of this phase is thus that of a Spin Liquid,
with no long range correlations in local observable,
unconfined spinons and a residual entropy of singlets at $T=0$, 
which is one
of the manifestations of an extraordinary large density of
states in each $S$ subspace.
\end{subsection}
\end{section}
\begin{section}{Next-neighbor Resonating Valence Bond
description of the spin-1/2 kagome antiferromagnet}
Considering a supposed-to-be large spin-gap, Zeng and
Elser~\cite{ze95} proposed a description of the ground-state and
low lying excitations of the kagome model in the basis of next
neighbor Valence Bonds. We know now that the spin gap is
certainly smaller than it was expected in 1995, nevertheless Mila
and Mambrini~\cite{m98,mm01} have convincingly shown that the
picture of a next neighbor resonating Valence Bond Spin liquid
captures some of the most perplexing features of this magnet and
specifically the absence of gap in the singlet sector and the
exponential number of singlets in {\it any given range of energy}.
This probably implies the absence of an intrinsic low energy
scale, which is consistent with the thermal behavior of the dynamic
spin susceptibility calculated by Georges and coll.~\cite{gsf01}.
This feature is typical of a critical state, but as seen in the
above discussion, the simple RK picture does not seem to fit
nicely to the exact diagonalization data: may be the available
sizes are too small or the behavior of this quantum system
corresponds to something definitely new.

More generally this picture of the ground-state and first
excitations as resonances between an exponential number of dimer
coverings gives a qualitatively interesting picture of the low
temperature physics of different oxides that can be described as
kagome antiferromagnets. The low temperature specific heat of
SrCr$_9$Ga$_{12}$0$_{19}$ is apparently dominated by local singlet
states~\cite{rhw00}. The magnetic excitations of this same
compound as seen by muons can be described as
spins 1/2 itinerant in a sea of singlets~\cite{ukkll94}. 
SrCr$_9$Ga$_{12}$0$_{19}$ exhibits at about 5 K a very large increase of
its non linear susceptibility reminiscent of spin
glasses~\cite{rec90}, but
neutrons and muons show that most of the spins are not frozen
below this temperature and exhibit still very rapid
fluctuations\cite{lbar96}. The same phenomena have been observed
in two jarosites that are equally good models of kagome antiferromagnets
with half-odd-integer spin per unit cell~\cite{kklllwutdg96,whmmt98}.
\end{section}
\begin{section}{ Haldane's conjecture}
Whereas the classical Heisenberg model on the kagome,
checker-board and pyrochlore lattices share the property of an
extensive entropy and disorder at $T=0$, their quantum
counterparts are quite different. As it has been explained in
Chapter 4, the Heisenberg model on the checker-board lattice has
an ordered Valence Bond Crystal with gaps to all excitations.

Less is known on the ground-state 
 of the Heisenberg model on the 3-dimensional pyrochlore lattice:
Canals and Lacroix~\cite{cl98} have shown that their spin-spin correlations
are short ranged and they have seen on a 16 sites spectrum 
 that the first excitations were singlet ones. Having done
the spectrum of a 32 sites pyrochlore sample, we confirm that the
first excitations are still singlets for this size. But the finite size
effects on these excitations between 16 and 32 are very large and
it remains possible that the gap in the singlet sector be larger
than the singlet-triplet gap in the thermodynamic
 limit~\cite{fmsl03,f03}: in any case there is definitely a gap
 in the singlet sector in this last model!\footnote{Tsutenegu has
recently developed an effective description of the singlet
sector, where he has a soft mode in the singlet sector. I am a
bit skeptical on his approach which strongly and artificially breaks the
symmetries at the mean-field level~\cite{t01} and then treats in a
semi-classical approach the fluctuations~\cite{t02}. The Core
approach of E. Berg and collaborators~\cite{baa03} seems more appropriate to
deal with these systems where the quantum dimerization is
probably the dominant phenomenon}

On the other hand, recent results from Hida~\cite{h00} show that there is a gap to all
excitations in the S=1 HK model.

All these results seem to confirm Haldane's conjecture: 
 among these strongly
frustrated systems with an extensive degeneracy in the classical
limit  the spin-1/2 kagome antiferromagnet is the only system to have
an half-odd-integer spin in the unit cell and gapless
excitations. The spin-1/2 Heisenberg model on the checkerboard lattice
or
on the pyrochlore lattice and the spin-1 Heisenberg model on the
kagome lattice have  integer spins in the unit cell and quantum
fluctuations lead to gapful
excitations.

\end{section}
\end{chapter}
\newpage
{\bf Acknowledgements} Ce texte repr\'esente des notes  accompagnant un
 cours 
donn\'e \`a l'automne 2002 \`a Lausanne dans le cadre des cours de troisi\`eme cycle de l'\'Ecole de Physique Romande sur invitation du Pr. F. Mila.
 Une partie notable des r\'esultats provient du travail de recherche fait \`a Paris
par mes coll\`egues, \'etudiants en th\`ese et post-doc: B. Bernu, J.-C. Domenge,
 J. B. Fouet, P. Lecheminant, G. Misguich, L. Pierre,  et d'une
 fructueuse collaboration avec C. Waldtmann et le Pr. H.-U. Everts. 
Que tous soient remerci\'es. 
 Un article de revue r\'edig\'e par G. Misguich et C. L. 
donnant un point de vue moins auto-centr\'e et plus r\'ecent est disponible
sur le serveur d'archives: cond-mat/0310405

\begin{thebibliography}{100}

\bibitem{m81}
D.~C. Mattis, {\em The Theory of Magnetism I}, Vol.~17 of {\em Springer Series
  in Solid-State Sciences} (Springer-Verlag, Berlin, Heidelberg, New York,
  Tokyo, 1981).

\bibitem{c89a}
W.~J. Caspers, {\em Spin Systems} (World Scientific, Singapore, 1989).

\bibitem{auer94}
A. Auerbach, {\em Interacting electrons and Quantum Magnetism}
  (Springer-Verlag, Berlin Heidelberg New York, 1994).

\bibitem{h38}
Hulthen, Ark. Mat. Astron. Fys. {\bf 26A},  1  (1938).

\bibitem{m55}
W. Marshall, Proc. Roy. soc. London {\bf A 232},  48  (1955).

\bibitem{lm62}
E.~H. Lieb and D. Mattis, J. of Math. Phys. {\bf 3},  749  (1962).

\bibitem{a52}
P. Anderson, Phys. Rev. {\bf 86},  694  (1952).

\bibitem{a63}
P. Anderson, {\em Concept in Solids} (Benjamin, New York, 1963).

\bibitem{a84}
P. Anderson, {\em Basic Notions of Condensed Matter Physics} (Benjamin, New
  York, 1984).

\bibitem{blp92}
B. Bernu, C. Lhuillier, and L. Pierre, Phys. Rev. Lett. {\bf 69},  2590
  (1992).

\bibitem{bllp94}
B. Bernu, P. Lecheminant, C. Lhuillier, and L. Pierre, Phys. Rev. B {\bf 50},
  10048  (1994).

\bibitem{le95}
P. Lecheminant, Ph.D. thesis, Universit\'e Pierre et Marie Curie, Paris, 95.

\bibitem{k52}
R. Kubo, Phys. Rev {\bf 87},  568  (1952).

\bibitem{kls88}
T. Kennedy, E. Lieb, and B. Shastry, J. Stat. Phys. {\bf 53},  1019  (1988).

\bibitem{dls78}
F. Dyson, E. Lieb, and B. Simon, J. Stat. Phys. {\bf 18},  335  (1978).

\bibitem{jnfp86}
E.~J. Neves and J.~F. Perez, Phys. Lett. {\bf 114A},  331  (1986).

\bibitem{aklt88}
I. Affleck, T. Kennedy, E.~H. Lieb, and H. Tasaki, Commun. Math. Phys. {\bf
  115},  477  (1988).

\bibitem{a73}
P. Anderson, Mater. Res. Bull. {\bf 8},  153,160  (1973).

\bibitem{fa74}
P. Fazekas and P. Anderson, Philos. Mag. {\bf 30},  423  (1974).

\bibitem{fsl01}
J.-B. Fouet, P. Sindzingre, and C. Lhuillier, Eur. Phys. J. B {\bf 20},  241
  (2001).

\bibitem{tr99}
P. Tomczak and J. Richter, Phys. Rev. B {\bf 59},  107  (1999).

\bibitem{tc90}
N. Trivedi and D. Ceperley, Phys. Rev. B {\bf 41},  4552  (1990).

\bibitem{web98}
C. Waldtmann {\it et~al.}, Eur. Phys. J. B {\bf 2},  501  (1998).

\bibitem{fmsl01}
J.-B. Fouet, M. Mambrini, P. Sindzingre, and C. Lhuillier, cond-mat/0108070
  (unpublished).

\bibitem{ukkll94}
Y. Uemura {\it et~al.}, Phys. Rev. Lett. {\bf 73},  3306  (1994).

\bibitem{gsvs89a}
M. Gross, E. Sanchez-Velasco, and E. Siggia, Phys. Rev. B {\bf 39},  2484
  (1989).

\bibitem{gsvs89b}
M. Gross, E. Sanchez-Velasco, and E. Siggia, Phys. Rev. B {\bf 40},  11328
  (1989).

\bibitem{nz89}
H. Neuberger and T. Ziman, Phys. Rev. B {\bf 39},  2608  (1989).

\bibitem{f89}
D. Fisher, Phys. Rev. B {\bf 39},  11783  (1989).

\bibitem{adm93}
P. Azaria, B. Delamotte, and D. Mouhanna, Phys. Rev. Lett. {\bf 70},  2483
  (1993).

\bibitem{hn93}
P. Hasenfratz and F. Niedermayer, Z. Phys. B. Condensed Matter {\bf 92},  91
  (1993).

\bibitem{d02}
J.~C. Domenge, rapport de DEA, Magistere des Sciences de la mati\`ere de Lyon,
  2002 (unpublished).

\bibitem{sf77}
A. S\"{u}t\"{o} and P. Fazekas, Phil. Mag. {\bf 35},  623  (1977).

\bibitem{lh95}
G. Levine and J. Howard, Phys. Rev. Lett. {\bf 75},  4142  (1995).

\bibitem{lblp95}
P. Lecheminant, B. Bernu, C. Lhuillier, and L. Pierre, Phys. Rev. B {\bf 52},
  6647  (1995).

\bibitem{f02}
J.-B. Fouet, Ph.D. thesis, Universit\'e Cergy Pontoise, 2002.

\bibitem{vbcc80}
J. Villain, R. Bidaux, J. Carton, and R. Conte, J. Phys. Fr. {\bf 41},  1263
  (1980).

\bibitem{s82}
E. Shender, Sov. Phys. J.E.T.P. {\bf 56},  178  (1982).

\bibitem{ont85}
T. Oguchi, H. Nishimori, and Y. Taguchi, J. Phys. Soc. Jpn. {\bf 54},  4494
  (1985).

\bibitem{h89}
C.~L. Henley, Phys. Rev. Lett. {\bf 62},  2056  (1989).

\bibitem{dm89}
E. Dagotto and A. Moreo, Phys. Rev. Lett. {\bf 63},  2148  (1989).

\bibitem{jdgb90}
T. Jolicoeur, E. Dagotto, E. Gagliano, and S. Bacci, Phys. Rev. B {\bf 42},
  4800  (1990).

\bibitem{cj92}
A. Chubukov and T. Jolicoeur, Phys. Rev. B {\bf 46},  11137  (1992).

\bibitem{k93}
S.~E. Korshunov, Phys. Rev. B {\bf 47},  6165  (1993).

\bibitem{de93}
R. Deutscher and H. Everts, Z. Phys. B. Condensed Matter {\bf 93},  77  (1993).

\bibitem{sz92}
H. Schultz and T. Ziman, Europhys. Lett. {\bf 8},  355  (1992).

\bibitem{zu96}
M.~E. Zhitomirsky and K. Ueda, Phys. Rev. B {\bf 54},  9007  (1996).

\bibitem{kosw99}
V.~N. Kotov, J. Oitmaa, O. Sushkov, and Z. Weihong, cond-mat/9912228
  (unpublished).

\bibitem{kosw99a}
V.~N. Kotov, J. Oitmaa, O. Sushkov, and Z. Weihong, Phys. Rev. B {\bf 60},
  14613  (1999).

\bibitem{cs00}
L. Capriotti and S. Sorella, Phys. Rev. Lett. {\bf 84},  3173  (2000).

\bibitem{cbps01}
L. Capriotti, F. Becca, A. Parola, and S. Sorella, Phys. Rev. Lett. {\bf 87},
  097201  (2001).

\bibitem{pc02}
S. Palmer and J. Chalker, Phys. Rev. B {\bf 64},  094412  (2002).

\bibitem{bh02}
W. Brenig and A. Honecker, Phys. Rev. B {\bf 65},  140407R  (2002),
  cond-mat/0111405.

\bibitem{baa02}
E. Berg, E. Altman, and A. Auerbach, cond-mat/0206384 (unpublished).

\bibitem{ss81a}
B. Shastry and B. Sutherland, Physica B (Amsterdam) {\bf 108},  1069  (1981).

\bibitem{mt00b}
T. Momoi and K. Totsuka, Phys. Rev. B {\bf 62},  15067  (2000).

\bibitem{ki94}
N. Katoh and M. Imada, J. Phys. Soc. Jpn. {\bf 63},  4529  (1994).

\bibitem{ttw94}
M. Troyer, H. Tsnunetsgu, and D. Wurte, Phys. Rev. B {\bf 50},  13515  (1994).

\bibitem{tnyk95}
S. Taniguchi {\it et~al.}, J. Phys. Soc. Jpn. {\bf 64},  2758  (1995).

\bibitem{fo96}
Y. Fukumoto and A. Oguchi, J. Phys. Soc. Jpn. {\bf 65},  1440  (1996).

\bibitem{am96a}
M. Albrecht and F. Mila, Europhys. Lett {\bf 34},  145  (1996).

\bibitem{am96b}
M. Albrecht and F. Mila, Phys. Rev. B {\bf 53},  2945  (1996).

\bibitem{my96}
T. Miyasaki and D. Yoshioka, J. Phys. Soc. Jpn. {\bf 65},  2370  (1996).

\bibitem{sr96}
S. Sachdev and N. Read, Phys. Rev. Lett. {\bf 77},  4800  (1996).

\bibitem{uksl96}
K. Ueda, H. Kontani, M. Sigrist, and P.~A. Lee, Phys. Rev. Lett. {\bf 76},
  1932  (1996).

\bibitem{khst96}
K. Kodama {\it et~al.}, J. Phys. Soc. Jpn. {\bf 65},  1941  (1996).

\bibitem{tku96}
M. Troyer, H. Kontani, and K. Ueda, Phys. Rev. Lett. {\bf 76},  3822  (1996).

\bibitem{khsk97}
K. Kodama {\it et~al.}, J. Phys. Soc. Jpn. {\bf 66},  28  (1997).

\bibitem{oyiu97}
T. Ohama, H. Yasuoka, M. Isobe, and Y. Ueda, J. Phys. Soc. Jpn. {\bf 66},  23
  (1997).

\bibitem{k99}
H. Kageyama {\it et~al.}, Phys. Rev. Lett. {\bf 82},  3168  (1999).

\bibitem{mu99}
S. Miyahara and K. Ueda, Phys. Rev. Lett. {\bf 82},  3701  (1999).

\bibitem{who99}
Z. Weihong, C.~J. Hamer, and J. Oitmaa, Phys. Rev. B {\bf 60},  6608  (19999).

\bibitem{ckrz99}
O. Cepas {\it et~al.}, J. Phys. Soc. Jpn. {\bf 68},  2906  (1999).

\bibitem{mu00}
S. Miyahara and K. Ueda, Phys. Rev. B {\bf 61},  3417  (2000).

\bibitem{kk00}
A. Koga and N. Kawakami, Phys. Rev. Lett. {\bf 84},  4461  (2000).

\bibitem{cms01}
C.~H. Chung, J.~B. Marston, and S. Sachdev, cond-mat/0102222 (unpublished).

\bibitem{msku00}
E. M\"uller-Hartmann, R.~R.~P. Singh, C. Knetter, and G.~S. Uhrig, Phys. Rev.
  Lett. {\bf 84},  1808  (2000).

\bibitem{ckrz01}
O. Cepas {\it et~al.}, Phys. Rev. Lett. {\bf 87},  167205  (2001).

\bibitem{rk88}
D. Rokhsar and S. Kivelson, Phys. Rev. Lett. {\bf 61},  2376  (1988).

\bibitem{cck89}
J. Chayes, L. Chayes, and S. Kivelson, Commun. Math. Physic. {\bf 123},  53
  (1989).

\bibitem{mlms02}
G. Misguich, C. Lhuillier, M. Mambrini, and P. Sindzingre, Eur. Phys. J. B {\bf
  26},  167  (2002), cond-mat/0112360.

\bibitem{ks88}
M. Kohmoto and Y. Shapir, Phys. Rev. B {\bf 37},  9439  (1988).

\bibitem{fs63}
M.~E. Fisher and J. Stephenson, Phys. Rev. {\bf 132},  1411  (1963).

\bibitem{s89}
S. Sachdev, Phys. Rev. B {\bf 40},  5204  (1989).

\bibitem{lcr96}
P. Leung, K. Chiu, and K. Runge, Phys. Rev. B {\bf 54},  12938  (1996).

\bibitem{f91}
E. Fradkin,  in {\em Field Theories of Condensed Matter Systems}, {\em
  Frontiers in Physics}, edited by D. Pines (Addison-Wesley, USA, 1991).

\bibitem{ms01}
R. Moessner and S.~L. Sondhi, Phys. Rev. Lett. {\bf 86},  1881  (2001).

\bibitem{am88}
I. Affleck and J. Marston, Phys. Rev. B {\bf 37},  3774  (1988).

\bibitem{rs89}
N. Read and S. Sachdev, Phys. Rev. Lett. {\bf 62},  1694  (1989).

\bibitem{lda88}
S. Liang, B. Doucot, and P. Anderson, Phys. Rev. Lett. {\bf 61},  365  (1988).

\bibitem{s88}
B. Sutherland, Phys. Rev. B {\bf 37},  3786  (1988).

\bibitem{s88a}
B. Sutherland, Phys. Rev. B {\bf 38},  7192  (1988).

\bibitem{rc89}
N. Read and B. Chakraborty, Phys. Rev. B {\bf 40},  7133  (1989).

\bibitem{cttt01}
R. Coldea, D.~A. Tennant, A.~M. Tsvelick, and Z. Tylczynski, Phys. Rev. Lett.
  {\bf 86},  1335  (2001).

\bibitem{s92}
S. Sachdev, Phys. Rev. B {\bf 45},  12377  (1992).

\bibitem{mblw98}
G. Misguich, B. Bernu, C. Lhuillier, and C. Waldtmann, Phys. Rev. Lett. {\bf
  81},  1098  (1998).

\bibitem{mlbw99}
G. Misguich, C. Lhuillier, B. Bernu, and C. Waldtmann, Phys. Rev. B {\bf 60},
  1064  (1999).

\bibitem{t65}
D. Thouless, Proc. Phys. Soc. {\bf 86},  893  (1965).

\bibitem{rhd83}
M. Roger, J. Hetherington, and J. Delrieu, Rev. Mod. Phys. {\bf 55},  1
  (1983).

\bibitem{h66}
C. Herring,  in {\em Magnetism}, edited by G.~T. Rado and H. Suhl (Academic
  Press, New York and London, 1966), p.\ vol IV.

\bibitem{rbbcg98}
M. Roger {\it et~al.}, Phys. Rev. Lett. {\bf 80},  1308  (1998).

\bibitem{collin01}
E. Collin {\it et~al.}, Phys. Rev. Lett. {\bf 86},  2447  (2001).

\bibitem{bcc01}
B. Bernu, L. Candido, and D. Ceperley, Phys. Rev. Lett. {\bf 86},  870  (2001).

\bibitem{chapfmcf01}
R. Coldea {\it et~al.}, Phys. Rev. Lett. {\bf 86},  5377  (2001).

\bibitem{kk01b}
A. Katanin and A. Kampf, cond-mat/0111533 (unpublished).

\bibitem{lst02}
A. L\"auchli, G. Schmid, and M. Troyer, cond-mat/0206153, and oral
  communication March Meeting 2002, Indianapolis (unpublished).

\bibitem{lmsl00}
W. LiMing, G. Misguich, P. Sindzingre, and C. Lhuillier, Phys. Rev. B {\bf 62},
   6372,6376  (2000).

\bibitem{m99}
G. Misguich, Ph.D. thesis, Universit\'e Pierre et Marie Curie. Paris. France,
  1999.

\bibitem{o00}
M. Oshikawa, Phys. Rev. Lett. {\bf 84},  1535  (2000).

\bibitem{k97}
A.~Y. Kitaev, quant-phys/9707021 (unpublished).

\bibitem{ifi02}
L.~B. Ioffe {\it et~al.}, Nature {\bf 415},  503  (2002), cond-mat/0111224.

\bibitem{k01}
S. Kivelson, cond-mat/0106126 (unpublished).

\bibitem{ns01}
C. Nayak and K. Shtengel, Phys. Rev. B {\bf 64},  064422  (2001).

\bibitem{ps02}
K. Park and S.Sachdev, Phys. Rev. B {\bf 65},  220405  (2002),
  cond-mat/0112003.

\bibitem{s01}
T. Senthil, cond-mat/0105104 and refs. therein (unpublished).

\bibitem{bfg02}
L. Balents, M.~P.~A. Fisher, and S.~M. Girvin, Phys. Rev. B {\bf 65},  224412
  (2002).

\bibitem{dnkks02}
E. Demler {\it et~al.}, Phys. Rev. B {\bf 65},  155103  (2002).

\bibitem{msf02}
R. Moessner, S.~L. Sondhi, and E. Fradkin, Phys. Rev. B {\bf 65},  024504
  (2002).

\bibitem{b89a}
N.~E. Bonesteel, Phys. Rev. B {\bf 40},  8954  (1989).

\bibitem{wen91}
X. Wen, Phys. Rev. B {\bf 44},  2664  (1991).

\bibitem{sp02}
S. Sachdev and K. Park, Annals of Physics (N.Y.) {\bf 58},  298  (2002),
  cond-mat/0108214.

\bibitem{h85}
F. Haldane, Phys. Rev. Lett. {\bf 55},  2095  (1985).

\bibitem{wn90}
X. Wen and Q. Niu, Phys. Rev. B {\bf 41},  9377  (1990).

\bibitem{ssgpt99}
G. Santoro {\it et~al.}, Phys. Rev. Lett. {\bf 83},  3065  (1999), we
  acknowledge a very interesting exchange on this subject in Trieste.

\bibitem{ms01b}
R. Moessner and S.~L. Sondhi, Phys. Rev. B {\bf 63},  224401  (2001).

\bibitem{k79}
J. Kogut, Rev. Mod. Phys. {\bf 51},  659  (1979).

\bibitem{lfs01}
C. Lannert, M.~P.~A. Fisher, and T. Senthil, Phys. Rev. B {\bf 63},  134510
  (2001).

\bibitem{sf01b}
T. Senthil and M.~P.~A. Fisher, Phys. Rev. B {\bf 63},  134521  (2001).

\bibitem{p38}
L. Pauling,  in {\em The nature of the chemical bond} (Cornell University
  Press, Ithaca, 1938).

\bibitem{kn53}
K. Kano and S. Naya, Progress in Theoretical physics {\bf 10},  158  (1953).

\bibitem{hr92}
D. Huse and A. Rutenberg, Phys. Rev. B {\bf 45},  7536  (1992).

\bibitem{msc00}
R. Moessner, S.~L. Sondhi, and P. Chandra, Phys. Rev. Lett. {\bf 84},  4457
  (2000).

\bibitem{chs92}
J. Chalker, P.~C.~W. Holdsworth, and E.~F. Shender, Phys. Rev. Lett. {\bf 68},
  855  (1992).

\bibitem{rcc93}
I. Richtey, P. Chandra, and P. Coleman, Phys. Rev. B {\bf 47},  15342  (1993).

\bibitem{e02}
M. Elhajal, Ph.D. thesis, Universit\'e Joseph Fourier. Grenoble. France, 2002.

\bibitem{ecl02}
M. Elhajal, B. Canals, and C. Lacroix, cond-mat/0202194 (unpublished).

\bibitem{e89}
V. Elser, Phys. Rev. Lett. {\bf 62},  2405  (1989).

\bibitem{ce92}
J. Chalker and J. Eastmond, Phys. Rev. B {\bf 46},  14201  (1992).

\bibitem{le93}
P. Leung and V. Elser, Phys. Rev. B {\bf 47},  5459  (1993).

\bibitem{ze95}
C. Zeng and V. Elser, Phys. Rev. B {\bf 51},  8318  (1995).

\bibitem{lblps97}
P. Lecheminant {\it et~al.}, Phys. Rev. B {\bf 56},  2521  (1997).

\bibitem{m98}
F. Mila, Phys. Rev. Lett. {\bf 81},  2356  (1998).

\bibitem{smlbpwe00}
P. Sindzingre {\it et~al.}, Phys. Rev. Lett. {\bf 84},  2953  (2000).

\bibitem{mm01}
M. Mambrini and F. Mila, Eur. Phys. J. B {\bf 17},  651,659  (2001).

\bibitem{lm01}
C. Lhuillier and G. Misguich,  in {\em High Magnetic Fields}, edited by C.
  Berthier, L. Levy, and G. Martinez (Springer, Berlin, 2002), pp.\ 161--190,
  cond-mat/0109146.

\bibitem{h01}
K. Hida, J. of the Phys. Soc. of Japan {\bf 70},  3673  (2001).

\bibitem{ey94}
N. Eltsner and A.~P. Young, Phys. Rev. B {\bf 50},  6871  (1994).

\bibitem{rhw00}
A.~P. Ramirez, B. Hessen, and M. Winkelmann, Phys. Rev. Lett. {\bf 84},  2957
  (2000).

\bibitem{mklmch00}
P. Mendels {\it et~al.}, Phys. Rev. Lett. {\bf 85},  3496  (2000).

\bibitem{mmpfa00}
T. Mondelli {\it et~al.}, Physica B {\bf 284},  1371  (2000).

\bibitem{gsf01}
A. Georges, R. Siddhartan, and S. Florens, Phys. Rev. Lett. {\bf 87},  277203
  (2001).

\bibitem{ls02}
C. Lhuillier and P. Sindzingre,  in {\em Quantum properties of Low dimensional
  antiferromagnets}, edited by Y. Ajiro and J.~P. Boucher (Kyushu University
  Press, Fukuoka, Japan, 2002), p.\ 111, iSBN 4 87378 740 8.

\bibitem{rec90}
A. Ramirez, G.~P. Espinosa, and A.~S. Cooper, Phys. Rev. Lett. {\bf 64},  2070
  (1990).

\bibitem{lbar96}
S.-H. Lee {\it et~al.}, Europhys. Lett {\bf 35},  127  (1996).

\bibitem{kklllwutdg96}
A. Keren {\it et~al.}, Phys. Rev. B {\bf 53},  6451  (1996).

\bibitem{whmmt98}
A.~S. Wills {\it et~al.}, Europhys. Lett {\bf 42},  325  (1998).

\bibitem{cl98}
B. Canals and C. Lacroix, Phys. Rev. Lett. {\bf 80},  2933  (1998).

\bibitem{t01}
H. Tsunetsgu, J. Phys. Soc. Jpn. {\bf 70},  640  (2001).

\bibitem{t02}
H. Tsunetsgu, Phys. Rev. B {\bf 65},  024415  (2002).

\bibitem{h00}
K. Hida, J. of the Phys. Soc. of Japan {\bf 69},  4003  (2000).

\end{thebibliography}

\end{document}